\documentclass[aps,prb,onecolumn,superscriptaddress,10pt]{revtex4-2}
\pdfoutput=1 
\usepackage[utf8]{inputenc}
\usepackage[english]{babel}
\usepackage[T1]{fontenc}
\usepackage[pdftex]{graphicx}
\usepackage{amsmath}
\usepackage{cancel}
\usepackage{amsthm}
\usepackage{algorithmic}
\usepackage{algorithm}
\usepackage{amssymb}
\usepackage{braket}
\usepackage{xcolor}
\usepackage{dcolumn}
\usepackage{bbm}
\usepackage{bbold}
\usepackage[normalem]{ulem}
\usepackage{footmisc}
\usepackage{xcolor}
\usepackage{booktabs}
\usepackage{comment}
\usepackage{multirow}
\usepackage[colorlinks=true,linkcolor=blue,citecolor=blue,urlcolor=blue,unicode]{hyperref}

\usepackage{amsmath}
\usepackage{bm}
\usepackage{graphicx}
\usepackage{subfigure}
\usepackage{geometry}
\usepackage{ulem}
\usepackage{indentfirst}
\usepackage{amsfonts,amssymb,dsfont}
\usepackage{accents}

\usepackage{setspace}
\usepackage{threeparttable}
\usepackage{amsmath,mathtools,amsthm}
\usepackage{array}
\usepackage{extarrows}
\usepackage{upgreek}
\usepackage{orcidlink}

\makeatletter
\renewcommand*\env@matrix[1][\arraystretch]{%
	\edef\arraystretch{#1}%
	\hskip -\arraycolsep
	\let\@ifnextchar\new@ifnextchar
	\array{*\c@MaxMatrixCols c}}
\makeatother

\usepackage{orcidlink}

\begin{document}
		\title{Non-Equilibrium Steady States and Emergent Stable Dephased Nonlinearity in a Driven-Dissipative Bose-Hubbard Chain}		
		\author{Chen-Huan Wu 
			\orcidlink{0000-0003-1020-5977} }
		\thanks{chenhuanwu1@gmail.com}
		\affiliation{College of Physics and Electronic Engineering, Northwest Normal University, Lanzhou 730070, China}
		

		%

		\begin{small}
		
	\begin{abstract}
		We investigate the non-equilibrium dynamics and steady-state properties of a driven-dissipative Bose-Hubbard chain using a self-consistent Gutzwiller mean-field (GMF) approach. By employing a robust Picard iteration scheme, we solve the non-linear master equation for the non-equilibrium steady state (NESS) in the presence of strong Kerr nonlinearity. While identification of quasilinear and chaotic regimes is common, we systematically focus on the intermediate stable nonlinear regime. The defining characteristics of this phase are the restoration of $U(1)$ symmetry in the steady-state phase space and an underdamped relaxation process toward the NESS. By calculating the operator out-of-time-order correlator (OTOC), we demonstrate that in this stable regime, the initial information scrambling is eventually overtaken by dissipative damping, causing the OTOC to decay exponentially to zero at long times. This behavior stands in stark contrast to the chaotic regime, where the OTOC exhibits persistent exponential growth and macroscopic saturation. Our results clarify the role of nonlinearity in stabilizing non-equilibrium phases and offer an efficient numerical framework to explore many-body correlations in larger photonic lattices.
	\end{abstract}
	
	\maketitle
	
	\section{Introduction}
	Driven-dissipative quantum many-body systems have emerged as a versatile platform for exploring novel phases of matter that have no equilibrium counterparts \cite{Schwingel, Ferrari}. Among these, the driven-dissipative Bose-Hubbard model, realizable in superconducting circuit QED \cite{Sliwa, Kamal} and coupled cavity arrays, serves as a paradigm for studying the competition between coherent drive, particle loss, and on-site nonlinearity \cite{Xie, Garcia}. Unlike closed systems governed by thermodynamics, these open systems settle into a NESS determined by the balance of gain and loss \cite{Uzdin, Tripathy}. A central open question in this field is characterizing the transition from regular, mean-field-like dynamics to complex behavior where quantum correlations and nonlinear instabilities play a dominant role \cite{Solanki, Mostafazadeh}.
	
	In this work, we address this problem by analyzing a one-dimensional Bose-Hubbard chain under coherent driving. We develop a numerical approach based on the Gutzwiller Mean-Field approximation, utilizing a Picard iteration technique to rigorously solve the non-linear self-consistency equations for the NESS. This method allows us to treat the on-site Kerr nonlinearity exactly, capturing high-order local correlations $\langle \hat{n} \hat{a} \rangle$ that are crucial for determining the stability of the steady state \cite{Wu, Polkovnikov}. By deriving the effective Liouvillian governing the fluctuations around the NESS, we identify an intermediate stable nonlinear regime. Unlike the quasilinear limit, this regime is characterized by a significant restoration of $U(1)$ symmetry, where nonlinear phase scrambling washes out the coherent influence of the drive.
	A significant feature for this regime is the slow deterministic transient drift toward the fixed point attractor in the complex phase space.
	Furthermore, we observe an underdamped relaxation toward the NESS, which can be interpreted as a transient limit-cycle behavior that eventually collapses into a stable fixed-point attractor \cite{Das, Chen}.
	
	To explicitly diagnose the dynamical nature of this phase and compare it with the chaotic regime, we compute the out-of-time-order correlator (OTOC), a hallmark of quantum information scrambling \cite{Patel, Braumüller}. While the chaotic regime is defined by exponential instability and OTOC saturation \cite{Zhou, Xu}, we show that the stable nonlinear regime exhibits a rapid short-time increase due to ballistic operator spreading \cite{Kim, Gopalakrishnan} followed by a definitive decay to zero. This decay signifies that dissipative damping eventually dominates the dynamics, erasing the memory of initial perturbations \cite{Zaburdaev}. 
	
	We specifically focus on a three-site Bose-Hubbard chain, which serves as a minimal but sufficient lattice architecture to distinguish between local driving, bulk transport, and terminal dissipation. This three-site configuration is strategically chosen as a critical threshold: while a two-site system (dimer) often lacks the intermediate bulk necessary to witness full phase-decoupling \cite{Garbin}, a larger lattice significantly increases the effective degrees of freedom, potentially triggering global many-body chaos that masks the subtle features of the stable dephased nonlinear phase. 
While dissipation tends to drive the system toward a stable attractor (fixed point) and induce phase-locking (similar to $F$), the nonlinearity $U$ triggers an amplitude-dependent Kerr shift. This shift leads to a phase-unlocked limit cycle, characterized by symmetry-induced phase degeneracy and a randomized, uniform phase distribution.
	Thus $U$ potentially increase the sensitivity to  initial conditions including the random initial phases which do not guarantee a uniform phase distribution ($U(1)$ rotational symmetry) by itself. But the absence of chaos implies the exponential sensitivity to perturbations is obscured by the dissipation and the small system size (see Appendix.A).
In semiclassical framework each trajectory is evolved by a deterministic ordinary differential equation and follows the nonlinear drift dictated by its own local field.
Despite $U$ enhance the sensitivity to initial conditions by converting amplitude fluctuations into phase randomization,
in the stable nonlinear regime, this sensitivity is transient in terms of operator growth. 
This also results in a transient uniform ensemble distribution due to the random initial phases.
The dynamical sensitivity (OTOC) eventually vanishes as the system's contractive Liouvillian spectrum ensures that all trajectories are damped by dissipation.
	By utilizing this model, we characterize how nonlinearity and dissipation compete to stabilize the system into a symmetric steady state, providing insights into the control of quantum coherence in open systems \cite{Messinger, Drummond}.

\section{model}
				We consider the system involving the interaction between local boson and nonlocal boson, which reads
		\begin{equation} 
			\begin{aligned}
				&H_{b}=\sum_{r=1}^{L} \left( \Delta \hat{a}^\dagger_r \hat{a}_r + \frac{1}{2} U \hat{a}^\dagger_r \hat{a}^\dagger_r \hat{a}_r \hat{a}_r \right) - \sum_{r=1}^{L-1} J \left(\hat{a}^\dagger_{l+1} \hat{a}_r + \hat{a}^\dagger_r \hat{a}_{l+1}\right) + F(\hat{a}^\dagger_1 + \hat{a}_1)
			\end{aligned}
		\end{equation}
		where 
			$\Delta$ is the pump-to-resonator detuning  between a drive frequency and the resonator's natural frequency ($\Delta = \omega_d - \omega_0$).
			$\sum_{r=1}^{L} \frac{1}{2} U \hat{a}^\dagger_r \hat{a}^\dagger_r \hat{a}_r \hat{a}_r = \frac{U}{2} \sum_{r=1}^{L} \hat{n}_r (\hat{n}_r - 1)$ is the term describe the on-site interaction.
			$U$ is the strength of the onsite Kerr nonlinearity.
		$J$	is the hopping amplitude.
	$F$ is the strength (amplitude) of coherent drive, which describes an external driving field that coherently pumps energy into the chain by creating ($\hat{a}^\dagger_1$) and destroying ($\hat{a}_1$) bosons.
	For $F>0$ and $\gamma>0$, the drive continuously pumps energy in and the loss continuously drains them. 
	The system settles into a stable state with a net flow of particles, which is the definition of a non-equilibrium (and non-thermal) state. As a result, it reach a
	unique NESS at long-time limit.
	
	\section{Gutzwiller mean-field approach}
	
		Using the Gutzwiller mean-field decoupling,
		we decompose the bath boson operator into the mean field part and fluctuational part,
		$a_{l}=\psi_{l}(t)+\delta a_{l}(t)$, where $\psi_r(t) = \langle \hat{b}_r \rangle$ is the classical complex condensate amplitude, and $\delta a_{l}(t)$ is the quantum fluctuation operator.
		Then the hopping interaction term becomes $\hat{a}_r^\dagger \hat{a}_{r'} = (\psi_r^* + \delta \hat{a}_r^\dagger) (\psi_{r'} + \delta \hat{a}_{r'})
= \psi_r^* \psi_{r'} + \psi_r^* \delta \hat{a}_{r'} + \delta \hat{a}_r^\dagger \psi_{r'} + \delta \hat{a}_r^\dagger \delta \hat{a}_{r'}$ where the quantum fluctuations described in last term is neglected in mean-field approximation.

For any site $r$ that is not the impurity site, the dynamics are governed by a single-site master equation
		\begin{equation} 
	\begin{aligned}
		\frac{d\hat{\rho}_r}{dt} = -i[H_{\text{MF}, r}, \hat{\rho}_r] + \mathcal{D}_r[\hat{\rho}_r]
					\end{aligned}
	\end{equation}
	The mean-field Hamiltonian for this site is
\begin{equation} 
\begin{aligned}
&	H_{\text{MF}, r} = \Delta \hat{a}_r^\dagger \hat{a}_r + \frac{U}{2} \hat{a}_r^\dagger \hat{a}_r^\dagger \hat{a}_r \hat{a}_r 
-J \left(\hat{a}^\dagger_r \hat{a}_{r-1} + \hat{a}^\dagger_{r-1} \hat{a}_r\right) - J \left(\hat{a}^\dagger_{r+1} \hat{a}_r + \hat{a}^\dagger_r \hat{a}_{r+1}\right)\\
&= \Delta \hat{a}_r^\dagger \hat{a}_r + \frac{U}{2} \hat{a}_r^\dagger \hat{a}_r^\dagger \hat{a}_r \hat{a}_r
-J \left[ \left( \Psi_{r-1} + \Psi_{r+1} \right) \hat{a}_r^\dagger + \left( \Psi_{r-1}^* + \Psi_{r+1}^* \right) \hat{a}_r \right]
\end{aligned}
\end{equation}
where we treat the neighbors ($r-1$ and $r+1$) as environment.
This is the single-site Hamiltonian where the quantum operators $\hat{a}_l$ and $\hat{a}_l^\dagger$ interact with a classical, external field $\left(\Psi_{l-1} + \Psi_{l+1}\right)$ that represents the average influence of all its neighbors.
where the operator $\hat{a}_l^\dagger$ is coupled to the classical field $\Psi_{l-1} + \Psi_{l+1}$.

In Gutzwiller approximation, 
the effective single-site Hamiltonian for a "bath" site $r(\neq j)$
\begin{equation} 
	\begin{aligned}
H_{\text{MF}, r}= &\Delta \hat{a}_r^\dagger \hat{a}_r + \frac{U}{2} \hat{a}_r^\dagger \hat{a}_r^\dagger \hat{a}_r \hat{a}_r\\
&
-J \left[ \left((1-\delta_{r,1}) \Psi_{r-1} + (1-\delta_{r,L})\Psi_{r+1} \right) \hat{a}_r^\dagger + \left( (1-\delta_{r,1})\Psi_{r-1}^* 
+ (1-\delta_{r,L})\Psi_{r+1}^* \right) \hat{a}_r \right]+F(a_{1}^{\dag}+a_{1})\delta_{r,1}.
	\end{aligned}
\end{equation}

The Gutzwiller mean-field approximation simplifies this by factorizing the total state. It assumes there are no quantum correlations (entanglement) between different sites. The total density matrix is approximated as a simple product of independent, single-site density matrices, $\hat{\rho}_{\text{total}}(t) \approx \hat{\rho}_1(t) \otimes \hat{\rho}_2(t) \otimes \dots \otimes \hat{\rho}_j(t) \otimes \dots \otimes \hat{\rho}_L(t)$.
Note that the vacuum state $|0\rangle_r$ is an eigenstate of the local Hamiltonian $H_{loc,r}:=\Delta \hat{a}_r^\dagger \hat{a}_r + \frac{U}{2} \hat{a}_r^\dagger \hat{a}_r^\dagger \hat{a}_r \hat{a}_r$ with also an eigenvalue of 0,
$|0\rangle_r$ is also an eigenstate of $a_r^\dagger a_r$ with an eigenvalue of 0.
Thus we have $H_{loc, r} |0\rangle_r=a_r |0\rangle_{r} = a_r^\dagger a_r |0\rangle_{r} = 0$,
$ a_r^\dagger a_r |n\rangle_{r} = n|n\rangle_{r}$ ($n\ge 0$),
$a_r |0\rangle_r'=|0\rangle_r'$ for $r\neq r'$.

Through a perturbative expansion of the density matrix evolution (detailed in Appendix A), we can characterize the initial spreading of the condensate amplitudes (classical complex amplitude of the bosonic field $\Psi_r(t) = \langle \hat{a}_r \rangle = \text{Tr}[\hat{a}_r \hat{\rho}_r(t)]$) across the chain. We summarize the resulting amplitudes for the first few time steps in Table I. Notably, the leading-order term for the $N$-th site exhibits a scaling of $\Psi_{N}(Ndt) \sim O((dt)^N)$. The coefficients of these leading terms grow polynomially, reflecting the sequential activation of each site as the coherent drive propagates from site 1. Keeping only the leading terms, we obtain
$\Psi_{1}(Ndt) = -iNFdt$, $\Psi_{2}(Ndt) = \frac{N(N-1)}{2}JF(dt)^2$, $\Psi_{3}(Ndt) = \frac{N(N-1)(N-2)}{6}iJ^2 F(dt)^3$.
Thus site $N$ requires at least $N$ cumulative actions of hopping ($J$) or driving ($F$) to develop a non-zero expectation value. This perturbative regime sets the stage for the full non-linear self-consistent iteration discussed in Sec. IV.

$\Psi_{3}(Ndt)=\frac{N(N-1)(N-2)}{6}iJ^2 F(dt)^3$.
	
	\begin{table}[h!]
		\centering
		\caption{Perturbative Expansion of Mean-Field Amplitudes $\Psi_l(N \cdot dt)$}
		\resizebox{\textwidth}{!}{
			\begin{tabular}{|l|l|l|l|}
				\hline
				\textbf{Time Step (N)} & \textbf{$\Psi_1(N \cdot dt)$} & \textbf{$\Psi_2(N \cdot dt)$} & \textbf{$\Psi_3(N \cdot dt)$} \\
				\hline
				\textbf{1} & $-iFdt$ & $0$ & $0$ \\
				\hline
				\textbf{2} & $-2iFdt - F\Delta(dt)^2$ & $JF(dt)^2$ & $0$ \\
				\hline
				\textbf{3} & \begin{tabular}[t]{@{}l@{}}
					$-3iFdt + (-3F\Delta + iF\gamma_1)(dt)^2$ \\
					$+ \left( i(J^2F + F\Delta^2 + 6F^3) + \frac{\gamma_1 F\Delta}{2} \right) (dt)^3$
				\end{tabular}
				& \begin{tabular}[t]{@{}l@{}}
					$3JF(dt)^2$ \\
					$+ \left(-2iJF\Delta - \frac{\gamma_2 JF}{2}\right) (dt)^3$
				\end{tabular}
				& $iJ^2F(dt)^3$ \\
				\hline
				\textbf{4} & \begin{tabular}[t]{@{}l@{}}
					$-4iFdt + (-6F\Delta + i\frac{5\gamma_1 F}{2})(dt)^2$ \\
					$+ \left(i(4J^2F + 4F\Delta^2 + 12F^3) + 3\gamma_1 F\Delta -\frac{i\gamma_{1}^2 F}{2}\right)(dt)^3$
				\end{tabular}
				& \begin{tabular}[t]{@{}l@{}}
					$6JF(dt)^2$ \\
					$+ \left(-8iJF\Delta - JF\gamma_1 - 2\gamma_2 JF\right) (dt)^3$
				\end{tabular}
				& $4iJ^2F(dt)^3$ \\
				\hline
			\end{tabular}%
		}
	\end{table}

The NESS requires $i[H_{MF,l}, \rho_l^{SS}] = \mathcal{D}_l[\rho_l^{SS}]$,
where the left-hand-side is the net coherent change in particle population at site $l$ and the right-hand-side is the incoherent loss of particles from site $l$ into the external environment (the bath).
Thus it is a balance between the effects of $\Psi_{l-1}$ and $\Psi_{l+1}$ which represent the particles hopping forward from $l-1$ to $l$ and
backwards from $l+1$ to $l$, respectively.

The Lindbladian dynamics of operator $a_{r}$ reads
\begin{equation} 
	\begin{aligned}
	&	\frac{d}{dt}\langle a_{r} \rangle = \text{Tr}\left(a_{r} \frac{d\rho}{dt} \right) = \text{Tr}\left( a_{r} \mathcal{L}(\rho) \right) 
		=\text{Tr}\left( a_{r}
\left( -i[\hat{H}, \hat{\rho}] + \sum_k \left( L_k \hat{\rho} L_k^\dagger - \frac{1}{2} \{ L_k^\dagger L_k, \hat{\rho} \} \right) \mathcal{D}(\hat{\rho})\right) \right) \\
&= \text{Tr}\left( \left( i[\hat{H},  a_{r}] + \sum_k \left( L_k^\dagger  a_{r} L_k - \frac{1}{2} \{ L_k^\dagger L_k,  a_{r} \} \right) \right) \hat{\rho} \right)
= \text{Tr}(\mathcal{L}^\dagger(a_{r}) \rho)
		= \left\langle i[H, a_{r}] + \mathcal{D}^{\dag}( a_{r}) \right\rangle\\
		&=\left\langle i \left( -\Delta \hat{a}_l - U \hat{n}_l \hat{a}_l\right)-iJ(-a_{r+1}-a_{r-1})
		-iF\delta_{1,r}
		-\frac{\gamma_l}{2} \hat{a}_l\right\rangle,
			\end{aligned}
	\end{equation}
	where we use the cyclic property of the trace
	$\text{Tr}(XYZ) = \text{Tr}(ZXY) = \text{Tr}(YZX)$ and the following results
\begin{equation} 
	\begin{aligned}
&i[\hat{H}_{\text{diag}}, a_r] = i \left[ \sum_{k} \left( \Delta n_k + \frac{U}{2} n_k (n_k - 1) \right), a_r \right]
= i \left[ \Delta n_r + \frac{U}{2} (n_r^2 - n_r), a_r \right]\\
&= i \left( \Delta [\hat{n}_l, \hat{a}_l] + \frac{U}{2} [\hat{n}_l^2, \hat{a}_l] - \frac{U}{2} [\hat{n}_l, \hat{a}_l] \right)
= i \left( \Delta (-\hat{a}_l) + \frac{U}{2} (-2\hat{n}_l \hat{a}_l - \hat{a}_l) - \frac{U}{2} (-\hat{a}_l) \right)\\
&= i \left( -\Delta \hat{a}_l - U \hat{n}_l \hat{a}_l - \frac{U}{2}\hat{a}_l + \frac{U}{2}\hat{a}_l \right) 
= i \left( -\Delta \hat{a}_l - U \hat{n}_l \hat{a}_l\right),\\
&i[\hat{H}_{\text{hop}}, a_r] = i \left[ -J  (\hat{a}^\dagger_{r+1} \hat{a}_r + \hat{a}_r^\dagger \hat{a}_{r+1})
 -J  (\hat{a}^\dagger_{r-1} \hat{a}_r + \hat{a}_r^\dagger \hat{a}_{r-1}), \hat{a}_r \right]
\\& =-iJ(-a_{r+1}-a_{r-1}),\\
&i[\hat{H}_{\text{drive}}, \hat{a}_l] = i \left[ F(\hat{a}^\dagger_1 + \hat{a}_1), \hat{a}_l \right] \delta_{l,1}=-iF\delta_{1,r},\\
&\mathcal{D}^{\dag}(a_r) = \gamma_l \hat{a}_l^\dagger \hat{a}_l \hat{a}_l - \frac{\gamma_l}{2} \{\hat{a}_l^\dagger \hat{a}_l, \hat{a}_l\}
= \gamma_l \hat{n}_l \hat{a}_l - \frac{\gamma_l}{2} (\hat{n}_l \hat{a}_l + \hat{a}_l \hat{n}_l)\\
&= \gamma_l \hat{n}_l \hat{a}_l - \frac{\gamma_l}{2} (\hat{n}_l \hat{a}_l + (\hat{n}_l + 1)\hat{a}_l)
= \gamma_l \hat{n}_l \hat{a}_l - \frac{\gamma_l}{2} (2\hat{n}_l \hat{a}_l + \hat{a}_l) = -\frac{\gamma_l}{2} \hat{a}_l.
	\end{aligned}
\end{equation}
Then for NESS $\frac{d}{dt}\langle a_{r} \rangle_{ss} = 0$ we can further obtain 
\begin{equation} 
	\begin{aligned}
		\langle \hat{a}_l \rangle \left( -i\Delta - \frac{\gamma_l}{2} \right) - iU \langle \hat{n}_l \hat{a}_l \rangle + iJ (\langle \hat{a}_{l-1} \rangle + \langle \hat{a}_{l+1} \rangle) = i F \delta_{l,1}
	\end{aligned}
\end{equation}
or equivalently,
\begin{equation} 
	\begin{aligned}
&\left( -i \Delta - \frac{\gamma_l}{2} \right) \Psi_l^{SS} - i U \langle \hat{n}_l \hat{a}_l \rangle_{ss}+i J (\Psi_{l-1}^{SS} + \Psi_{l+1}^{SS}) =  i F \delta_{l,1}.
	\end{aligned}
\end{equation}
where $\langle \hat{n}_l \hat{a}_l \rangle_{SS}$ is itself a function of $\Psi_{l\pm 1}^{SS}$.
Thus for the drive, bulk, and drain sites, the NESS equations are
\begin{equation} 
	\begin{aligned}
&-\left( i\Delta + \frac{\gamma_1}{2} \right) \Psi_1^{SS} - iU \langle \hat{n}_1 \hat{a}_1 \rangle_{SS} + iJ \Psi_2^{SS} = i F,\\
&-\left( i\Delta + \frac{\gamma_2}{2} \right) \Psi_2^{SS} - iU \langle \hat{n}_2 \hat{a}_2 \rangle_{SS} + iJ (\Psi_1^{SS} + \Psi_3^{SS}) = 0,\\
&-\left( i\Delta + \frac{\gamma_3}{2} \right) \Psi_3^{SS} - iU \langle \hat{n}_3 \hat{a}_3 \rangle_{SS} + iJ \Psi_2^{SS} = 0
	\end{aligned}
\end{equation}
The NESS is a highly excited state with $\langle n_r \rangle \sim 1$, and satisfies, $\frac{d\Psi_r^{ss}}{dt}=\text{Tr}\left[a_r  \mathcal{L}_r(\rho_r^{ss})\right] = 0$.

The analytical solutions are available only for $U=0$
(quadratic models),
which read
\begin{equation} 
	\begin{aligned}
&	\Psi_2^{SS} = \frac{JF \Lambda_3}{\Lambda_1 \Lambda_2 \Lambda_3 + J^2 \Lambda_3 + J^2 \Lambda_1},\\
&	\Psi_1^{SS} = \frac{i F}{\Lambda_1} - \frac{iJ}{\Lambda_1} \Psi_2^{SS},\\
&	\Psi_3^{SS} = \frac{-iJ}{\Lambda_3} \Psi_2^{SS},
\label{26}
	\end{aligned}
\end{equation}
where $\Lambda_{l}=-( i\Delta + \frac{\gamma_l}{2}) $.
The system becomes a driven-dissipative Harmonic oscillator when $U=0$, in which case $\hat{\rho}_l^{SS} = |\Psi_l^{SS}\rangle \langle \Psi_l^{SS}|$ is a coherent state and
the local nonlinear correlation $\langle \hat{n}_l \hat{a}_l \rangle_{SS} = \langle \hat{a}_l^\dagger \hat{a}_l \hat{a}_l \rangle_{SS}
= \langle \hat{a}_l^\dagger \rangle_{SS} \langle \hat{a}_l \rangle_{SS} \langle \hat{a}_l \rangle_{SS} = (\Psi_l^{SS})^* (\Psi_l^{SS})^2$ can be calculated as
\begin{equation} 
	\begin{aligned}
	\langle \hat{n}_l \hat{a}_l \rangle
	 & = \text{Tr}[\hat{\rho}_l \hat{a}_l^\dagger \hat{a}_l \hat{a}_l]= \sum_{n, m} \langle m | (\hat{a}_l^\dagger \hat{a}_l \hat{a}_l) | n \rangle \langle n | \hat{\rho}_l | m \rangle \\
	&= \sum_{n, m} c_{n, m} \langle m | (\hat{a}_l^\dagger \hat{a}_l \hat{a}_l) | n \rangle
= \sum_{n=1}^{N_{\max}} c_{n-1, n} \cdot (n-1)\sqrt{n}
	\end{aligned}
\end{equation}
where we use $\hat{a}_l^\dagger \hat{a}_l \hat{a}_l |n\rangle = (n-1)\sqrt{n} |n-1\rangle$.
Note that the complex number $\Psi_l^{SS} = \langle \hat{a}_l \rangle_{SS}$ describes the amplitude and phase of boson field, with $\langle \hat{n}_l \rangle \approx |\Psi_l^{SS}|^2$ at small $U$ (as shown in the TWA discussed below), and $\arg(\Psi_l^{SS})$ determines the coherent coupling between sites.

\section{Numerical Solution for NESS in the Non-Linear Regime ($U \neq 0$)}

For nonzero on-site Kerr interaction $U\neq 0$, and the nonlinear correlation is
$\hat{\rho}_l^{SS} = \sum_{n, m=0}^{N_{\text{max}}} c_{n, m} |n\rangle_l \langle m|_l$
and $\langle \hat{n}_l \hat{a}_l \rangle = \text{Tr}[\hat{\rho}_l \hat{a}_l^\dagger \hat{a}_l \hat{a}_l] 
 = \sum_{n} c_{n,n-1} \langle n-1 | (\hat{a}_l^\dagger \hat{a}_l \hat{a}_l) | n \rangle
=\sum_{n} c_{n, n-1} \cdot \sqrt{n} \cdot (n-1)$
where $ c_{n, n-1} =\langle n | \rho_r | n-1 \rangle$ and we use $\hat{a}_l^\dagger \hat{a}_l \hat{a}_l |n\rangle = \sqrt{n} (n-1) |n-1\rangle$.
The amplitudes of NESS can be solved by the above-introduced Gutzwiller mean-field method.
The steady-state mean-field amplitudes $\Psi_l^{\text{SS}}$ of the driven-dissipative Bose-Hubbard chain are determined self-consistently using the Gutzwiller mean-field (GMF) approximation.

The effective local Hamiltonian $\hat{H}_l^{\text{eff}}$ for site $l$ depends on the mean-field amplitudes $\Psi_{l\pm 1}$ of its neighbors,
\begin{equation} 
	\begin{aligned}
\hat{H}_l^{\text{eff}}(\Psi) = \hat{H}_l^{\text{loc}} + \hat{H}_l^{\text{drive}} + \hat{H}_l^{\text{MF}}(\Psi),
\label{28}
	\end{aligned}
\end{equation}
where the hopping term is linearized to the mean-field drive term $\hat{H}_l^{\text{MF}}(\Psi) = -J \left[ (\Psi_{l-1} + \Psi_{l+1}) \hat{a}_l^\dagger + 
(\Psi_{l-1}^* + \Psi_{l+1}^*) \hat{a}_l
 \right]$, $H_{l}^{{\rm drive}}=F\delta_{l,1}(a^{\dag}_{l}+a_{l})$.
 Here $\hat{H}_l^{\text{eff}}(\Psi)$ is a $N\times N$ matrix with $N=N_{max}+1$ and $N_{max}$ is the truncated number of photons.
For our three-site model, $N_{max}=3$, $N=4$, thus the single-site density $\hat{\rho}_l$ is $N\times N$ matrix, and the dimension of total Hilbert space ($\rho_{tot}$) is $N^{N_{max}}=64$.
The local steady-state density matrix $\hat{\rho}_l^{\text{SS}}$ is the solution to the time-independent master equation,
\begin{equation} 
	\begin{aligned}
	\mathcal{L}_l(\hat{\rho}_l^{\text{SS}}) = -i [\hat{H}_l^{\text{eff}}(\Psi^{(k)}), \hat{\rho}_l^{\text{SS}}] + \gamma_l \left( \hat{a}_l \hat{\rho}_l^{\text{SS}} \hat{a}_l^\dagger - \frac{1}{2} \{ \hat{a}_l^\dagger \hat{a}_l, \hat{\rho}_l^{\text{SS}} \} \right) = 0.
	\label{29}
	\end{aligned}
\end{equation}
Using column-stacking Roth's Lemma ($\text{vec}(ABC) = (C^T \otimes A)\text{vec}(B)$), the Liouvillian reads
\begin{equation} 
	\begin{aligned}
		\mathbf{L}\hat{\rho}_l^{\text{SS}} 
=-i(I \otimes \hat{H}_l^{\text{eff}}(\Psi^{(k)}) - [\hat{H}_l^{\text{eff}}(\Psi^{(k)})]^T \otimes I)
+\gamma_{l} (\hat{a}_l^* \otimes \hat{a}_l - \frac{1}{2}(I \otimes \hat{a}_l^\dagger \hat{a}_l + (\hat{a}_l^\dagger \hat{a}_l)^T \otimes I)).
	\end{aligned}
\end{equation}
The self-consistency condition is enforced by requiring the output amplitude $\Psi_l^{\text{new}}$ derived from the solution must equal the input amplitude $\Psi_l^{\text{old}}$ (appears in Eqs.\ref{28},\ref{29}),
$\Psi_l^{\text{new}} = \langle a_{l}\rangle_{SS}=\text{Tr}[\hat{a}_l \hat{\rho}_l^{\text{SS}}(\Psi_l^{\text{old}})]$,
and the nonlinear effect from $U$ is contained within the $\langle n_{r} \hat{a}_l \rangle = \text{Tr}\left[ n_{r} \hat{a}_l \hat{\rho}_l^{\text{SS}} \right]$
as well as $\hat{\rho}_l^{\text{SS}}$ solved from Eq.\ref{29}.
The algorithm iteratively using the Picard method
is shown in ALGORITHM 1, where we repeatedly applying the nonlinear mapping function $\mathcal{F}(\Psi)$: $\Psi_{\text{new}} = \mathcal{F}(\Psi_{\text{old}})$, until convergence in NESS.
Instead of solving the analytical expression of $\langle \hat{n}_l \hat{a}_l \rangle$, 
the effect of $\langle \hat{n}_l \hat{a}_l \rangle$ is captured by the matrix structure of $\Psi_{l\pm 1}^{\text{SS}}$.
The Picard algorithm solve the $\hat{\rho}_l^{\text{SS}}$ under the effects of $U$ and $\Psi_{l\pm 1}^{\text{SS}}$.
Here the coefficients of operator $\mathbf{L}$ depend on the solution $\hat{\rho}_{l\pm 1}$ (or $\Psi_{l\pm 1}$), thus $	\mathbf{L}(\hat{\rho}_{l\pm 1})\hat{\rho}_l^{\text{SS}}=0$ is linear equation once variable of $\mathbf{L}$ is fixed by the previous iteration. Picard iteration is required to linearize the problem temporarily by fixing a guess for the neighbor fields' amplitude.

Unlike simple mean-field theories that approximate $\langle \hat{n} \hat{a} \rangle \approx \langle \hat{n} \rangle \langle \hat{a} \rangle$, GMF handles the non-linearity ($U \neq 0$) locally 
by solving $\mathcal{L}_l(\hat{\rho}_l) = 0$.
The non-linearity is moved from the operators (Liouvillian) to the iteration loop, i.e., the nonlinear mapping $\Psi_{\text{old}} \to \Psi_{\text{new}}$ as $\hat{\rho}_l$ depends on $\Psi_{\text{old}}$ in a complex and non-proportional way (e.g., through the photon blockade).

\begin{algorithm}[H]
	\caption{Self-Consistent GMF NESS Solver (Picard Iteration)}
	\begin{algorithmic}[1]
		\STATE \textbf{Input:} Physical parameters $(J, U, \Delta, F, \{\gamma_l\})$; Convergence tolerance $\epsilon$; Max Iterations $M$.
		\STATE \textbf{Initialize:} Initial guess $\Psi_{\text{old}}^{(0)} = \{\Psi_1, \Psi_2, \Psi_3\}$; Iteration counter $k \gets 0$.
		\WHILE{$k < M$}
		\STATE \textbf{Compute New Amplitudes $\Psi_{\text{new}} \leftarrow \mathcal{F}(\Psi_{\text{old}})$}:
		\FOR{each site $l = 1$ to $L$}
		\STATE $\hat{H}_l^{\text{eff}} \gets$ Compute $\hat{H}_l^{\text{eff}}$ using $\Psi_{l\pm 1}^{\text{old}}$.
		\STATE $\mathbf{L}_l \gets$ Construct the vectorized Liouvillian matrix for $\hat{H}_l^{\text{eff}}$.
		\STATE $\vec{\rho}_l^{\text{SS}} \gets$ Solve $\mathbf{L}_l \vec{\rho}_l = 0$ subject to $\text{Tr}(\hat{\rho}_l) = 1$ (which becomes $\mathbf{L}'_l \vec{\rho}_l\neq 0 $ under trace trick).
		\STATE $\Psi_l^{\text{new}} \gets$ Compute $\text{Tr}[\hat{a}_l \hat{\rho}_l^{\text{SS}}]$.
		\ENDFOR
		
		\STATE \textbf{Check Convergence:} Compute Residual $R \gets ||\Psi_{\text{new}} - \Psi_{\text{old}}||$.
		\IF{$R < \epsilon$}
		\STATE \textbf{break} (Solution $\Psi_{\text{SS}} = \Psi_{\text{new}}$ Found)
		\ENDIF
		
		\STATE \textbf{Update:} $\Psi_{\text{old}} \gets \Psi_{\text{new}}$; $k \gets k+1$.
		\ENDWHILE
	\end{algorithmic}
\end{algorithm}

The trace conservation requires $\frac{d}{dt}\text{Tr}(\hat{\rho}) = \text{Tr}\left( \frac{d\hat{\rho}}{dt} \right) = \langle\langle \mathbb{1} | \frac{d}{dt} \rho \rangle\rangle = \langle\langle \mathbb{1} | \mathbf{L} | \rho \rangle\rangle = 0$,
where we use the Hilbert-Schmidt inner product of the form $\text{Tr}(\hat{A}) = \text{Tr}(\hat{\mathbb{1}}^\dagger \hat{A}) \equiv \langle\langle \mathbb{1} | A \rangle\rangle$ with the vectorized density $| A \rangle\rangle$ (according to Roth's Lemma).
Thus $\langle\langle \mathbb{1} | \mathbf{L} = \mathbf{0}^T$,
$\det(\mathbf{L}) = \prod \lambda_i = 0$, and consequently $\mathbf{L}$ is a singular matrix.
Meanwhile the singularity of $\mathbf{L}$ guarantees that $\mathbf{L} | \rho_{ss} \rangle\rangle = 0$ has nontrivial solution (nonzero steady state).
Further, since $\frac{d}{dt}\text{Tr}(\hat{\rho}) =
\langle\langle\mathbb{1}|\frac{d}{dt}|\rho\rangle\rangle=\langle\langle\mathbb{1}|(\mathbf{L}|\rho\rangle\rangle)=(\langle\langle\mathbb{1}|\mathbf{L})|\rho\rangle\rangle=0$,
where the vector $\langle\langle\mathbb{1}|$ is equivalents to $|\rho\rangle\rangle^{T}$ but replace the diagonal element $\rho_{ii}$ by 1 and replace the off-diagonal element $\rho_{ij}$ by 0.
Then we have $\langle\langle\mathbb{1}| \mathbf{L} = \vec{0}^T$,
which means that for each column of $\mathbf{L}$,
sum of elements corresponding to the positions of nonzero elements of 
$\langle\langle\mathbb{1}|$ must be zero, such that the sum of the corresponding rows of $\mathbf{L}$ is zero (and thus those rows are linearly dependent):
\begin{equation} 
	\begin{aligned}
L_{1,j}+L_{N+2,j}+L_{2N+3,j}+\cdots+L_{N(N-1)+N,j}=0,
\label{1}
	\end{aligned}
\end{equation}
 with $N$ the dimension of matrix $\rho$.
This is equivalents to the trace conservation condition
\begin{equation} 
	\begin{aligned}
\frac{d}{dt}\text{Tr}(\rho) =\sum_{i} \frac{d\rho_{ii}}{dt}=\sum_{i} \frac{d|\rho\rangle\rangle_{i}}{dt}=\sum_{i}(\mathbf{L}|\rho\rangle\rangle)_{i}=\sum_{i} \sum_{j=1}^{N^2} L_{i, j} |\rho\rangle\rangle_{j}
= \sum_{j=1}^{N^2} |\rho\rangle\rangle_j \left( \sum_{i} L_{i,j} \right)=0,
 	\end{aligned}
\end{equation}
where here $i=1,N+2,2N+3,\dots,N(N-1)+N$ corresponds to the position of the diagonal elements.
 Multiple with $|\rho\rangle\rangle$ in Eq.\ref{1},
 we can again obtain
 $\frac{d\rho_{11}}{dt} + \dots  \frac{d\rho_{NN}}{dt} = 0$.
For singular matrix $\mathbf{L}$,
$\mathbf{L}\vec{\rho} = \mathbf{0}$ describes a homogeneous linear system, thus 
introducing the condition $\text{Tr}(\hat{\rho}) = 1$
is necessary to find the sole solution.
This can be realized by replacing the first row of $\mathbf{L}$ by the vector satisfying 
$\langle\langle \mathbb{1}|\rho\rangle\rangle=
1 \cdot \rho_{11} + 0 \cdot \rho_{12} + \dots + 1 \cdot \rho_{22} + \dots = \sum_i \rho_{ii} ={\rm Tr}\rho=1$
(guarantees the normalization condition), while for other rows the product with $|\rho\rangle\rangle$ remains zero.
The modification of first row of $\mathbf{L}$ realize 
\begin{equation}
	\frac{d|\rho\rangle\rangle}{dt} = \mathbf{L} |\rho\rangle\rangle=
\begin{pmatrix}
	\frac{d\rho_{11}}{dt} \\
	\frac{d\rho_{12}}{dt} \\
	\vdots \\
	\frac{d\rho_{NN}}{dt}
\end{pmatrix}
\rightarrow
\mathbf{L}' |\rho\rangle\rangle =
\begin{pmatrix}
	\sum_{i} \rho_{ii} \\
	\frac{d\rho_{12}}{dt} \\
	\vdots \\
	\frac{d\rho_{NN}}{dt}
\end{pmatrix}=
\begin{pmatrix}
1\\
	\frac{d\rho_{12}}{dt} \\
	\vdots \\
	\frac{d\rho_{NN}}{dt}
\end{pmatrix}=
\begin{pmatrix}
1\\
0\\
\vdots \\
0
\end{pmatrix}.
\end{equation}
For full rank invertible (nonsinglular) matrix $\mathbf{L}'$, the non-homogeneous linear system
$\mathbf{L}' |\rho\rangle\rangle$ has sole solution:
$\mathbf{L}' |\rho\rangle\rangle=0$ has only one solution $|\rho\rangle\rangle=0$,
$\mathbf{L}' |\rho\rangle\rangle=b\neq 0$ has only one solution $|\rho\rangle\rangle=\mathbf{L}^{'-1}b$.
For singlular matrix $\mathbf{L}$,
the homogeneous linear system
$\mathbf{L} |\rho\rangle\rangle$ has infinite solutions:
$\mathbf{L} |\rho\rangle\rangle=0$ has a solution $|\rho\rangle\rangle=0$, and infinite other nonzero solutions $c|\rho\rangle\rangle_{ss}$ ($c$ is arbitary constant), and all these solutions form the so-called null space or kernel.
$\mathbf{L} |\rho\rangle\rangle=b\neq 0$ has infinite nonzero solutions if $\vec{b}$ lies in the column space of $\mathbf{L}$.
Thus for homogeneous linear system
$\mathbf{L} |\rho\rangle\rangle$,
if solutions $|\rho_{1}\rangle\rangle=c|\rho_{1}\rangle\rangle$,
then ${\rm Tr}\rho_{1}=c{\rm Tr}\rho_{2}$.
Note that $\mathbf{L}$ in our system (or other ergodic open quantum system) always has rank $(N^2-1)$ which means the dimension of nullspace is 1 (this dimension will be larger than 1 when there are dark states or symmetry-protected/decoherence-free subspaces), corresponding to the number of unique steady state.
Further, for singlular matrix $\mathbf{L}$,
the nonzero solution for $\mathbf{L} |\rho\rangle\rangle$ could have arbitary trace ${\rm Tr}\rho_{ss}$ as long as $||\rho_{ss}\rangle\rangle|^2=1$.
${\rm Tr}\rho_{ss}$ could even be zero when there 
are dark states or symmetry-protected/decoherence-free subspaces (like the case $J=0$).
For $\mathbf{L}$ with rank $(N^2-1)$,
there exists one vector $\langle\langle\mathbb{1}|$ satisfying $\langle\langle\mathbb{1}| \mathbf{L} = \vec{0}^T$,
for special $\mathbf{L}$ with rank $(N^2-k)$,
there are $k$ linearly independent steady states,
i.e., expect vector $\langle\langle\mathbb{1}|$, there are additionally $k-1$ vector $v$ satisfying $v \mathbf{L} = \vec{0}^T$,
and the final state depends on initial conditions.
 In this case, there exist $k$ linearly independent left eigenvectors with eigenvalue 0, representing $k$ independent conservation laws (one being the total trace, others being e.g. population in decoupled subspaces).

For the special $\mathbf{L}$ with rank $(N^2-k)$ ($k>1$), there is a $k$-dimensional nullspace spanned by $d$ linearly independent basis vectors.
We consider the conservation law of a observable $O$ 
\begin{equation}
\frac{d}{dt} \langle \hat{O} \rangle = \text{Tr}\left( \hat{O} \frac{d\hat{\rho}}{dt} \right) = \text{Tr}\left( \hat{O} \mathcal{L}(\hat{\rho}) \right)
= \text{Tr}(\hat{O}^\dagger \mathcal{L}(\hat{\rho})) = \langle\langle O | \mathcal{L}(\hat{\rho}) \rangle\rangle
= \langle\langle O | \left( \mathbf{L} |\rho\rangle\rangle \right)
= \left( \langle\langle O | \mathbf{L} \right) |\rho\rangle\rangle
= \vec{v} \mathbf{L} |\rho\rangle\rangle,
\end{equation}
where we using the rule of inner product, $\langle\langle A | B \rangle\rangle \equiv \text{Tr}(\hat{A}^\dagger \hat{B})$.
This contains the both the Schrödinger picture ($\mathbf{L} |\rho\rangle\rangle$) and Heisenberg picture ($\langle\langle O| \mathbf{L}$), where $\langle\langle O| \mathbf{L} = 0$ when $O$ is a conserved quantity.
For system has two decoupled subspaces $A$ and $B$, probability is conserved within $A$ and within $B$ separately. The projectors onto these subspaces can be written as $\vec{P}_A$ and $\vec{P}_B$.
Then $\langle\langle 1_{Tr}| = \vec{P}_A + \vec{P}_B$ for $\vec{P}_A$ and $\vec{P}_B$ corresponding to independent conservation laws,
$	\vec{P}_A \mathbf{L} = \vec{0} \quad \text{and} \quad \vec{P}_B \mathbf{L} = \vec{0}$, such that $\frac{d}{dt} P_A = \text{Tr}(P_A \mathcal{L}\hat{\rho}) = \vec{P}_A \mathbf{L} |\rho\rangle\rangle = 0$ and $\frac{d}{dt} P_B = \text{Tr}(P_B \mathcal{L}\hat{\rho}) = \vec{P}_B \mathbf{L} |\rho\rangle\rangle = 0$ as long as no population flows between subspace $A$ and subspace $B$.
The steady state in this case is not unique. The final state depends on the initial condition (i.e., how much probability was initially in the two subspace).


In GMF approximation, the total density matrix reads
$	\hat{\rho}_{\text{tot}}^{\text{GMF}} = \hat{\rho}_1 \otimes \hat{\rho}_2 \otimes \dots \otimes \hat{\rho}_L$.
The interaction between site $l$ and its neighbor $l+1$ is replaced by an effective field
$	-J \hat{a}_l^\dagger \hat{a}_{l+1} \approx -J \hat{a}_l^\dagger \langle \hat{a}_{l+1} \rangle$,
where $\langle \hat{a}_{l+1} \rangle = \text{Tr}(\hat{a}_{l+1} \hat{\rho}_{l+1})$ depends on the density matrix $\hat{\rho}_{l+1}$ solved self-consistency loop $\mathbf{L}(\hat{\rho}) \cdot \hat{\rho} = 0$.
Thus the hopping term coefficient $J\Psi_{l+1}=J{\rm Tr}[\hat{a}_{l+1} \hat{\rho}_{l+1}]$ depends on $\hat{\rho}_{l+1}$.
Even though $U, J, F, \gamma$ are constant, the coefficients of the hopping operator in the matrix $\mathbf{L}$ depend on the solution $\hat{\rho}$ of the adjacent sites, i.e., to solve for $\rho_l$, $\Psi_{l+1}$ comes from $\rho_{l+1}$ is needed,
and vise-vase.
Picard iteration breaks this circle by fixing $\rho$ from the previous step to calculate the hopping coefficients for the current step.


For a comparasion, we also apply the exact solver use the global basis space with size $N^{N_{max}}$,
where the NESS solver constructs the Liouvillian in the full $(N^{N_{max}})^2$ space. 
The interaction term $\frac{U}{2} \hat{n}(\hat{n}-1)$ is represented as a static $(N^{N_{max}}) \times (N^{N_{max}})$ matrix operator.
The hopping term $-J \hat{a}_l^\dagger \hat{a}_{l+1}$ is also a static $(N^{N_{max}}) \times (N^{N_{max}})$ matrix operator.
In the master equation
\begin{equation}
	\frac{d\hat{\rho}_{\text{tot}}}{dt} = -i[\hat{H}_{\text{tot}}, \hat{\rho}_{\text{tot}}] + \sum \mathcal{D}[\hat{L}](\hat{\rho}_{\text{tot}}),
\end{equation}
with constant coefficients $U, J, F, \gamma$ that do not depend on the value of $\hat{\rho}_{\text{tot}}$. 
Thus, the equation is a standard linear system
	$\mathbf{L}_{\text{tot}} |\rho_{\text{tot}}\rangle\rangle = 0$.
Despite the non-linearity $U$, the density matrix equation itself is linear in the state space. Therefore the NESS can be solved without iteration.

\section{OTOC}

\subsection{local OTOC}

In the Gutzwiller mean-field framework linearized around the NESS, this OTOC is related to the squared magnitude of the retarded Green's function (propagator) governing the fluctuations $\delta \hat{a}$
\begin{equation}
	D_{1,L}(t) = |G^R_{L,1}(t)|^2 = \left| \left( e^{\mathbf{L}_{\text{tot}} t} \right)_{L,1} \right|^2,
\end{equation}
where 
the retarded Green function reads $G^R_{L,1}(t)=\text{Tr}\{ \hat{a}_L e^{\mathcal{L}t} [\hat{a}_1^\dagger, \rho_{SS}] \}$.
To find the change in $\langle \hat{a}_L \rangle$ due to a perturbation at site 1, we consider the vectorized Master Equation $\frac{d}{dt} |\rho\rangle\rangle = \mathbf{L}_{tot} |\rho\rangle\rangle$.
A perturbation at site 1 at $t=0$ creates a deviation
$\delta \rho(0) = [\hat{a}_1^\dagger, \hat{\rho}_{SS}]$
such that the instantaneous state reads  $\hat{\rho}(0^+) = \hat{\rho}_{SS} + \epsilon \delta \hat{\rho}(0)$.
Then the linear response reads
\begin{equation}
	\begin{aligned}
	&\delta \langle \hat{a}_L(t) \rangle = \text{Tr} \{ \hat{a}_L e^{\mathcal{L}t} \delta \hat{\rho}(0) \} = \langle\langle a_L | e^{\mathbf{L}_{tot}t} \delta| \rho(0) \rangle\rangle,\\
	 \end{aligned}
\end{equation}
	The sensitivity is the projection of the evolved deviation onto the target operator
\begin{equation}
	\begin{aligned}
& D_{1,L}(t) = \left| \langle\langle a_L | e^{\mathbf{L}_{tot}t} \delta|\rho(0)\rangle\rangle \right|^2
	 = \left| \text{Tr} \{ \hat{a}_L e^{\mathcal{L}t} [\hat{a}_1^\dagger, \hat{\rho}_{SS}] \} \right|^2 
	 \end{aligned}
\end{equation}
$\mathbf{L}_{\text{tot}}$ is the collective effective Liouvillian ($(N^{N_{max}})^2\times (N^{N_{max}})^2$) for the full system reads
\begin{equation}
	\mathbf{L}_{\text{tot}} = -i ( \mathbb{I} \otimes \hat{H}_{tot} - \hat{H}_{tot}^T \otimes \mathbb{I} ) + \sum_l \gamma_l \left( \hat{a}_l^* \otimes \hat{a}_l - \frac{1}{2} \mathbf{I} \otimes \hat{a}_l^\dagger \hat{a}_l - \frac{1}{2} \hat{a}_l^T \hat{a}_l^* \otimes \mathbf{I} \right).
\end{equation}
where $\hat{a}_l(0) = \hat{I}_1 \otimes \hat{I}_2 \otimes \dots \otimes \hat{a}_l \otimes \dots \otimes \hat{I}_L$ with $\hat{I}$ the $4\times 4$ identity matrix.
 In the expression $|\hat{a}_1(t)\rangle\rangle = e^{\mathbf{L}_{Heis} t} |\hat{a}_1(0)\rangle\rangle$, the matrix exponential acts as a rotation operator in the 4096-dimensional basis space. For example, the non-zero off-diagonal blocks generated by $J$ allow the weight of the operator $\hat{a}_1$ to leak into the indices corresponding to $\hat{a}_2$, $\hat{a}_{3}$, $\hat{n}_1\hat{a}_2$, etc., simulating the spatial spread of information.
 Eventually, the evolved operator $\hat{a}_i(t)$ has non-zero components across almost all $D^2$ basis operators.
 Here in the Lindblad framework, the adjoint Liouvillian $\mathcal{L}^\dagger$
 \begin{equation}
 	\frac{d}{dt} \hat{a}_l(t) = \mathcal{L}^\dagger \hat{a}_l(t) \implies \frac{d}{dt} |\hat{a}_l(t)\rangle\rangle = \mathbf{L}_{Heis} |\hat{a}_l(t)\rangle\rangle
 \end{equation}
$L_{Heis}$ accounts for the global system ($(N^L)^2 \times (N^L)^2 = 4096^2$). It contains the hopping $J$ as internal coupling elements, allowing site 1 information to leak into site 3.
While the local one $\mathbf{L}_{local}$ ($N^2 \times N^2 = 16^2$ matrix) replaces hopping with a static drive, severing the dynamic causal link between sites.

For a given site $l$, using the equivalence between the Schrödinger and Heisenberg pictures
$\text{Tr}(\hat{A} e^{\mathcal{L}t}\hat{\rho}) = \text{Tr}((e^{\mathcal{L}^{\dag}t} \hat{A}) \hat{\rho})$, the evolution of the vectorized operator $|\hat{a}_l(\tau)\rangle\rangle$ in Liouville space is
\begin{equation}
\begin{aligned}
&	\frac{d}{dt} |\hat{a}_l(t)\rangle\rangle = \mathbf{L}_{local}^\dagger |\hat{a}_l(t)\rangle\rangle ,\\ 
&|\hat{a}_l(t)\rangle\rangle = e^{\mathbf{L}_{local}^\dagger \tau} |\hat{a}_l(0)\rangle\rangle
\end{aligned}
\end{equation}
where $\mathbf{L}_{local}^{\dag}$ is the local Heisenberg-Liouvillian,
$\mathbf{L}_{local}$ is the 
local Schrodinger Liouvillian ($N^2 \times N^2$) is constructed using the Kronecker product of local operators $\hat{H}_l^{\text{eff}}(\Psi)$ and $\hat{a}$
\begin{equation}
	\mathbf{L}_{local} = -i ( \mathbb{I} \otimes \hat{H}_l^{\text{eff}}(\Psi) - (\hat{H}_l^{\text{eff}}(\Psi) )^T \otimes \mathbf{I} ) + \gamma \left( \hat{a}_{l}^* \otimes \hat{a}_{l} - \frac{1}{2} \mathbf{I} \otimes \hat{a}_{l}^\dagger \hat{a}_{l} - \frac{1}{2} \hat{a}_{l}^T \hat{a}_{l}^* \otimes \mathbf{I} \right)
\end{equation}
where we use the conventions $(A \otimes B) \text{vec}(X) = \text{vec}(A X B^T)$,
$\text{vec}(AXB) = (A \otimes B^T) \text{vec}(X)$
$L^* \rho L^T = (L \rho L^\dagger)^T$,
$\rho H^T - H^T \rho = -(H \rho - \rho H)^T$.
In GMF approximation, $\hat{H}_l^{\text{eff}}(\Psi)$
(Eq.(\ref{28})) contains the static fields $\Psi_{l\pm 1}$ from neighbors
where $\Psi_{l \pm 1} = \langle \hat{a}_{l \pm 1} \rangle_{SS}$ are fixed complex numbers obtained from Picard iteration, effectively treating the hopping as a constant local drive.
The OTOC at site $l$ is calculated using the commutator evaluated in the local Hilbert space
\begin{equation}
	D_l(t) = \text{Tr} \left\{ \hat{\rho}_{SS,l} \left[ \hat{a}_l(t), \hat{a}_l^\dagger(0) \right]^\dagger \left[ \hat{a}_l(t), \hat{a}_l^\dagger(0) \right] \right\}
\end{equation}
where $\hat{a}_l(t)$ is reconstructed from the above vectorized evolution. This single-site OTOC represents the local decoherence and sensitivity to local perturbations under a static mean-field environment.
At $t=0$, the cross term 
$- \hat{a}_i(0) \hat{a}_i^\dagger(t) \hat{a}_i^\dagger(0) \hat{a}_i(t) 
- \hat{a}_i^\dagger(t) \hat{a}_i(0) \hat{a}_i(t) \hat{a}_i^\dagger(0)
 =-2\langle \hat{n}_i^2 + \hat{n}_i \rangle_{SS}$ is at its maximal magnitude with perfect phase correlation. For $t>0$, chaotic dynamics smear the phase. The cross term decays from its initial value toward zero.
 The incoherent terms at $t=0$ reads
 $ \langle \hat{a}_i(0) \hat{a}_i^{\dag}(t) \hat{a}_i(t) \hat{a}_i^\dagger(0) 
 + \hat{a}_i^\dagger(t) \hat{a}_i(0) \hat{a}_i^\dagger(0) \hat{a}_i(t) \rangle=
 \langle 2\hat{n}^2 + 2\hat{n} + 1 \rangle_{SS}$.
 Thus in an infinite-dimensional Hilbert space, the commutation relation $[\hat{a}, \hat{a}^\dagger] = \hat{I}$ holds exactly
 and the single site OTOC reads $\mathbf{D}_{ii}(0) = \text{Tr}\{ \hat{\rho}_{SS;i} [\hat{a}_i, \hat{a}_i^\dagger]^\dagger [\hat{a}_i, \hat{a}_i^\dagger] \} = \text{Tr}\{ \hat{\rho}_{SS;i} \cdot \hat{I} \} = 1$.
While for simulation shown in Fig.\ref{otoc},
we consider a truncated Fock basis of dimension $N=N_{max}+1=4$, thus $\hat{a}^\dagger |N-1\rangle = 0$.
$\hat{a}^\dagger \hat{a} 
= \sum_{m,n} n \delta_{m,n} |m\rangle\langle n| 
= \sum_{n=0}^{N-1} n|n\rangle\langle n|= \text{diag}(0, 1, 2, \dots, N-1)$,
$\hat{a} \hat{a}^\dagger
=\sum_{n=0}^{N-2} (n+1)|n\rangle\langle n|
=\text{diag}(1, 2, \dots, N-1, 0)$.
The commutator becomes $[\hat{a}, \hat{a}^\dagger] = \text{diag}(1, 1, \dots, 1, -(N-1))$.
The expectation value at $t=0$ is
\begin{equation}
\mathbf{D}_{ii}(0)
=\text{Tr}\{\hat{\rho}_{SS;i} [\hat{a}_i, \hat{a}_i^\dagger]^\dagger [\hat{a}_i, \hat{a}_i^\dagger]\} 
= \sum_{n=0}^{N-1} \langle n| \hat{\rho}_{SS;i} [\hat{a}_i, \hat{a}_i^\dagger]^\dagger [\hat{a}_i, \hat{a}_i^\dagger] |n\rangle
 = \sum_{n=0}^{N-2} P_n  + P_{N-1} (N-1)^2 
 = (1 - P_{N-1}) + (N-1)^2 P_{N-1}
 = 1 + 8 P_{3},
\label{45}
\end{equation}
where $P_{n}
=\langle n| \hat{\rho}_{SS;i} |n\rangle
={\rm Tr}[\hat{\rho}_{SS;i}|n \rangle\langle n|]=\langle n| \text{Tr}_{j \neq i} [\hat{\rho}_{SS}] |n \rangle$, and thus $P_{N-1}$ is the population of the highest Fock state $|N-1\rangle=|3\rangle$ at site i. 
The identity OTOC can be recovered by increase $N_{max}$ until $P_{N_{max}} \to 0$.

Fig.\ref{otoc} show the OTOC as a function of time in unit of inverse hopping amplitude ($1/J$;
such that the value $t=1$ corresponds to the characteristic time required for a particle to tunnel between neighboring sites).
The OTOC does not saturate smoothly but exhibits large-amplitude oscillations due to the strong coherent interactions.
The wave packet reflects off the boundaries, causing recurrences. Unlike the Drain site, the OTOC for the Drive and Bulk sites does not decay to zero. Instead, it fluctuates around a finite value ($O(1)$ to $O(3)$) at long times ($t=5$).
Saturation to a finite value at long time signaling significant dephasing.
The high peak in the bulk site suggests strong parametric amplification of fluctuations due to nonlinearity before dissipation takes over.
The fact that the signal remains non-zero suggests the system is in the chaotic regime, where operators $\hat{a}(t)$ and $\hat{a}^\dagger(0)$ fail to commute over long periods, signifying sustained scrambling despite the presence of dissipation.
The competition between unitary scrambling (driven by nonlinearity $U$) and non-unitary dissipation (driven by $\gamma$).


In GMF solver,
each site $l$ is solved using a local Liouvillian $\mathbf{L}_l$ where neighbors are treated as constant $c$-numbers ($\Psi_{l\pm 1}$) and the perturbation never spread along the chain.
As shown in Fig.\ref{otoc}(a),
$D_{ii}(0) \neq 0$
and $[\hat{a}, \hat{a}^\dagger]=\text{diag}(1, 1, 1, -3)$. 
The squared commutator is $\text{diag}(1, 1, 1, 9)$. 
The OTOC at site 3 decays to zero at long time, implying that the information of initial perturbation is lost to the environment quickly due to the dissipation.
Sites 1 and 2 have much lower dissipation ($\gamma = 0.1$) and the damped nonlinear oscillations represent the effects of coherent tunneling ($J$) and the nonlinear energy shifts ($U$). 
The OTOC tracks how information of initial $a^\dagger$ operator is scrambled across the local Hilbert space before dissipation eventually takes over.
For stable (nonchaotic) NESS, the Heisenberg operator $\hat{a}(t)$ eventually loses its operator character and converges to a $c$-number steady-state value $\langle a \rangle_{ss}$, such that
 $[\langle a \rangle_{ss} \mathbb{1}, \hat{a}^\dagger] = 0$.
Any initial perturbation from $\hat{a}_1^\dagger$ is eventually dampened by the dissipation
and the sensitivity $\partial a(t) / \partial a(0)\sim e^{-\gamma t/2}$ vanishes as $t \to \infty$. This is in contrast with the chaotic regime where the OTOC would saturate at a high value ($\sim 2n^2$) rather than returning to zero.

For exact solver as shown in Fig.\ref{otoc}(b),
using Eq.\ref{45}, for $F=2.5$,
$D_{1,1}(0) \approx 1.9$ implies for site 1 $P_{N-1}\approx (D_{1,1}(0)-1)/8=0.11$, reflecting the local quantum fluctuations.
Here the deviation for $D_{1,1}(0)$ from 1 is due to the strong drive which push the occupation toward the cutoff value.
Sites 2 and 3 start at 0 because $[\hat{a}_{2,3}(0), \hat{a}_1^\dagger(0)] = 0$ due to spatial commutation relations. 
The subsequent rise $D_{2,2},D_{3,3}\sim (Jt)^2$ reflects the hopping-induced ballistic or perturbative operator spreading due to the information scrambling\cite{Kim,Gopalakrishnan,Zaburdaev}.
The eventual decay to zero is the signature of a stable NESS, where local dissipation removes the phase memory of the initial perturbation.
At short times the Hamiltonian (coherent) part of the dynamics dominates over the dissipators. Even in a stable system, the local operator $\hat{a}_1(0)$ begins to spread via tunneling $J$.
As $\hat{a}_1(t)$ evolves, it becomes a multi-body operator 
with nonzero  commutator $[\hat{a}_l(t), \hat{a}_j(0)]$.
The OTOC increases as the operator grows in complexity and spatial extent. 

The system enters into the nonlinear regime for lower dissipation ($J, U \gg \gamma$) and higher $F$ to pushes the system into highly excited states where the nonlinearity $U$ becomes more dominant.
In linear regime the decay of OTOC 
is dominated by dissipative damping (loss channel) and finite Liouvillian gap.
is the global decoherence where particles and information are removed by the dissipators, and the perturbation propagates like a coherent wave through the chain and flows out of the drain site. 
For chaotic regime the decay cause is the local decoherence (scrambling) where the information is hidden in high-order multi-site correlations.
 The nonlinearity $U$ causes trajectories to diverge exponentially. The perturbation is randomized across the entire Hilbert space. 
In the intermediate dephased nonlinear phase, there is dissipation-induced thermalization
(local phase randomization due to the dissipation and interaction-induced dephasing)
and the perturbation is suppressed by dissipative damping before it can evolve into many-body complexity. 
OTOC decays toward zero in a stable NESS and
 operator spreading is followed by strong interaction-induced dephasing and phase diffusion.

The exponential growth of OTOC in chaotic regime should related to the dynamical instability 
where the quantum memory is hidden by the complexity of many-body degrees of freedom 
instead of dissipation, i.e., $D(\tau) \sim e^{\lambda_L \tau}$ (where $\lambda_L$ is the quantum Lyapunov exponent) before saturating to a finite value $O(1)$ (independent of $L$) determined by the local Hilbert space dimension, and the information exponentially populating the entire Hilbert space.
The saturation should be $D(t \to \infty) \sim 2\langle n \rangle^2$ for our bosonic chain where the memory of the initial local perturbation is randomized (scrambled) across all degrees of freedom of Hilbert space and hidden within many-body correlations.
While in strongly dissipative Lindblad dynamics with a stable NESS, the gapped Liouvillian spectrum is contractive (negative real part) without positive Lyapunov exponent, where the OTOC exhibit a rapid short-time increase due to operator spreading.

The inset in right panel of Fig.\ref{otoc} displays $\log D_{1,2}$ and $\log D_{1,3}$. This further prove that the system is in a stable (non-chaotic) NESS regime rather than a chaotic one.
Also, the inset shows that after the initial peak (driven by ballistic operator spreading), the curves exhibit a linear downward slope on a logarithmic scale. This indicates an exponential decay toward zero ($D(t) \to 0$ as $t \to \infty$), which is the hallmark of a stable attractor where local dissipation ($\Gamma$) dominates the long-time dynamics.
Contractive Liouvillian: This behavior confirms that the Liouvillian spectrum is gapped and contractive, lacking a positive Lyapunov exponent. Any initial perturbation is lost into the environment by global decoherence before it can evolve into many-body complexity.

\subsection{nonlocal OTOC}
For nonlocal case, the commutator $[\hat{a}_i(t), \hat{a}_j^\dagger(0)]$ measures how much a perturbation applied to site $j$ initially fails to commute with a measurement performed at site $i$ at $t$. 
In linear and stable regime, the growth of OTOC is bounded or oscillatory and the OTOC may exhibit a rapid short-time increase due to operator spreading.
Thus chaos is defined as the exponential instability of trajectories within the NESS manifold where
the exponential growth of OTOCs is commonly associated with dynamical instability and quantum chaos in closed or weakly dissipative systems.

For non-Hermitian operator $[a_{i}(t), a^{\dag}_{j}(0)]$, the operator OTOC can be written as
\begin{equation} 
	\begin{aligned}
&D_{ij}(t) 
= \text{Tr}\left[ \hat{\rho}_{SS} [a_{i}(t), a^{\dag}_{j}(0)]^{\dag} [a_{i}(t), a^{\dag}_{j}(0)] \right]
= \langle [\hat{a}_i(t), \hat{a}_j^\dagger(0)]^\dagger [\hat{a}_i(t), \hat{a}_j^\dagger(0)] \rangle_{SS}\\
&= \langle \hat{a}_j(0) \hat{a}_i^{\dag}(t) \hat{a}_i(t) \hat{a}_j^\dagger(0) 
- \hat{a}_j(0) \hat{a}_i^\dagger(t) \hat{a}_j^\dagger(0) \hat{a}_i(t) 
- \hat{a}_i^\dagger(t) \hat{a}_j(0) \hat{a}_i(t) \hat{a}_j^\dagger(0) 
+ \hat{a}_i^\dagger(t) \hat{a}_j(0) \hat{a}_j^\dagger(0) \hat{a}_i(t) \rangle\\
&=2\langle \hat{a}_i^\dagger(t) \hat{a}_i(t) \rangle
(\langle \hat{a}_j^\dagger(0) \hat{a}_j(0) \rangle +1)- 2\text{Re}(\langle \hat{a}_i^\dagger(t) \hat{a}_j(0) \hat{a}_i(t) \hat{a}_j^\dagger(0) \rangle)
	\end{aligned}
\end{equation}
where in GMF framework the incoherent classical correlation terms are $\langle \hat{a}_j(0) \hat{a}_i^{\dag}(t) \hat{a}_i(t) \hat{a}_j^\dagger(0) \rangle
=\langle \hat{a}_j(0) \hat{a}_j^\dagger(0) \rangle
\langle \hat{a}_i^{\dag}(t) \hat{a}_i(t) \rangle
=(\langle n_{j}\rangle+1)\langle n_{i}\rangle$,
 $\langle\hat{a}_i^\dagger(t) \hat{a}_j(0) \hat{a}_j^\dagger(0) \hat{a}_i(t) \rangle
 =\langle\hat{a}_i^\dagger(t) \hat{a}_i(t) \rangle
 \langle \hat{a}_j(0) \hat{a}_j^\dagger(0) \rangle
 =\langle n_{i}\rangle(\langle n_{j}\rangle+1)$.
In a fully scrambled system, the phase information is randomized such that the coherent cross-terms (transport of non-local quantum interference) vanish asymptotically, leaving only the incoherent density-density (classical) correlations that saturate at $D_{ij} \approx 2 \langle n_i \rangle_{SS} (\langle n_j \rangle_{SS} + 1)$ in a fully scrambled state and inaccessible for a local observer. 
While in stable nonlinear regime where the system undergoes an underdamped convergence toward a stable fixed-point attractor (see Sec.VII),
$\hat{a}_i(t) \approx \alpha_i e^{-i\omega t} \hat{I}$,
$\hat{a}_i^\dag(t)\approx\alpha_i^* e^{i\omega t})$,
 (the Gutzwiller mean value $\alpha_i = \langle \hat{a}_i \rangle_{SS} = \text{Tr}(\hat{\rho}_i^{SS} \hat{a}_i)$ is nonzero unless perfect U(1) symmetry, as shown in Fig.\ref{poin}(c)-(e)) and $[\hat{a}_i(t), \hat{a}_j^\dagger(0)] =[\alpha_i \hat{I}, \hat{a}_j^\dagger(0)] = 0$ in GMF stable state,
the coherent cross term becomes
$\langle \hat{a}_i^\dagger(t) \hat{a}_j(0) \hat{a}_i(t) \hat{a}_j^\dagger(0) \rangle \xrightarrow{t \to \infty} \langle \hat{a}_i^\dagger \hat{a}_i \rangle \langle \hat{a}_j \hat{a}_j^\dagger \rangle = n_i (n_j + 1)$.
At the long-time limit, the memory of the initial operator $\hat{a}_j(0)$ is erased by dissipative contraction, leading to the statistical factorization of the four-point correlation function,
 the coherent and incoherent components cancel each other out, resulting in $D_{ij}(t\to \infty)\sim [\hat{a}_i(t), \hat{a}_j^\dagger(0)] \to 0$. 
The nonvanishing coherent cross terms implies that dissipative time scale is shorter than the scrambling time scale, and the operator complexity does not simply convert into incoherent noise intensity distributed across the chain. 
The global dissipative contraction of the phase space contracts the operator norm itself and $U$ causes  mean-field $\langle \hat{a}_i \rangle=0$ (restoring $U(1)$ symmetry). 
Thus for local observer, there is local information dissipation (local dephasing), rather than being hidden within complex multi-particle correlations. 

The OTOC measures sensitivity of the field at site $i$ to a perturbation at site $j$ at $t=0$, probing the scrambling of phase information across the chain.
The normalized OTOC (phase OTOC) reads
\begin{equation} 
	\begin{aligned}
		&\tilde{D}_{ij}(t) 
=\frac{2\langle \hat{a}_i^\dagger(t) \hat{a}_i(t) \rangle
			(\langle \hat{a}_j^\dagger(0) \hat{a}_j(0) \rangle +1)- 2\text{Re}(\langle \hat{a}_i^\dagger(t) \hat{a}_j(0) \hat{a}_i(t) \hat{a}_j^\dagger(0) \rangle)}{2\langle \hat{a}_i^\dagger(t) \hat{a}_i(t) \rangle \langle \hat{a}_j(0) \hat{a}_j^{\dag}(0) \rangle}\\
		&=
 1 - \text{Re}\left[ \frac{\langle a_{i}^\dagger(t) a_{j} a_{i}(t) a^{\dag}_{j} \rangle_{SS}}{\langle a_{i}^\dagger a_{i} \rangle_{SS} \langle a_{j} a^{\dag}_{j} \rangle_{SS}} \right].
	\end{aligned}
\end{equation}
In this normalized form, 
$\tilde{D} \to 0$ means the phases are perfectly correlated (regular),
$\tilde{D} \to 1$ means the phase information is completely scrambled.
The Cauchy-Schwarz inequality $|\text{Tr}(\hat{\rho}\hat{A})| \le ||\hat{\rho}||_2 ||\hat{A}||_2$ and the Hilbert-Schmidt norm
$||\hat{A}||_2 = \sqrt{\text{Tr}(\hat{A}^\dag \hat{A})}$,
$||\hat{\rho}||_2 = \sqrt{\text{Tr}(\rho^2)} \le 1$ (with equality only for pure states) imply that the OTOC is bounded by the norm of the squared commutator\cite{Uzdin}. 
For steady state $\rho_{SS}$, the theoretical bound is related to the purity of the state. In a maximally mixed state in a finite dimension $N=N_{max}+1=4$, $||\rho||_2 = 1/\sqrt{N}=1/2$.
In a finite Hilbert space the operator OTOC cannot grow indefinitely and the Cauchy-Schwarz inequality provides the theoretical maximum growth. Saturation occurs when the operator $\hat{a}_i(t)$ becomes uniformly distributed over the available operator basis. 
The operator OTOC is theoretically bounded by the Hilbert-Schmidt norm of the system's density matrix, $D_{ij}(t) \le 2||\hat{\rho}||_2$. In our three-site Bose-Hubbard model, the Gutzwiller truncation effectively confines the dynamics to a finite-dimensional local Hilbert space, preventing indefinite growth. 
This maximum saturation in chaotic regime signaling a uniform spreading of the operator basis, 
while the vanishing OTOC in stable nonlinear regime indicates that the dissipative contraction effectively collapses the available operator space into a stable, low-dimensional fixed-point manifold.

Note that the above dynamics of density matrices are related to the Liouvillian in Schrodinger picture,
such that $\frac{d\vec{\rho}}{dt} = \mathbf{L}_{Schro} \vec{\rho}$.
 While the evolution of $a_{i}(\tau)$ is related to the Liouvillian in Heisenberg picture,
$\frac{da_{i}(\tau)}{d\tau}= \mathcal{L}^\dagger(a_{i}) = i[H,a_{i}] + \mathcal{D}^\dagger(a_{i}) = \mathcal{L}^\dagger(a_{i}(\tau))$,
and thus $a_{i}(\tau) = e^{\mathcal{L}^\dagger \tau} a_{i}(0)$ or in the vectorized form
$\mathbf{a}_{i}(\tau) = e^{\mathbf{L}^\dagger \tau} \mathbf{a}_{i}(0)$.
Since $\text{Tr}(\hat{A} \mathcal{L}(\hat{\rho})) = \text{Tr}(\mathcal{L}^\dagger(\hat{A}) \hat{\rho})$,
we have $\mathbf{L}_{Heis} = \mathbf{L}_{Schro}^\dagger$.
Saturation of $D_{1,\ell}(\tau \to \infty) \approx 1$ as the hallmark of quantum chaos happen
when the driving force $F$ is strong enough to push the local photon number into the non-linear regime where $U$ dominates.
This saturation value implies that the operator has fully scrambled across the system, and the number fluctuation is large (follows the super-Poissonian distribution) and compatible with thermalization.
The high dissipation rate prevents the operator from fully scrambling across the system,
and the system settles into a contractive steady state where all eigenvalues of the Liouvillian have strictly negative real parts. The decay of the OTOC to zero effectively signals that the dissipative damping rate outpaces the internal scrambling rate.

In the stable nonlinear regime, the OTOC exhibits a short-term increase as the Kerr nonlinearity $U$ initially spreads the perturbations, followed by a monotonic decay toward zero. This signals that the system’s memory is governed by dissipative contraction rather than chaotic scrambling.
Any initial local perturbation at the first site is asymptotically damped before it can evolve into a global chaotic entanglement. The information regarding the initial state is not hidden within many-body correlations but is instead leaked to the environment. 
Zero OTOC at long time reflects that the contractive NESS 
rendering the system’s long-term memory inaccessible due to environmental decoherence rather than internal randomization.
As depicted in Fig.\ref{otoc}, the OTOC vanishes in the long-time limit under both the GMF and exact frameworks. For stable systems, the local perturbation generated by $\hat{a}_j^\dagger$ is completely dissipated by the local loss channel $\gamma$, leading to a vanishing asymptotic sensitivity, $\lim_{t \to \infty} \partial a(t) / \partial a(0) = 0$. 
During the underdamped transient, $U$ causes the phase to spread 
 but the dynamics remain deterministic and phase-coherent in the operator space.
The coherent cross-terms evolve to exactly match the magnitude of the incoherent terms. At the fixed point, the operator $\hat{a}(t)$ becomes a c-number $\alpha_{ss}$, The OTOC goes to zero because the cross-terms cancel the density terms.

An alternative expression of OTOC is related to the semiclassical Jacobian
\begin{equation}
	D_{1,L}(\tau) = \left| \frac{\partial \alpha_L(\tau)}{\partial \alpha_1(0)} \right|^2 \approx |G^R_{L,1}(\tau)|^2 = \left| \left( e^{\mathbf{L}_{\text{chain}} \tau} \right)_{L,1} \right|^2,
\end{equation}
where the coupled chain Jacobian (a $2L \times 2L$ matrix) includes the effect of coupling $J$. 
In TWA, the deterministic evolution is given by $\dot{\alpha}_l = \mathcal{A}_l(\{\alpha\})$ with complex amplitudes $\alpha_l$.
Let $\mathbf{V}(t) = (\delta \alpha_1, \delta \alpha_1^*, \dots, \delta \alpha_L, \delta \alpha_L^*)^T$ be the fluctuation vector. Linearizing around the steady state
\begin{equation}
	\frac{d\mathbf{V}(t)}{dt} = \mathbf{J} \mathbf{V}, 
\end{equation}
where $J_{ij} = \frac{\partial \dot{V}_i}{\partial V_j}$ and the solution is $\mathbf{V}(t) = e^{\mathbf{J}t} \mathbf{V}(0)$. The sensitivity of site $L$ to a perturbation at site 1 is
\begin{equation}
	\frac{\partial \alpha_L(\tau)}{\partial \alpha_1(0)} = \left( e^{\mathbf{J}\tau} \right)_{2L-1, 1}
\end{equation}
with $\mathbf{J}$ the $2L \times 2L=6\times 6$ Jocobian matrix.

For the deterministic evolution of complex fields $\dot{\alpha}_l = \mathcal{A}_l(\alpha_1, \alpha_1^*, \dots, \alpha_L, \alpha_L^*)$,
the steady state at equilibrium $\{\alpha_{SS}\}$ is defined as $\mathcal{A}_l(\{\alpha_{SS}\}) = 0 \quad \text{for all } l$.
We assume the system is slightly perturbed from the steady state $\alpha_l(t) = \alpha_{l,SS} + \delta \alpha_l(t)$,
$\dot{\alpha}_l(t) = \frac{d}{dt}(\alpha_{l,SS} + \delta \alpha_l(t)) = \delta \dot{\alpha}_l(t)$.
Expand the function $\mathcal{A}_l$ around the steady state value using a first-order Taylor series
$\mathcal{A}_l(\{\alpha\}) \approx \mathcal{A}_l(\{\alpha_{SS}\}) + \sum_{j} \left. \frac{\partial \mathcal{A}_l}{\partial \alpha_j} \right|_{SS} \delta \alpha_j + \sum_{j} \left. \frac{\partial \mathcal{A}_l}{\partial \alpha_j^*} \right|_{SS} \delta \alpha_j^* + \mathcal{O}(\delta^2)$.
Thus $\delta \dot{\alpha}_l \approx \sum_{j} \left( \frac{\partial \mathcal{A}_l}{\partial \alpha_j} \delta \alpha_j + \frac{\partial \mathcal{A}_l}{\partial \alpha_j^*} \delta \alpha_j^* \right)$.
We define the column vector $\mathbf{V}$ as the collection of all fluctuations and their conjugates
$\mathbf{V} = (\delta \alpha_1, \delta \alpha_1^*, \dots, \delta \alpha_L, \delta \alpha_L^*)^T$.
The entries of the Jacobian matrix $\mathbf{J}$ are precisely the partial derivatives calculated at the steady state $J_{ij} = \frac{\partial \dot{V}_i}{\partial V_j}$.
If $V_i = \delta \alpha_l$ and $V_j = \delta \alpha_k$, then $J_{ij} = \frac{\partial \mathcal{A}_l}{\partial \alpha_k}$.
If $V_i = \delta \alpha_l$ and $V_j = \delta \alpha_k^*$, then $J_{ij} = \frac{\partial \mathcal{A}_l}{\partial \alpha_k^*}$.

In the TWA or GMF limit, the system is described by complex amplitudes $\alpha_l$. The sensitivity $D_{1,L}(\tau) = |\partial \alpha_L(\tau) / \partial \alpha_1(0)|^2$ is obtained from the $2L \times 2L$ Jacobian matrix $\mathbf{J}$,
\begin{equation}
	\mathbf{J} = 
	\begin{pmatrix} 
		\mathbf{M}_1 & \mathbf{J}_{hop} & 0 \\
		\mathbf{J}_{hop} & \mathbf{M}_2 & \mathbf{J}_{hop} \\
		0 & \mathbf{J}_{hop} & \mathbf{M}_3 
	\end{pmatrix}
\end{equation}
where the local blocks $\mathbf{M}_l$ and hopping blocks $\mathbf{J}_{hop}$ are
\begin{equation}
	\mathbf{M}_l = \begin{pmatrix} -i(\Delta + 2U|\Psi_l|^2 - U) - \frac{\gamma_l}{2} & -iU\Psi_l^2 \\ iU(\Psi_l^*)^2 & i(\Delta + 2U|\Psi_l|^2 - U) - \frac{\gamma_l}{2} \end{pmatrix}, \quad \mathbf{J}_{hop} = \begin{pmatrix} iJ & 0 \\ 0 & -iJ \end{pmatrix}
\end{equation}
The sensitivity is the element $(\exp(\mathbf{J}t))_{2L-1, 1}$. 

	\begin{figure}
	\centering
	\includegraphics[width=0.4\linewidth]{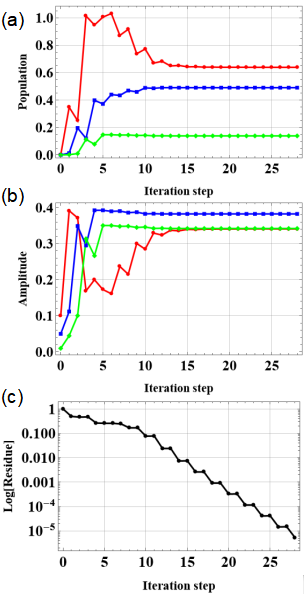}
	\caption{
Numerical solution of the NESS and quantum chaos diagnostics.
The system is solved using the self-consistent Gutzwiller Mean-Field method with Picard iteration.
(a) Dynamic convergence of the local particle population $\langle \hat{n}_l \rangle$ for the drive (Site 1, blue), bulk (Site 2, green), and drain (Site 3, red) sites versus iteration steps. The system settles into a non-uniform density profile driven by the source-drain bias.
(b) Convergence of the mean-field order parameter magnitude $|\Psi_l|$, showing the stabilization of the coherent field background.
(c) The residual error $\epsilon_k = ||\Psi^{(k)} - \Psi^{(k-1)}||$ on a logarithmic scale. The linear slope indicates the exponential convergence of the Picard iteration algorithm to the unique NESS.
}
\end{figure}

	\begin{figure}
	\centering
	\includegraphics[width=0.9\linewidth]{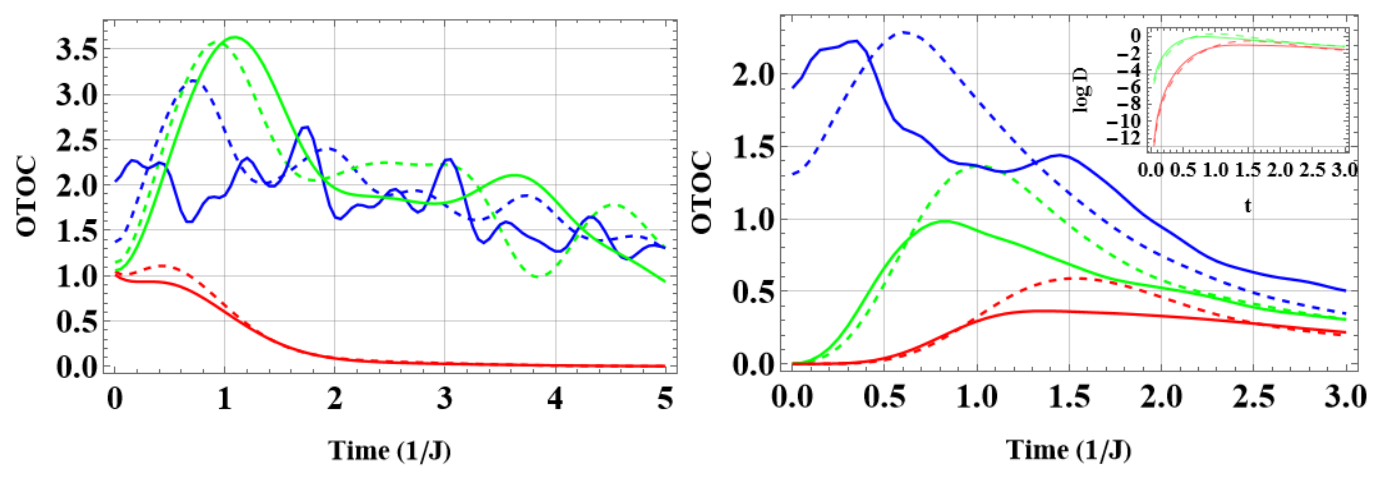}
	\caption{Operator OTOC governed by the adjoint Liouvillian $e^{\mathcal{L}^{\dag}t}$ where phase diffusion in a trajectory picture does not imply persistent operator growth in $\mathcal{L}^{\dag}$
and dissipation dominate the long-time behavior.
(Left) Time evolution of the local operator OTOC $D_{i,i}(t)$ calculated on NESS background using the Picard iteration self-consistent solver. 
Blue, green, and red correspond to site 1, site 2, and site 3, respectively. 
The real line and dashed line correspond to setting $F=2.5$ and $F=1$, respectively.
Our parameters are setted as $J = 1,\ U = 2,\ \Delta = -0.5,\ \gamma_1 = 0.1=\gamma_2 = 0.1,\  \gamma_3 = 1$.
The rapid initial growth and subsequent saturation at a non-zero value indicate significant information scrambling and the onset of local quantum chaos induced by the Kerr nonlinearity.
(Right) Operator OTOC $D_{i,1}(t)$ solved by exact solver.
Different to the above GMF approximation which assumes the total state is a product state $\hat{\rho}_{tot} = \hat{\rho}_1 \otimes \hat{\rho}_2 \otimes \hat{\rho}_3$ and thus forbids any entanglement or quantum correlations between sites and only captures local quantum effects (like onsite squeezing or photon blockade),
the TWA allows trajectories of different sites to be correlated 
and captures classical-like correlations as well as the scrambling. However, it cannot represent purely non-local quantum entanglement (such as EPR pairs or Bell states) because 
it is still based on point-like trajectories in phase space.
The exact solver acts on the full many-body Hilbert space and it captures every possible quantum correlation, including multi-site entanglement and non-local phase coherence.
The light cone structure can be observed in the nonlocal OTOC (for site 2 and site 3) as shown in the right panel\cite{Zhou,Xu,Tripathy,Braumüller}, in consistent with the perturbative expansion in Table I.
The inset in right panel displays the time evolution of $\log D$ for the cross-site (nonlocal) OTOC. 
	}
	\label{otoc}
\end{figure}
	
		\begin{figure}
		\centering
		\includegraphics[width=0.9\linewidth]{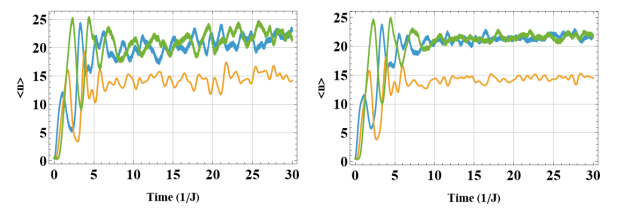}
		\caption{The population base on
TWA for 100 trajectories (left) and 500 trajectories (right). The blue, orange, and green correspond to the site 1, site 2, and site 3, respectively.
At long time, $\langle n_{1}\rangle\approx \langle n_{3}\rangle\approx \frac{3}{2}\langle n_{2}\rangle$.
		}
	\end{figure}
	
	\section{Effective Liouvillian $\mathbf{L}_{eff}$}
	
	To explicitly study the stability, we consider quantum fluctuations $\delta \hat{a}_l(t)$ around the mean-field steady state. We apply the linearization ansatz $\hat{a}_l(t) = \Psi_{SS} + \delta \hat{a}_l(t)$ and $\hat{a}^{\dag}_l(t) = \Psi^{*}_{SS} + \delta \hat{a}^{\dag}_l(t)$, where $\Psi_{SS}$ is the classical solution to the NESS equation.
For $\hat{H}_{loc} = \Delta \hat{n} + \frac{U}{2}\hat{n}(\hat{n}-1)$, 
	the detuning term $\Delta \hat{a}^\dagger \hat{a}
	=\Delta (\Psi_{SS}^* + \delta \hat{a}^\dagger)(\Psi_{SS} + \delta \hat{a}) =\Delta \left( |\Psi_{SS}|^2 + \Psi_{SS}^* \delta \hat{a} + \Psi_{SS} \delta \hat{a}^\dagger + \delta \hat{a}^\dagger \delta \hat{a} \right)$ contains the
	constant terms which don't affect dynamics and the linear terms which cancel out for system in equilibrium, and the quadratic term.
and	the interaction term expands as
	\begin{equation}
		\frac{U}{2} (\Psi_{SS}^* + \delta \hat{a}^\dagger)^2 (\Psi_{SS} + \delta \hat{a})^2 \approx \frac{U}{2} \left[ (\Psi_{SS}^*)^2 + 2\Psi_{SS}^*\delta \hat{a}^\dagger + (\delta \hat{a}^\dagger)^2 \right] \left[ \Psi_{SS}^2 + 2\Psi_{SS}\delta \hat{a} + (\delta \hat{a})^2 \right].
	\end{equation}
The constant terms only contributes to the energy offset and does not affect dynamics.
The linear terms ($\propto \delta \hat{a}$) vanish identically because $\Psi_{SS}$ is defined as the steady state where the net force should be zero (corresponding to the saddle point of the action).
If the sum of all linear driving forces is finite, 
$\Psi_{SS}$ would change over time, violates the condition of NESS.

Only the quadratic leading terms determine the linearized dynamics (small oscillations) and determine stability.
Higher order terms describing the interactions between the small fluctuations are also ignored.
	Retaining terms up to quadratic order in fluctuations $\delta \hat{a}$
	which approximates the potential energy surface around the steady state as a harmonic oscillator, the fluctuation dynamics reads
	\begin{equation}
		\hat{H}_{\text{fluct}} \approx (\Delta + 2U|\Psi_{SS}|^2) \delta \hat{a}^\dagger \delta \hat{a} + \frac{U}{2} \Psi_{SS}^2 (\delta \hat{a}^\dagger)^2 + \frac{U}{2} (\Psi_{SS}^*)^2 (\delta \hat{a})^2.
	\end{equation}
	The term $2U|\Psi_{SS}|^2$ represents the Hartree-Fock energy shift, while the terms proportional to $\Psi_{SS}^2$ represent parametric driving terms\cite{Xie,Garcia,	Sliwa,Kamal} that allow the creation or annihilation of pairs of fluctuations from vacuum driven by the coherent field.
Higher orders terms representing the interactions between fluctuations are neglected to obtain a solvable linear system (Gaussian approximation) for stability analysis.
	
	The time evolution is governed by the Heisenberg-Langevin equation, $\frac{d}{dt} \delta \hat{a} = -i [\delta \hat{a}, \hat{H}_{\text{fluct}}] - \frac{\gamma}{2} \delta \hat{a} + \hat{\xi}(t)$,
	$\frac{d}{dt} \delta \hat{a}^{\dag} = -i [\delta \hat{a}^{\dag}, \hat{H}_{\text{fluct}}] - \frac{\gamma}{2} \delta \hat{a}^{\dag} + \hat{\xi}(t)$. 
	$\hat{\xi}(t)$ is quantum noise (Langevin noise) arising from the coupling to the environment
as required by the fluctuation-dissipation theorem. Without this noise term, the canonical commutation relation $[\delta \hat{a}(t), \delta \hat{a}^\dagger(t)]$ would decay to zero due to the damping $-\frac{\gamma}{2} \delta \hat{a}$, violating quantum mechanics. The noise correlation $\langle \hat{\xi}(t) \hat{\xi}^\dagger(t') \rangle = \gamma \delta(t-t')$ continuously replenishes the quantum fluctuations, ensuring $[\delta \hat{a}(t), \delta \hat{a}^\dagger(t)] = 1$ at all times.
	
	It is necessary to preserve the commutation relation $[\hat{a}(t), \hat{a}^\dagger(t)] = 1$ as the system decays.
	Calculating the commutators $[\delta \hat{a}, \delta \hat{a}^\dagger \delta \hat{a}] = \delta \hat{a}$ and $[\delta \hat{a}, (\delta \hat{a}^\dagger)^2] = 2\delta \hat{a}^\dagger$, we obtain the linearized equation of motion
for the deterministic drift (neglecting the noise term $\hat{\xi}(t)$ for the eigenvalue stability analysis)
	\begin{equation}
			\begin{aligned}
	&	\frac{d}{dt} \delta \hat{a} = -i \left( (\Delta + 2U|\Psi_{SS}|^2)\delta \hat{a} + U\Psi_{SS}^2 \delta \hat{a}^\dagger \right) - \frac{\gamma}{2} \delta \hat{a},\\
	&	\frac{d}{dt} \delta \hat{a}^{\dag} = i \left( (\Delta + 2U|\Psi_{SS}|^2)\delta \hat{a}^{\dag} + U(\Psi_{SS}^{*})^2 \delta \hat{a} \right) - \frac{\gamma}{2} \delta \hat{a}^{\dag}.
		\end{aligned}
	\end{equation}

The above linearized equation of motion describes the deterministic drift of the fluctuations where the noise term $\hat{\xi}(t)$ was averaged out.
$\hat{\xi}(t)$ disappears in the matrix form because the matrix $\mathbf{L}_{eff}$ describes the drift (deterministic) part of the evolution. Stability is determined by the eigenvalues of this drift matrix. 
If the drift causes fluctuations to decay (stable eigenvalues), the noise maintains a finite variance. If the drift causes growth (unstable eigenvalues), the noise is amplified exponentially.
Only the homogeneous part of the differential equation $\mathbf{L}_{eff}$ determine if the NESS is stable.
If the eigenvalues of deterministic matrix have negative real parts, any perturbation decays.
If they have positive real parts, the perturbation grows.
The random noise $\hat{\xi}(t)$ acts as an inhomogeneous driving term and does not determine whether those modes are stable or unstable, and thus it is dropped when calculating the eigenvalues.

The stability of a NESS is only determined by the homogeneous part of the differential equation ($\mathbf{L}_{eff}$).
If the eigenvalues of the deterministic matrix have negative real parts, any perturbation decays.
If they have positive real parts, the perturbation grows.
The quantum expectation value of the Langevin equation reads \begin{equation}
	\begin{aligned}
\frac{d}{dt} \langle \delta \hat{a} \rangle& = -i \left( (\Delta + 2U|\Psi_{SS}|^2)\delta \hat{a} + U\Psi_{SS}^2 \delta \hat{a}^\dagger \right) - \frac{\gamma}{2} \langle \delta \hat{a} \rangle + \langle \hat{\xi}(t) \rangle\\
&= -i \left( (\Delta + 2U|\Psi_{SS}|^2)\delta \hat{a} + U\Psi_{SS}^2 \delta \hat{a}^\dagger \right) - \frac{\gamma}{2} \langle \delta \hat{a} \rangle
		\end{aligned}
\end{equation}
since $\langle \hat{\xi}(t) \rangle = 0$).

	Different to above nonlinear equation $\frac{d}{dt}\hat{a}_r$,
analytical solutions are available here, which can be expressed as the superposition of two exponential modes corresponding to the eigenvalues $\lambda_{\pm}$,
\begin{equation}
	\begin{aligned}
\delta \hat{a}(t) = u(t) \delta \hat{a}(0) + v(t) \delta \hat{a}^\dagger(0)
		\end{aligned}
\end{equation}
with 
\begin{equation}
\begin{aligned}
	u(t) &= e^{-\frac{\gamma}{2}t} \left[ \cos(\Omega t) - i \frac{ \Delta + 2U|\Psi_{SS}|^2}{\Omega} \sin(\Omega t) \right] \\
	v(t) &= e^{-\frac{\gamma}{2}t} \left[ -i \frac{U\Psi_{SS}^2}{\Omega} \sin(\Omega t) \right],\\
	&\Omega = \sqrt{ (\Delta + 2U|\Psi_{SS}|^2)^2-| U\Psi_{SS}^2|^2}
		\end{aligned}
\end{equation}
where $ |\Delta + 2U|\Psi_{SS}|^2|$ is the effective detuning, $ U\Psi_{SS}^2$ is the parametric gain,
and $\Omega $ is the growth rate or oscillation frequency.
Then for $(\Delta + 2U|\Psi_{SS}|^2)^2 > |U\Psi_{SS}^2|^2$ (stable and oscillatory),
$\Omega$ is real and $\sinh(\Omega t)\sim \frac{1}{2}e^{\Omega t}$. 
The solution is a damped oscillation decaying at rate $\gamma/2$.
For $|U\Psi_{SS}^2|^2 > (\Delta + 2U|\Psi_{SS}|^2)^2$ (unstable and parametric),  $\Omega$ is imaginary. 
If this growth rate $\Omega$ is faster than the decay $\gamma/2$ (i.e., $\Omega > \gamma/2$), the fluctuation $\delta \hat{a}(t)$ will explode exponentially.
The off-diagonal term $v(t)$ and $v^*(t)$ critically depends on the ononlinear strength $U$.
In linear limit $U=0$, $U\Psi_{SS}^2=0$,
$\Delta + 2U|\Psi_{SS}|^2=\Omega=\Delta$, and 
$u(t) = e^{-\frac{\gamma}{2}t} (\cos \Delta t - i \sin \Delta t) = e^{-i\Delta t - \frac{\gamma}{2}t}$,
$v(t) = 0$.
Thus $\delta\hat{a}(t) = \delta\hat{a}(0) e^{-i\Delta t - \frac{\gamma}{2}t}$, without squeezing.

The commutator of the deterministic part of the fluctuation reads 
\begin{equation}
[\delta\hat{a}(t),\delta\hat{a}^\dagger(t)] = \left[ u(t) \delta\hat{a}(0) + v(t) \delta\hat{a}(0)^\dagger,  u^*(t) \delta\hat{a}(0)^\dagger + v^*(t) \delta\hat{a}(0) \right]=|u(t)|^2 - |v(t)|^2 = e^{-\gamma t}.
\end{equation}
This shows that the system loss the quantum uncertainty of its initial state due to dissipation.
For closed system with $\gamma=0$, $|u(t)|^2 - |v(t)|^2 = 1$ corresponds to unitary evolution
without loss of quantum information and preserved commutation relation.
The full time evolution of the operator in the linearized regime reads
\begin{equation}
	\begin{aligned}
\hat{a}(t) = \Psi_{SS} + u(t) [\hat{a}(0) - \Psi_{SS}]+ v(t) [\hat{a}^\dagger(0) - \Psi_{SS}^*] + \hat{\mathcal{F}}(t)
		\end{aligned}
\end{equation}
where the second and third terms in the right-hand-side represent the deterministic part\cite{Subramanyan} (memory of initial state).
$\hat{\mathcal{F}}(t)$ is the accumulated noise integral.
$u(t)$ and $v(t)$ are the entries of the matrix propagator $  e^{\mathbf{L}_{\text{eff}}t}$.
The coherent amplitude $\Psi_{SS}$ is the constant classical background of NESS.
$u(t)$ is the normal evolution (rotation and decay),
$v(t)$ is the anomalous evolution (mixing creation and annihilation operators due to the non-linearity $U$).
$\hat{\mathcal{F}}(t)$ the accumulated quantum noise (integral of $\hat{\xi}(t')$), which is necessary to keep $[\hat{a}(t), \hat{a}^\dagger(t)] = 1$ as the deterministic parts $u(t)$ and $v(t)$ decay.
The total operator commutator must remain 1 for all time, 
\begin{equation}
	[\hat{a}(t), \hat{a}^\dagger(t)] = [\delta\hat{a}(t) + \hat{\mathcal{F}}(t), \delta\hat{a}^\dagger(t) + \hat{\mathcal{F}}^\dagger(t)] = [\delta\hat{a}(t), \delta\hat{a}^\dagger(t)] + [\hat{\mathcal{F}}(t), \hat{\mathcal{F}}^\dagger(t)]
	=e^{-\gamma t} + [\hat{\mathcal{F}}(t), \hat{\mathcal{F}}^\dagger(t)]=1,
\end{equation}
where the cross terms $[\delta\hat{a}(t), \hat{\mathcal{F}}^\dagger(t)]$ vanish because initial operators commute with future noise operators).
According to fluctuation-dissipation theorem,
this implies $[\hat{\mathcal{F}}(t), \hat{\mathcal{F}}^\dagger(t)] = 1 - e^{-\gamma t}$.
The accumulated noise $\hat{\mathcal{F}}(t)$ exactly replenishes the commutator as it decays. Without $\hat{\mathcal{F}}(t)$, the commutator would go to zero, which is a violation of the laws of quantum mechanics\cite{Wu,Wu2}.

With spinor $(\delta \hat{a}, \delta \hat{a}^\dagger)^T$ and the effective Liouvillian matrix $\mathbf{L}_{eff}$ (Bogoliubov-de Gennes matrix), the deterministic evolution reads
	\begin{equation}
		\frac{d}{dt} \begin{pmatrix} \delta \hat{a} \\ \delta \hat{a}^\dagger \end{pmatrix} = 
		\underbrace{
			\begin{pmatrix} 
				-i(\Delta + 2U|\Psi_{SS}|^2) - \frac{\gamma}{2} & -iU\Psi_{SS}^2 \\
				iU(\Psi_{SS}^*)^2 & i(\Delta + 2U|\Psi_{SS}|^2) - \frac{\gamma}{2}
			\end{pmatrix}
		}_{\mathbf{L}_{eff}}
		\begin{pmatrix} \delta \hat{a} \\ \delta \hat{a}^\dagger \end{pmatrix}.
		\label{48}
	\end{equation}
The complex eigenvalues $\lambda_{\pm}$ of $\mathbf{L}_{eff}$ determine the system's stability and oscillation frequencies,
\begin{equation}
	\lambda_{\pm} = -\frac{\gamma}{2} \pm i \sqrt{(\Delta + 2U|\Psi_{SS}|^2)^2 - |U\Psi_{SS}^2|^2}.
\end{equation}
The real part of the eigenvalues are
\begin{equation}
	\text{Re}(\lambda_{\pm}) = 
	\begin{cases} 
		-\frac{\gamma}{2} & \text{if } (\Delta + 2U|\Psi_{SS}|^2)^2 \ge |U\Psi_{SS}^2|^2 \quad \text{(stable/oscillatory regime)} \\
		-\frac{\gamma}{2} \mp \sqrt{|U\Psi_{SS}^2|^2 - (\Delta + 2U|\Psi_{SS}|^2)^2} & \text{if } (\Delta + 2U|\Psi_{SS}|^2)^2 < |U\Psi_{SS}^2|^2 \quad \text{(prarametric regime)}
	\end{cases}
\end{equation}
Note that $\gamma>0$,
thus the NESS is stable if and only if the largest real part $\text{Re}(\lambda_{-}) < 0$. If the parametric gain $|U\Psi_{SS}^2|$ is sufficiently strong such that $\sqrt{|U\Psi_{SS}^2|^2 - (\Delta + 2U|\Psi_{SS}|^2)^2} > \frac{\gamma}{2}$, the real part becomes positive,
\begin{equation}
	|U\Psi_{SS}^2|^2 > (\Delta + 2U|\Psi_{SS}|^2)^2 + \frac{\gamma^2}{4}.
\end{equation}
The NESS is stable if and only if all eigenvalues satisfy $\text{Re}(\lambda) < 0$. If the term under the square root becomes sufficiently negative (or if the parametric gain dominates the damping).
Positive $\text{Re}(\lambda)$ signaling an exponential growth of fluctuations and a transition to a dynamical instability (parametric oscillation) and chaotic regime.

 Since $\det(e^{\mathbf{A}}) = e^{\text{Tr}(\mathbf{A})}$ and $\text{Tr}(\mathbf{L}_{\text{eff}}) = (-i(\Delta + 2U|\Psi_{SS}|^2) - \frac{\gamma}{2}) + (i(\Delta + 2U|\Psi_{SS}|^2) - \frac{\gamma}{2}) = -\gamma$,
we have $\det(e^{\mathbf{L}_{\text{eff}}t}) =
u(t)u^*(t)-v(t)v^*(t)=|u(t)|^2-|v(t)|^2= e^{\text{Tr}(\mathbf{L}_{\text{eff}})t} = e^{-\gamma t}$.
Using Cayley-Hamilton theorem,
the 2-by-2 matrix 
$\mathbf{M}:=\mathbf{L}_{eff}+\frac{\gamma}{2}{\bf I}$ satisfy $\mathbf{M}^2 - \text{tr}(\mathbf{M})\mathbf{M} + \det(\mathbf{M})\mathbf{I} = \mathbf{0}$,
thus $\mathbf{M}^2 =- \det(\mathbf{M})\mathbf{I}
=-\Omega^2 {\bf I} $.
Note that in oscillatory stable regime,
 $ \det(\mathbf{M}) = (-i(\Delta+2U|\Psi_{SS}|^2))(i(\Delta+2U|\Psi_{SS}|^2)) - (-iU\Psi_{SS}^2)(iU(\Psi_{SS}^*)^2) = (\Delta+2U|\Psi_{SS}|^2)^2 - |U\Psi_{SS}^2|^2=\Omega^2\ge 0$.
\begin{equation}
\begin{aligned}
&e^{\mathbf{M}t} = \sum_{n=0}^{\infty} \frac{(\mathbf{M}t)^n}{n!} = \mathbf{I} + \mathbf{M}t + \frac{\mathbf{M}^2 t^2}{2!} + \frac{\mathbf{M}^3 t^3}{3!} +\frac{{\bf M}^4 t^4}{4!}+ \dots\\
& = \mathbf{I} + \mathbf{M}t + \frac{
 	- \Omega^2{\bf I} t^2}{2!} + \frac{- \Omega^2{\bf I}\mathbf{M} t^3}{3!} +\frac{\Omega^4 {\bf I} t^4}{4!}+ \dots\\
&= \sum_{k=0}^\infty \frac{t^{2k}}{(2k)!} (-\Omega^2)^k \mathbf{I} + \mathbf{M} \sum_{k=0}^\infty \frac{t^{2k+1}}{(2k+1)!} (-\Omega^2)^k \\
	&= \left( \sum_{k=0}^\infty \frac{(-1)^k (\Omega t)^{2k}}{(2k)!} \right) \mathbf{I} + \frac{\mathbf{M}}{\Omega} \left( \sum_{k=0}^\infty \frac{(-1)^k (\Omega t)^{2k+1}}{(2k+1)!} \right) \\
	&= \cos(\Omega t) \mathbf{I} + \frac{\sin(\Omega t)}{\Omega} \mathbf{M}\\
& = \begin{pmatrix} \cos(\Omega t) & 0 \\ 0 & \cos(\Omega t) \end{pmatrix} + \frac{\sin(\Omega t)}{\Omega} \begin{pmatrix} -i(\Delta+2U|\Psi_{SS}|^2) & -iU\Psi_{SS}^2 \\ i(U\Psi_{SS}^2)^* & i(\Delta+2U|\Psi_{SS}|^2) \end{pmatrix}
\end{aligned}
\end{equation}
Then we can obtain
\begin{equation}
	\begin{aligned}
		&u(t) = e^{-\frac{\gamma}{2}t} \left[ \cos(\Omega t) - i \frac{\Delta+2U|\Psi_{SS}|^2}{\Omega} \sin(\Omega t) \right],\\
		&v(t) =  e^{-\frac{\gamma}{2}t} \left[ -i \frac{U\Psi_{SS}^2}{\Omega} \sin(\Omega t) \right],\\
		&v^{*}(t) =  e^{-\frac{\gamma}{2}t} \left[ i \frac{(U\Psi_{SS}^2)^*}{\Omega} \sin(\Omega t) \right],\\
		&u^*(t) = e^{-\frac{\gamma}{2}t} \left[ \cos(\Omega t) + i \frac{\Delta+2U|\Psi_{SS}|^2}{\Omega} \sin(\Omega t) \right].
	\end{aligned}
\end{equation}

While in unstable/parametric regime ($\det(\mathbf{M}) < 0$) where the parametric gain dominates ($(\Delta+2U|\Psi_{SS}|^2)^2 < |U\Psi_{SS}^2|^2$), we define the growth rate $\Lambda = \sqrt{-\det(\mathbf{M})} = \sqrt{|U\Psi_{SS}^2|^2 - (\Delta+2U|\Psi_{SS}|^2)^2}$.
In this case, $\mathbf{M}^2 = \Lambda^2 \mathbf{I}$. The Taylor expansion yields a hyperbolic solution:
\begin{align}
	e^{\mathbf{M}t} &= \sum_{k=0}^\infty \frac{t^{2k} \Lambda^{2k}}{(2k)!} \mathbf{I} + \sum_{k=0}^\infty \frac{t^{2k+1} \Lambda^{2k}}{(2k+1)!} \mathbf{M} \\
	&= \cosh(\Lambda t) \mathbf{I} + \frac{\sinh(\Lambda t)}{\Lambda} \mathbf{M}.
\end{align}
This corresponds to the exponential growth of fluctuations (parametric amplification) which can lead to instability if the gain $\Lambda$ exceeds the damping rate $\gamma/2$.

Since $\mathbf{L}_{\text{eff}}$ is time-independent
(as a function of constants $\Psi_{SS}, \Delta, U, \gamma$),
thus the propagator $e^{\mathbf{L}_{\text{eff}}t}$ is defined by the power series of the constant matrix $\mathbf{L}_{\text{eff}}$ scaled by the scalar $t$, and it is a linear time translation invariant system.
Next we focus on the noise accumulation integral.
The inhomogeneous integral term provides the required drift ($\mathbf{L}_{\text{eff}}$) and the instantaneous noise ($\boldsymbol{\xi}$) to satisfy the inhomogeneous Langevin equation
(which is the Bogoliubov-de Gennes linear equation
describing the fluctuation)
		\begin{equation}
		\frac{d}{dt} \begin{pmatrix} \delta \hat{a}(t) \\ \delta \hat{a}^\dagger(t) \end{pmatrix} = 
\mathbf{L}_{eff}
		\begin{pmatrix} \delta \hat{a}(t) \\ \delta \hat{a}^\dagger(t) \end{pmatrix}+
		\begin{pmatrix}
			\xi(t)\\
			\xi^{\dag}(t)\end{pmatrix}
	\end{equation}
	the solution is 
		\begin{equation}
		 \begin{pmatrix} \delta \hat{a}(t) \\ \delta \hat{a}^\dagger(t) \end{pmatrix} = e^{\mathbf{L}_{eff}t} 
		  \begin{pmatrix} \delta \hat{a}(0) \\ \delta \hat{a}^\dagger(0) \end{pmatrix}
		   + \int_0^t e^{\mathbf{L}_{eff}(t-\tau)} 	\begin{pmatrix}
		   	\xi(\tau)\\
		   	\xi^{\dag}(\tau)\end{pmatrix} d\tau,
				\end{equation}
where the first temr is homogeneous solution and the second term is inhomogeneous solution.
Note that using Leibniz integral rule,
$\frac{d}{dt} \left( \int_{a(t)}^{b(t)} f(t, \tau) d\tau \right) = f(t, b(t)) \cdot b'(t) - f(t, a(t)) \cdot a'(t) + \int_{a(t)}^{b(t)} \frac{\partial}{\partial t} f(t, \tau) d\tau$,
we have
		\begin{equation}
\frac{d}{dt}  \int_0^t e^{\mathbf{L}_{eff}(t-\tau)} 	\begin{pmatrix}
	\xi(\tau)\\
	\xi^{\dag}(\tau)\end{pmatrix} d\tau
	 = \mathbf{L}_{eff} 
 \int_0^t e^{\mathbf{L}_{eff}(t-\tau)} 	\begin{pmatrix}
	\xi(\tau)\\
	\xi^{\dag}(\tau)\end{pmatrix} d\tau
	+ \begin{pmatrix} \xi(t)\\ \xi^{\dag}(t)\end{pmatrix},
				\end{equation}
where $\mathbf{L}_{eff}(t-\tau)$ is the result of integral in the exponent
$\int_\tau^t \mathbf{L}_{\text{eff}} dt' = \mathbf{L}_{\text{eff}} \cdot (t - \tau)$, in constract to the time-ordered exponential $\mathcal{T} \exp \left( \int_\tau^t \mathbf{L}_{\text{eff}}(t') dt' \right)$.

Using the Bogoliubov structure $e^{\mathbf{L}_{eff} t} = \begin{pmatrix} u(t) & v(t) \\ v^*(t) & u^*(t) \end{pmatrix}$
and $e^{\mathbf{L}_{eff}(t-\tau)} = \begin{pmatrix} u(t-\tau) & v(t-\tau) \\ v^*(t-\tau) & u^*(t-\tau) \end{pmatrix}$,
we have the following noise accumulate integral
		\begin{equation}
			\begin{pmatrix} \hat{\mathcal{F}}_1(t) \\ \hat{\mathcal{F}}_2(t) \end{pmatrix} = \int_0^t \begin{pmatrix} u(t-\tau) & v(t-\tau) \\ v^*(t-\tau) & u^*(t-\tau) \end{pmatrix} \begin{pmatrix} \hat{\xi}(\tau) \\ \hat{\xi}^\dagger(\tau) \end{pmatrix} d\tau
							\end{equation}
where $u(t-\tau) \hat{\xi}(\tau)$ and $u^{*}(t-\tau) \hat{\xi}^{\dag}(\tau)$ represent the normal noise that injected at time $\tau$ propagates to time $t$ behaving like a regular particle.
$v(t-\tau) \hat{\xi}^\dagger(\tau)$ and $v^*(t-\tau) \hat{\xi}(\tau)$ represent the anomalous noise. Due to nonlinear pumping, the hole noise (and particle noise) injected at time $\tau$ is converted and mixed into the particle's (and hole's) evolution, respectively. This term is essential for the squeezing.

Performing the matrix multiplication explicitly:
\begin{equation}
	\begin{aligned}
	\hat{\mathcal{F}}_1(t) &= \int_0^t \left[ u(t-\tau) \hat{\xi}(\tau) + v(t-\tau) \hat{\xi}^\dagger(\tau) \right] d\tau \\
	\hat{\mathcal{F}}_2(t) &= \int_0^t \left[ v^*(t-\tau) \hat{\xi}(\tau) + u^*(t-\tau) \hat{\xi}^\dagger(\tau) \right] d\tau
		\end{aligned}
\end{equation}
where $(\hat{\mathcal{F}}_1(t))^\dagger = \hat{\mathcal{F}}_2(t)$.
Thus $\hat{\mathcal{F}}_2$ represents the fluctuations of the creation operator $\delta \hat{a}^\dagger$.

At zero-temperature (vacuum state),
we use the commutation relation $[\hat{\xi}, \hat{\xi}^\dagger] = \gamma \delta$ and $\langle \hat{\xi} \hat{\xi}^\dagger \rangle = \langle \hat{\xi}^\dagger \hat{\xi} \rangle + \langle [\hat{\xi}, \hat{\xi}^\dagger] \rangle = 0 + \gamma \delta$, the Markovian noise satisfies the relations
\begin{equation}
	\begin{aligned}
		\langle \hat{\xi}(\tau) \hat{\xi}^\dagger(\tau') \rangle = \gamma \delta(\tau-\tau'), \quad \langle \hat{\xi}^\dagger(\tau) \hat{\xi}(\tau') \rangle =  \langle \hat{\xi}(\tau) \hat{\xi}(\tau') \rangle =  \langle \hat{\xi}(\tau) \rangle=0 .
		\end{aligned}
\end{equation}
Thus the particle number fluctuations ($\langle \hat{\mathcal{F}}_1^\dagger \hat{\mathcal{F}}_1 \rangle$) corresponds to the normal-ordered correlation $\langle \delta \hat{a}^\dagger \delta \hat{a} \rangle$.
\begin{equation}
\begin{aligned}
	\langle \hat{\mathcal{F}}_1^\dagger \hat{\mathcal{F}}_1 \rangle 
&= \int_0^t d\tau_1 \int_0^t d\tau_2 \langle \hat{\mathcal{F}}_2(\tau_1) \hat{\mathcal{F}}_1(\tau_2) \rangle \\
& = \int_0^t d\tau_1 \int_0^t d\tau_2 \langle [v^*(t-\tau_{1}) \hat{\xi}_1 + u^*(t-\tau_{1}) \hat{\xi}_1^\dagger] [u(t-\tau_{2}) \hat{\xi}_2 + v(t-\tau_{2}) \hat{\xi}_2^\dagger] \rangle\\
&= \int_0^t d\tau_1 \int_0^t d\tau_2 v^*(t-\tau_1) v(t-\tau_2) \langle \hat{\xi}(\tau_1) \hat{\xi}^\dagger(\tau_2) \rangle
 = \gamma \int_0^t |v(t-\tau)|^2 d\tau
\end{aligned}
\end{equation}
where $\langle \hat{\xi}(\tau_1) \hat{\xi}^\dagger(\tau_2) \rangle=\gamma \delta(\tau_1-\tau_2)$ the last step is because only the term containing $\langle \hat{\xi}_1 \hat{\xi}_2^\dagger \rangle$ survives.
Using $|v(t)| = e^{-\gamma t/2} \frac{|U\Psi_{SS}^2|}{\Omega} |\sin(\Omega t)|$,
we have
\begin{equation}
	\begin{aligned}
 \gamma \int_0^t |v(t-\tau)|^2 d\tau
& = \frac{\gamma |P|^2}{\Omega^2} \int_0^t e^{-\gamma(t-\tau)} \sin^2(\Omega(t-\tau)) d\tau\\
 &= \frac{\gamma |P|^2}{\Omega^2} e^{-\gamma t} \int_0^t e^{\gamma \tau} \sin^2(\Omega(t-\tau)) d\tau\\
  &= \frac{\gamma |P|^2}{\Omega^2} e^{-\gamma t} 
 \int_0^t e^{\gamma \tau} \frac{1 - \cos(2\Omega(t-\tau))}{2} d\tau\\
&=  \frac{\gamma |P|^2}{\Omega^2} e^{-\gamma t} 
\left[\frac{1}{2} \int_0^t e^{\gamma \tau} d\tau - \frac{1}{2} \int_0^t e^{\gamma \tau} \cos(2\Omega(t-\tau)) d\tau\right]\\
& = \frac{\gamma |P|^2}{\Omega^2} e^{-\gamma t} \cdot \frac{1}{2} \left[ \frac{e^{\gamma t} - 1}{\gamma} - \text{Re}\left( \frac{e^{\gamma t} - e^{i2\Omega t}}{\gamma - 2i\Omega} \right) \right]\\
& = \frac{|P|^2}{2\Omega^2} \left[ (1 - e^{-\gamma t}) - \gamma \cdot \text{Re}\left( \frac{1 - e^{-(\gamma - 2i\Omega)t}}{\gamma - 2i\Omega} \right) \right]
\end{aligned}
\end{equation}

The hole fluctuations $\langle \hat{\mathcal{F}}_2^\dagger \hat{\mathcal{F}}_2 \rangle$ corresponds to the anti-normal correlation $\langle \delta \hat{a} \delta \hat{a}^\dagger \rangle = \langle \hat{\mathcal{F}}_1 \hat{\mathcal{F}}_1^\dagger \rangle$,
\begin{equation}
\begin{aligned}
	\langle \hat{\mathcal{F}}_1 \hat{\mathcal{F}}_1^\dagger \rangle &= \int_0^t d\tau_1 \int_0^t d\tau_2 \langle [u(t-\tau_1)\hat{\xi}_1 + v(t-\tau_1)\hat{\xi}_1^\dagger] [v^*(t-\tau_2)\hat{\xi}_2 + u^*(t-\tau_2)\hat{\xi}_2^\dagger] \rangle
	= \gamma \int_0^t |u(t-\tau)|^2 d\tau,
\end{aligned}
\end{equation}
where the only surviving term is $\langle \hat{\xi}(\tau_1) \hat{\xi}^\dagger(\tau_2) \rangle$ multiplied by $u(\tau_1) u^*(\tau_2)$.

Using the Bogoliubov identity $|u|^2 - |v|^2 = e^{-\gamma \tau}$, we have
\begin{equation}
	\langle \hat{\mathcal{F}}_2^\dagger \hat{\mathcal{F}}_2 \rangle - \langle \hat{\mathcal{F}}_1^\dagger \hat{\mathcal{F}}_1 \rangle =[\hat{\mathcal{F}}_1,\hat{\mathcal{F}}_1^{\dag}]= \gamma \int_0^t (|u|^2 - |v|^2) d\tau = \gamma \int_0^t e^{-\gamma \tau} d\tau = 1 - e^{-\gamma t}
\end{equation}
Thus $\langle \hat{\mathcal{F}}_2^\dagger \hat{\mathcal{F}}_2 \rangle$ is the particle number plus the commutation relation term which decays to 1 (restoring the vacuum commutator) as $t \to \infty$.

\section{Truncated Wigner Approximation (TWA)}

The truncated Wigner approximation (TWA) provides a semi-classical framework to study the many-body dynamics of the driven-dissipative Bose-Hubbard chain by mapping the Lindblad master equation onto a set of coupled stochastic differential equations (SDEs).
In TWA the boson operator $\hat{a}_l$ is replaced by complex field $\alpha_l$ which can be decomposed into its ensemble average and its stochastic fluctuation
$\alpha_l(t) = \Psi_l(t) + \delta \alpha_l(t)$
where $\Psi_l(t) = \langle \alpha_l(t) \rangle$ is the coherent amplitude and $\delta \alpha_l(t)$ represents the quantum fluctuations.
The total intensity measured in the Wigner representation is
\begin{equation}
	\langle |\alpha_l|^2 \rangle = |\Psi_l|^2 + \langle |\delta \alpha_l|^2 \rangle
\end{equation}
where $|\Psi_l|^2$ is the coherent power,
$\langle |\delta \alpha_l|^2 \rangle = \langle |\alpha|^2 \rangle - |\langle \alpha \rangle|^2$ is the total variance of $\alpha_l$.
The physical photon number is then $\langle n_l \rangle= |\Psi_l|^2 + (\langle |\delta \alpha_l|^2 \rangle - 1/2)$.
During a deterministic (ordering) process, the mean value $\Psi_l$ grows from zero driven by the external field $F$.
$\langle |\delta \alpha_l|^2 \rangle =\int |\alpha|^2 W(\alpha) d^2\alpha= 1/2$ if the system is in a coherent state and $\langle n_l \rangle= |\Psi_l|^2$,
where $W(\alpha) = \frac{2}{\pi} e^{-2|\alpha|^2}$.
In chaotic regime (bulk site and drain site), the system nonlinearly scrambles the phase of the drive, effectively restoring the $U(1)$ symmetry locally.
The nonlinearity $U$ causes trajectories to diverge exponentially\cite{Patel} and the variance increases ($\langle |\delta \alpha_l|^2 \rangle > 1/2$) due to the incoherent thermal excitations. This leads to the loss of phase coherence and the restoration of $U(1)$ symmetry.
Kerr nonlinearity and hopping term are invariant under a global phase shift $\alpha_l \to \alpha_l e^{i\theta}$, which corresponds to photon number conservation. But the drive term $F(\hat{a}_1 + \hat{a}_1^\dagger)$ at site 1 breaks the U(1) symmetry by fixing a phase. 

The bulk site is only coupled to the drive through hopping $J$. In dephased stable nonlinear regime ($U\gg \gamma$), the nonlinear scrambling of phases washes out the drive's influence.
In a dephased NESS, the distribution of $\alpha_{2}$ in phase space ($Re[\alpha]$ vs $Im[\alpha]$) is a transient ring-shaped manifold and thus independent of the phase $\theta={\rm arg}\ \alpha$, signifying the restoration of $U(1)$ symmetry despite the presence of a phase-fixed drive at the boundary.
Site 3 also in scrambled phase but the high dissipation rate $\gamma_3$ keeps the average amplitude $|\alpha_3|$ smaller than in the bulk site.

As shown in Fig.\ref{phase}(a),
the external drive $F$ fixes the average phase (determined by the driving laser) and prevents the distribution from closing into a ring.
The Kerr nonlinearity $U$ makes the rotation frequency dependent on the amplitude ($|\alpha|^2$). 
The external drive $F$ at Site 1 acts as a phase anchor, fixing $\langle \phi \rangle$. In a nonlinear system, the effective resonance frequency (angular velocity in phase space) depends on the photon number: $\Delta_{\text{eff}} = \Delta + U(|\alpha|^2-1)=-\dot{\theta}$. In the phase plane, this means points with a larger amplitude ($|\alpha|$) rotate at a higher angular velocity than points with a smaller amplitude. 


In Fig.\ref{phase}(a) we set $U=2.5, F=6$. 
In site 1 the drive $F$ is strong enough to keep the coherent offset from the origin with broken symmetry.
Sites 2 and 3, the nonlinearity $U$ in the bulk is strong enough to scramble the phase information, restoring the $U(1)$ symmetry ($\langle \alpha \rangle = 0$).
The local nonlinearity at site 2 is strong enough to eliminate the phase information inherited from the drive as long as the phase diffusion rate ($\sim U$) is larger than the coherent transfer rate ($\sim J$). 
We show the final complex amplitudes $\{\alpha_l^{(m)}\}$ for a large ensemble of $M=800$ trajectories. We use a Monte-Carlo sampling that reconstructs the steady-state local Wigner function $W_l(\alpha)$.
The transition from a nonsymmetry distribution
in site 1 to a symmetry distribution in site 2 and site 3 confirms the system's transition from a driven-coherent state to a many-body chaotic state where phase coherence is lost due to the scrambling but energy (particle number) is transported
and the system has thermalized and become chaotic locally, restoring the $U(1)$ symmetry $\langle\alpha\rangle=0$.
The phase $\theta = \text{arg}(\alpha)$ becomes uniformly distributed over $[0, 2\pi]$. 
The vanishing expectation value $\langle \alpha \rangle$ can be 
the Wigner-weighted integral
\begin{equation}
	\langle \alpha \rangle = \int \alpha W(\alpha) d^2\alpha = \int_0^\infty \int_0^{2\pi} r e^{i\theta} W(r, \theta) r dr d\theta=0.
\end{equation}
where $\alpha=r e^{i\theta}$ is the Weyl symbol of $\hat{a}$.
 In the chaotic regime with ring-shape phase distribution, the distribution $W(r, \theta)$ becomes phase-independent ($W(r)$), and since $\int_0^{2\pi} e^{i\theta} d\theta = 0$, the integral vanishes.
Although the average field $\langle \alpha \rangle$ is zero, the points are distributed far from the origin ($r \gg 0$). The  photon number $\langle n \rangle = \langle |\alpha|^2 \rangle - 1/2$ remains large, representing a phase-scrambled high-density state.
In (b) we set $U=0, F=1$.
In this linear limit where the coherent information is preserved, the system behaves as a collection of driven-damped harmonic oscillators. The ring-shape distribution observed are the circular uncertainty clouds of vacuum noise with $1/2$ width centered at the coherent amplitude $\alpha_{SS}$. 
 The trajectories form a Gaussian cloud centered at the origin with a width of $1/2$. This represents the vacuum state.
 The drift away from the origin signifies broken $U(1)$ symmetry due to the drive.
In TWA, the vacuum state has variance $\langle |\alpha|^2 \rangle = \frac{1}{2}$ as a result of the commutation relation $[\hat{a}, \hat{a}^\dagger] = 1$, play the role of quantum noise and zero-point fluctuations which  allows the system to explore the phase space and trigger non-linear effects or chaos\cite{Ferrari}.


The Monte Carlo sampling in TWA applies $\langle \alpha \rangle = \frac{1}{M} \sum_{m=1}^M \alpha^{(m)}$.
As $M$ increases, the random phases of the trajectories $\alpha^{(m)}$ cancel out pairwise in the complex plane, leading to $\langle \alpha \rangle \to 0$. 
The TWA works by mapping the quantum master equation to a set of mutually independent stochastic differential equations (Langevin equations).
The sampling start with $M$ copies of the system by using an ensemble of $M$ classical-like trajectories. Each copy's initial state $\alpha(0)$ is sampled from the Wigner distribution of the vacuum (a Gaussian cloud of width $1/\sqrt{2}$).
At $t=0$, the vacuum state $|0\rangle$ in Wigner representation is a Gaussian with variance $1/2$,
\begin{equation}
	\alpha_l^{(m)}(0) = \frac{1}{\sqrt{2}} (\eta_{1,l}^{(m)} + i\eta_{2,l}^{(m)}), \quad \eta \sim \mathcal{N}(0,1).
\end{equation}
Each trajectory is evolved independently using the Heun integrator. The stochastic noise term $\sqrt{\gamma/2}dW$ simulates the quantum fluctuations entering from the environment.
 Each trajectory $m$ evolves according to a stochastic differential equation (Langevin equation)
$\alpha(t+dt) = \alpha(t) + \mathcal{A}(\alpha)dt+ \sqrt{\frac{\gamma}{2}} d\mathcal{W}$.
 The noise $d\mathcal{W}$ ensures that the diffusion of trajectories satisfy the uncertainty principle.
Because the underlying Fokker-Planck equation (FPE) describes the evolution of the probability density $W(\alpha, t)$, the set of $M$ points at any time $t$ serves as a statistical histogram of the quantum state in phase space.  
As $M \to \infty$, the density of these points perfectly reconstructs the continuous density $W(\alpha, t)$ defined by the FPE.

The symmetric (Weyl) ordering\cite{Polkovnikov} of the Hamiltonian leads to the following correspondence for the Kerr nonlinearity
\begin{equation}
	\begin{aligned}
&		\hat{a}^\dagger \hat{a}^\dagger \hat{a} \hat{a}=\hat{n}(\hat{n}-1) = \hat{n}^2 - \hat{n} \to
		(|\alpha|^2 - \frac{1}{2})^2 - \frac{1}{4}
		-(|\alpha|^2-\frac{1}{2})\\
&		=(|\alpha|^4 - |\alpha|^2) - (|\alpha|^2 - 1/2) = |\alpha|^4 - 2|\alpha|^2 + \frac{1}{2}.
	\end{aligned}
\end{equation}
This results in a renormalized nonlinearity in the drift equations. The quantum fluctuations are incorporated through the initial conditions and the stochastic noise terms derived from the dissipators.
In the Wigner-Weyl correspondence, the classical variable $|\alpha|^2$ does not map to particle number $\hat{n} = \hat{a}^\dagger \hat{a}$ but maps to the symmetrically ordered product
$|\alpha|^2 = \frac{1}{2}(\hat{a}^\dagger \hat{a} + \hat{a} \hat{a}^\dagger) = \hat{a}^\dagger \hat{a} + \frac{1}{2} [\hat{a}, \hat{a}^\dagger] = \hat{n} + \frac{1}{2}$.
Thus the averaged population can be obtained by the stochastic average of $|\alpha|^2$ subtract a $1/2$ quantum bias,
$\langle n \rangle = \overline{|\alpha|^2} - 1/2$,
and the particle number operator in Wigner space is projected as $\hat{a}^\dagger \hat{a} \to |\alpha|^2 - \frac{1}{2}$.
At vacuum state ($|0\rangle$) where $\langle \hat{n} \rangle = 0$, the variance of the complex field is exactly 1/2, i.e., the minimal uncertainty state.
Here the minimal variance $1/2$ for the complex variable $\alpha = X + iP$ directly corresponds to the Heisenberg minimal uncertainty state.
For commutator $[\hat{X}, \hat{P}] = \frac{i}{2}$ with $\hat{X} = \frac{\hat{a} + \hat{a}^\dagger}{2}={\rm Re}\alpha$ and $\hat{P} = \frac{\hat{a} - \hat{a}^\dagger}{2i}={\rm Im}\alpha$, the Heisenberg uncertainty principle states
\begin{equation}
	\Delta X \Delta P \ge \frac{1}{2} |\langle [\hat{X}, \hat{P}] \rangle| = \frac{1}{4}.
\end{equation}
For a symmetric vacuum state, the variance of the complex variable is $\langle |\alpha|^2 \rangle = \langle X^2 \rangle + \langle P^2 \rangle = 1/2$,
with $\langle X^2 \rangle = \langle P^2 \rangle = 1/4$ and $\Delta X = \Delta P = 1/2$ (which is possibel only for the vacuum or a coherent/unsqueezed state).
The product of uncertainties in linear system with Gaussian state is $\Delta X \Delta P = \frac{1}{4}$,
corresponds to the minimal uncertainty state and minimal variance ($1/2$) in TWA.
This is consistent with above symmetric ordering requirement $\overline{ |\alpha|^2 } = \langle \hat{n} \rangle + 1/2$.
The noise $\sqrt{\gamma/2} dW$ which inject randomness is isotropic while the nonlinearity ($U$) is anisotropic and cause non-Gaussian state with $\Delta X \Delta P > \frac{1}{4}$.
As the system relaxes from the vacuum to a driven NESS, the total variance $|\alpha|^2$ increases due to the fluctuation part of TWA which suppressing the coherent amplitude and restoring the $U(1)$ symmetry for a local observer ($\langle\hat{a}_{i}\rangle=0$). 
While the sum of variances is bounded by the noise term, the individual quadrature variance (e.g., $\Delta X^2$) can be compressed below 1/4. 

The Kerr term $\hat{n}(\hat{n}-1)$ maps to the Wigner symbol $H_{int} = \frac{U}{2}(|\alpha|^4 - 2|\alpha|^2 + 1/2)$. 
The drift $\mathcal{A}$ in Langevin equation is derived from the derivative (Poisson bracket) $\dot{\alpha}=-i \frac{\partial H_{int}}{\partial \alpha^*}$,
\begin{equation}
\frac{\partial}{\partial \alpha^*} \left[ \frac{U}{2}(|\alpha|^2\alpha\alpha^* - 2\alpha\alpha^*) \right] = \frac{U}{2}(2|\alpha|^2\alpha - 2\alpha) = U(|\alpha|^2 - 1)\alpha.
\end{equation}
where the quantum bias $-1$ is related to the above bias 1/2 in the operator $\alpha$ in symmetric (Weyl) ordering and ensures the classical trajectories obeys the quantum commutation relations during evolution.
To find $(\hat{a}^\dagger \hat{a} \cdot \hat{a})_W$, we apply the Moyal product
\begin{equation}
 (|\alpha|^2 - 1/2) \star \alpha
 = (|\alpha|^2 - 1/2) \alpha + \frac{1}{2} \left( \frac{\partial (|\alpha|^2 - 1/2)}{\partial \alpha} \frac{\partial \alpha}{\partial \alpha^*} - \frac{\partial (|\alpha|^2 - 1/2)}{\partial \alpha^*} \frac{\partial \alpha}{\partial \alpha} \right)
 = (|\alpha|^2 - 1)\alpha
\end{equation}
where
$\frac{\partial \alpha}{\partial \alpha^*} = 0$,
$\frac{\partial (|\alpha|^2 - 1/2)}{\partial \alpha^*} = \frac{\partial (\alpha \alpha^* - 1/2)}{\partial \alpha^*} = \alpha$.
The term $U(|\alpha|^2 - 1)\alpha$ in the Langevin equation is the exact Weyl symbol of the quantum operator $U \hat{a}^\dagger \hat{a} \hat{a}$. 
The $-1$ is a quantum correction arising from the non-commutativity of the bosonic operators, ensuring that the semiclassical trajectories correctly account for the underlying quantum statistics.

The evolution of complex amplitudes $\alpha_l$ for a chain of length $L$ is governed by the Langevin equation (in the Itô sense)
\begin{equation}
d\alpha_l =\dot{\alpha}_{l}= \mathcal{A}_l(\{\alpha\}) dt + \sum_k \mathcal{B}_{lk} d\mathcal{W}_k
\end{equation}
where the first term in right-hand-side is the deterministic drift term and second term is the 
stochastic diffusion term.
The deterministic drift term $\mathcal{A}_l$ reads
\begin{equation}
\mathcal{A}_l = -i \left[ \Delta \alpha_l + U(|\alpha_l|^2 - 1)\alpha_l - J(\alpha_{l-1} + \alpha_{l+1}) + F\delta_{l,1} \right] - \frac{\gamma_l}{2} \alpha_l
\end{equation}
Note that in any cases, we consider the deturning term $\propto \Delta n<0$ (i.e., $-i\Delta\alpha$ in Heisenberg equation of motion), in consistent with the content in above sections.
From the Heisenberg equation $\dot{\hat{a}} = -i[\hat{a}, \hat{H}]$, the interaction part is $
	-i [\hat{a}, \frac{U}{2} \hat{a}^\dagger \hat{a}^\dagger \hat{a} \hat{a}] = -i U \hat{a}^\dagger \hat{a} \hat{a} = -i U \hat{n} \hat{a}$.
The TWA variables $\alpha$ in symmetric (Weyl) ordering can be used to map the operator product $\hat{n}\hat{a} = \hat{a}^\dagger \hat{a} \hat{a}$ to phase space, by the Weyl symbol of $\hat{a}^\dagger \hat{a} \hat{a}$ is $|\alpha|^2 \alpha - \alpha = (|\alpha|^2 - 1)\alpha$,
where $-1$ is a quantum correction (ordering bias) that accounts for the commutation relation $[\hat{a}, \hat{a}^\dagger]=1$. 
The term $|\alpha|^2 \alpha$ represents the classical mean-field interaction.
The term $-\alpha$ represents the self-interaction with vacuum noise that subtracts the over-counted zero-point energy inherent in the Wigner representation.
The hopping term $-J(\alpha_{l-1} + \alpha_{l+1})$ describes the coherent exchange of particles between neighboring sites in the chain.
The dissipative damping $-\frac{\gamma_l}{2} \alpha_l$ comes from the $\mathcal{D}[\sqrt{\gamma}\hat{a}]$ term in the Lindblad equation. It causes the amplitude to decay exponentially toward zero in the absence of a drive.
The stochastic term represents the vacuum noise entering through the loss channels, with the Markovian white noise $\langle d\mathcal{W}_l(t) d\mathcal{W}_k^*(t') \rangle = \delta_{lk} \delta(t-t') dt$.
Since the loss is local to each site, 
$\mathcal{B}$ is a diagonal matrix with $\mathcal{B}_{ll} = \sqrt{\gamma_l/2}$ and scales the random noise by the square root of the dissipation rate.
The complex wiener process $d\mathcal{W}_l$ represents independent random variables sampled at each time step.
The evolution of the fluctuation spinor $\mathbf{v} = (\delta \hat{a}, \delta \hat{a}^\dagger)^T$ is governed by the linearized equation $\dot{\mathbf{v}} = \mathbf{L}_{\text{eff}} \mathbf{v}$. The matrix $\mathbf{L}_{\text{eff}}$ is defined as the Jacobian matrix of the total deterministic drift function $\dot{\alpha} = \mathcal{A}(\alpha, \alpha^*)$,
\begin{equation}
	\mathbf{L}_{\text{eff}} = 
	\begin{pmatrix} 
		\frac{\partial \dot{\alpha}}{\partial \alpha} & \frac{\partial \dot{\alpha}}{\partial \alpha^*} \\
		\frac{\partial \dot{\alpha}^*}{\partial \alpha} & \frac{\partial \dot{\alpha}^*}{\partial \alpha^*}
	\end{pmatrix}
\end{equation}
Based on the stochastic differential equation
\begin{equation}
	d\alpha_l = \left[ -i(\Delta \alpha_l + U(|\alpha_l|^2-1)\alpha_l - J \sum_{k=l\pm 1} \alpha_k + F \delta_{l,1}) - \frac{\gamma_l}{2} \alpha_l \right] dt + \sqrt{\frac{\gamma_l}{2}} d\mathcal{W}_l(t),
\end{equation}
the deterministic drift term for a site is
$\dot{\alpha} = -i\Delta \alpha - iU(\alpha^2 \alpha^* - \alpha) + iJ \sum \alpha_k - iF - \frac{\gamma}{2}\alpha$.

 \begin{equation}
 	\begin{aligned}
&	L_{11} = \frac{\partial \dot{\alpha}}{\partial \alpha} = -i\Delta - iU(2|\alpha|^2 - 1) - \frac{\gamma}{2} = -i(\Delta + 2U|\alpha|^2 - U) - \frac{\gamma}{2},\\
&	L_{12} = \frac{\partial \dot{\alpha}}{\partial \alpha^*} = -iU\alpha^2,\\
&	L_{21} = \frac{\partial \dot{\alpha}^*}{\partial \alpha} = iU(\alpha^*)^2,\\
&	L_{22} = \frac{\partial \dot{\alpha}^*}{\partial \alpha^*} = i\Delta + iU(2|\alpha|^2 - 1) - \frac{\gamma}{2} = i(\Delta + 2U|\alpha|^2 - U) - \frac{\gamma}{2}.
 	\end{aligned}
\end{equation}
Thus the $\mathbf{L}_{\text{eff}}$ in Eq.(\ref{48}) can be rewritten as
\begin{equation}
	\mathbf{L}_{\text{eff}} = 
	\begin{pmatrix} 
		-i(\Delta + 2U|\alpha|^2 - U) - \frac{\gamma}{2} & -iU\alpha^2 \\
		iU(\alpha^*)^2 & i(\Delta + 2U|\alpha|^2 - U) - \frac{\gamma}{2}
		\label{77}
	\end{pmatrix}
\end{equation}
A discussion on quantum metric framework base on $\mathbf{L}_{\text{eff}}$ is presented in Sec.VIII.

In linear limit $U=0$, $L_{11} = -i\Delta - \frac{\gamma}{2}=L^*_{22}$,
The total equation of motion for $\alpha$ (which includes the hopping and drive terms that are treated as constants in the local Jacobian derivative) reads
$\dot{\alpha} = (-i\Delta - \frac{\gamma}{2})\alpha + iJ \sum \alpha_k - iF$.
For a single driven resonator ($J=0$), 
$\alpha=\frac{iF}{-i\Delta-\frac{\gamma}{2}}$ and $|\alpha| = \frac{|F|}{\sqrt{\Delta^2 + \frac{\gamma^2}{4}}}$.
The center of phase space ($\text{Re}[\alpha] =\text{Im}[\alpha]=0$) represents the vacuum state with zero photon and zero drive. A finite coherent drive $F$ breaks U(1) symmetry and causes a displacement $\alpha_{SS}$ where $|\alpha_{SS}|$ shows coherent photons number and $\text{arg}(\alpha_{SS})$ shows the phase of the light that is locked to the phase of the driving laser.
The steady-state displacement $\alpha_l$ is found by solving $\dot{\alpha}_l = 0$. Define the complex impedance $z_l = -\Lambda_{l}$, we have
\begin{equation}
	\begin{aligned}
	z_1 \alpha_1 - iJ \alpha_2 &= -iF \\
	z_2 \alpha_2 - iJ(\alpha_1 + \alpha_3) &= 0 \\
	z_3 \alpha_3 - iJ \alpha_2 &= 0
		\end{aligned}
\end{equation}
Similar to Eq.(\ref{26}), we have
\begin{equation}
		\begin{aligned}
&\alpha_1 = \frac{-iF}{z_1 + \frac{J^2}{z_2 + J^2/z_3}},\\
&\alpha_2 = \left( \frac{iJ}{z_2 + J^2/z_3} \right) \alpha_1,\\
&\alpha_3 = \left( \frac{iJ}{z_3} \right) \alpha_2
		\end{aligned}
\end{equation}
where $J\alpha_1$ and $J\alpha_2$ play the role of effective drive that allows the coherence propagation through the chain. In phase space, this means the centers of the Wigner distributions for all three sites are shifted away from $(0,0)$.
\begin{equation}
	\begin{aligned}
&|\alpha_3| = \frac{|F J^2|}{|z_1 z_2 z_3 + J^2(z_1 + z_3)|},\\
&\frac{|\alpha_3|}{|\alpha_2|} = \frac{|J|}{\sqrt{\Delta^2 + \frac{\gamma_3^2}{4}}},\\ &\frac{|\alpha_2|}{|\alpha_1|} = \frac{|J| \sqrt{\Delta^2 + \frac{\gamma_3^2}{4}}}{\sqrt{\left( J^2 - \Delta^2 + \frac{\gamma_2 \gamma_3}{4} \right)^2 + \Delta^2 \frac{(\gamma_2 + \gamma_3)^2}{4}}}
		\end{aligned}
\end{equation}

Despite the absence of $F$ in site 2 and site 3, all sites are connected by the hopping $J$: Since $\alpha_1 \propto F$, and since $\alpha_2$ is proportional to $\alpha_1$, and $\alpha_3$ is proportional to $\alpha_2$, all site amplitudes are linearly scaled by $F$, and the energy and the specific phase of the drive will transport through  $J$.
This is why in Fig.\ref{phase}(b), all three sites are drifted away from the origin. They have all inherited a portion of the drive's coherent phase and amplitude.
We set $\Delta = -1$ to consider the case that the pump frequency is higher than the cavity frequency.
As long as $|\Delta| < 1$, the bulk site has larger steady state amplitude than the drive site.

		\begin{figure}
	\centering
	\includegraphics[width=0.9\linewidth]{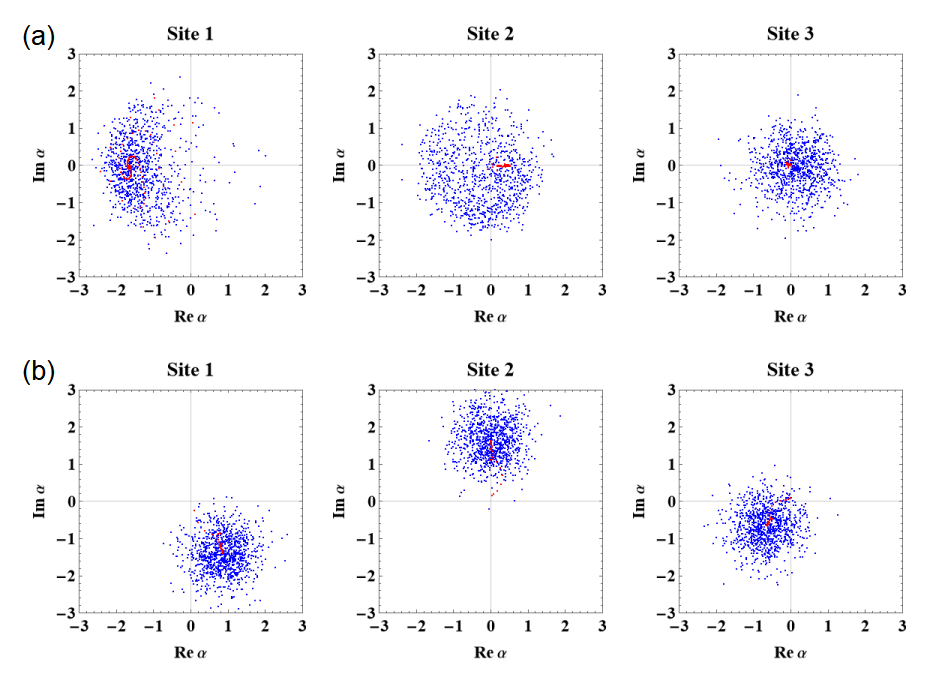}
	\caption{
Distribution of final-time amplitudes $\alpha_{l}(t)$
across the trajectory ensemble in phase space.
(a) The dephasing nonlinear regime ($U = 2.5,F=6,J=0.8$) where nonlinearity introduces phase scrambling. 
This reflects a intermediate phase between the linear phase and chaotic regime: the dephased nonlinear regime, which can also be referred to as symmetric regular phase or dissipatively stabilized phase.
At site 1, the drive $F$ is strong enough to maintain a coherent offset.
At sites 2 and 3, the scrambling is stronger than the coherence transfer $J$ and cause strong dephasing
where the operator OTOC decays to zero at long time by dissipation and by phase averaging
This generates diffusion and randomizes the phase, and the Wigner distribution is centered at (0,0), effectively restoring the $U(1)$ symmetry.
The nonlinear phase-diffusive ring distribution can be observed only at site 2 and site 3.
(b) The linear regime ($U=0,F=1,J=0.8$) where all sites are displaced from the origin. The phase is pinned and U(1) symmetry is broken by the drive because the coherence $J$ is dominate over the randomizing force.
Site 2 and site 3 signify the thermal state with small local chemical potential such that the photons are not numerous enough to overcome vacuum fluctuations, so the distribution is centered at the origin and filled in by noise. This explain the absence of photon blockade and the scrambled prethermal phase.
	}
	\label{phase}
\end{figure}

		\begin{figure}
	\centering
	\includegraphics[width=0.9\linewidth]{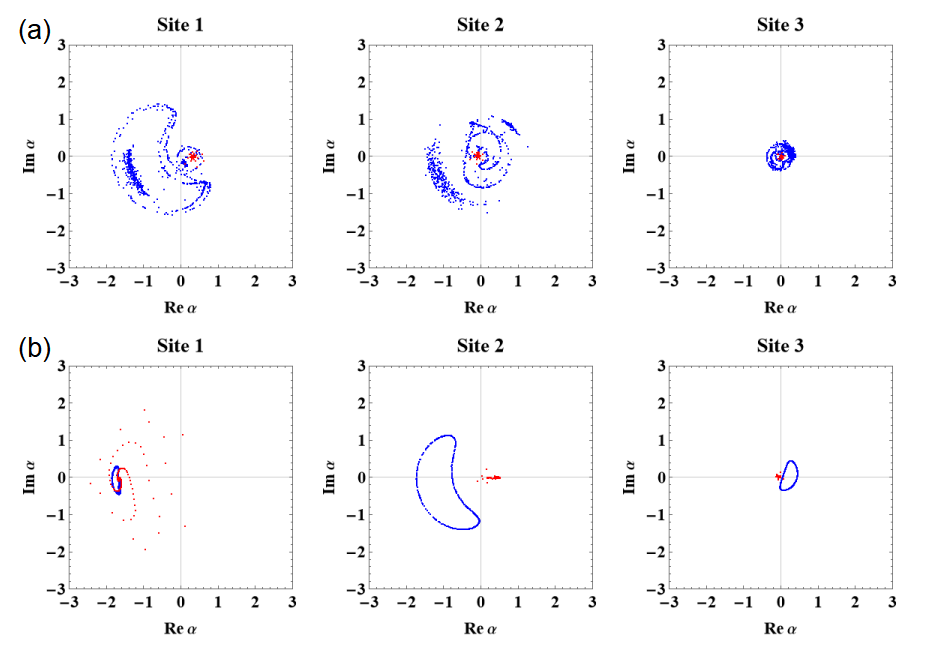}
	\caption{
Phase-space distributions under noiseless evolution. (a) shows the response under a weak coherent drive ($F=1.0$), while (b) represents a strong drive ($F=6.0$). Blue points represent the ensemble endpoint distribution of 1,000 trajectories at $t_{\text{max}}=50$, each initialized with random vacuum Wigner fluctuations to simulate initial quantum uncertainty. Red trajectories illustrate the deterministic path of a single initial condition.
At weak drive, site 1 exhibits a diffused, phase-nonuniform limit cycle where nonlinearity shears the initial Wigner phase uncertainty into a broad arc. Site 3 remains a stable fixed-point near the origin due to high local dissipation.
At strong drive, site 1 converges to a stable fixed-point attractor. Site 2 displays a deterministic, phase-nonuniform limit cycle, appearing as a phase-nonuniform transient limit cycle as shown in blue due to the competition between nonlinear phase-shearing and the coherent phase-locking of the drive. 
This periodic pattern reflects the absence of chaos.
	}
\label{noiseless}
\end{figure}

In Fig.\ref{noiseless}, we set the dynamical noise as zero.
The cloud-like distribution vanish
in the absence of noise-driven phase diffusion,
where there is only the initial noise from the vacuum fluctuations (i.e., deterministic evolution).
The blue points show an steady-state 
ensemble snapshot distribution at long time, where we consider 1000 trajectories each corresponds to a unique initial condition. 
The spread of blue points reflects phase diffusion and noise-induced mixing in the stochastic semiclassical framework.
The red points show a single-run time-parametrized path in phase space (a time series from one single initial condition).
This proves that the decay is due to the classical phase diffusion and damping instead of deterministic chaos.
Thus our result shows that the random initial phases and dynamical noise cannot guarantees the uniform phase mixing and hence a uniformly filled ring by itself.

Site 2 for $F=6.0$ exhibits a phase-nonuniform attractor due to the phase-locking.
The increased field amplitude $|\alpha|\propto F/\gamma$ strengthens the deterministic transient limit-cycle dynamics\cite{Garbin} where system's amplitude and phase oscillate deterministically and phase along the cycle depends on the random initial condition.
For $F=1$ at site 1, there is a diffused phase-nonuniform transient limit cycle.
Here the spreading arises from nonlinear phase shearing of the initial Wigner phase uncertainty by $U|\alpha|^2$ 
(deterministic nonlinear mapping of initial quantum fluctuations into a nonlinear phase space) as well as the finite-time deterministic sampling\cite{Drummond}, rather than environmental noise.
For $F=6$ at site 1, the red trajectory spirals toward a stable fixed point attractor (equilibrium steady state).
For $F=1.0$ at site 2, there is a diffused ring
(partially phase-locked) where the phase is relatively undetermined.
 Because the drive is weaker, the phase-locking is less dominant, allowing the noise to spread the points more effectively around the circle.
The spread of the noisy (TWA/Langevin) ensemble around the deterministic attractor quantifies noise-induced diffusion phase diffusion. 

The nonlinearity $U$ causes different amplitudes to rotate at different angular velocities, stretching the initial Wigner distribution into a ring or arc.
The larger dissipation at site 3 suppresses the amplitude and operator growth, results in a small phase-nonuniform limit-cycle (large $F$) or cluster (small $F$).
Transition from $F=1$ to $F=6$ shows the system moving from a regime dominated by quantum fluctuations to one dominated by nonlinear dynamical instabilities ($U|\alpha|^2$) with phase nonuniform ring.
The drive $F$ increases the steady-state amplitude $|\alpha|$ and nonlinear shift $U|\alpha|^2$
but meanwhile suppresses diffusion.
This can be evidenced by considering the drive $F$ with randomized drive phase across trajectories.
It randomize the phase to an extend independent of site.
While larger nonlinearity $U$ will cause larger difference of $|\alpha|$ between sites.

The Poincaré section analysis in Fig.\ref{poin}(a) reveals that the deterministic dynamics at site 2 for $F=6.0$ correspond to a stable fixed-point attractor undergoing an underdamped approach to the NESS. Although the trajectory initially appears as a transient limit cycle (Fig.\ref{noiseless}(b)) due to the large ratio $U/\gamma$, the sequential Poincaré crossings exhibit a damped transient deterministic drift toward a stable fixed point attractor. 
This drift confirms a slow dissipative underdamped convergence, where the coherent oscillations are gradually suppressed by the environment until the system settles into a phase-locked steady state. The nonuniform transient limit cycle observed in the TWA ensemble is thus a intermediate state of this slow relaxation process, capturing the distribution of initial Wigner fluctuations before fully collapsed onto the fixed point (fully damped transient oscillations).
Also, $Im \alpha_{2}(n)(\sim 1.02 Im \alpha_{2}(n+1))$ in Fig.\ref{poin}(b) is monotonically approaching the diagonal (from right to left), such that $Im \alpha_{2}(n) = Im \alpha_{2}(0) e^{-\kappa n}$ ($\kappa>0$).
For deterministic (noiseless) evolution toward its unique NESS, the crossings sequence (upper inset in (b)) reveals an exponential decay in the amplitude of oscillations, signifying a damped spiral approach to the attractor rather than a sustained limit cycle.
The temporal evolution of the field amplitudes $|\alpha_i(t)|$ in Fig.\ref{poin}(c) shows that all sites exhibit an underdamped relaxation that asymptotically saturates to a constant NESS, with $|\alpha_{1,ss}| \approx 1.65$, $|\alpha_{2,ss}| \approx 0.48$, and $|\alpha_{3,ss}| \approx 0.11$. This saturation effectively rules out the existence of permanent limit cycles or chaotic dynamics (where return map becomes a scatter cloud) or quasiperiodicity (where return map forms a curve or band without a monotonic approach) in this parameter regime.
This confirms that the system is in a stable nonlinear regime where the dissipative gap of the Liouvillian ensures that all initial perturbations—including the  fluctuations simulated by the TWA ensemble—are eventually lost. 

		\begin{figure}
	\centering
	\includegraphics[width=0.8\linewidth]{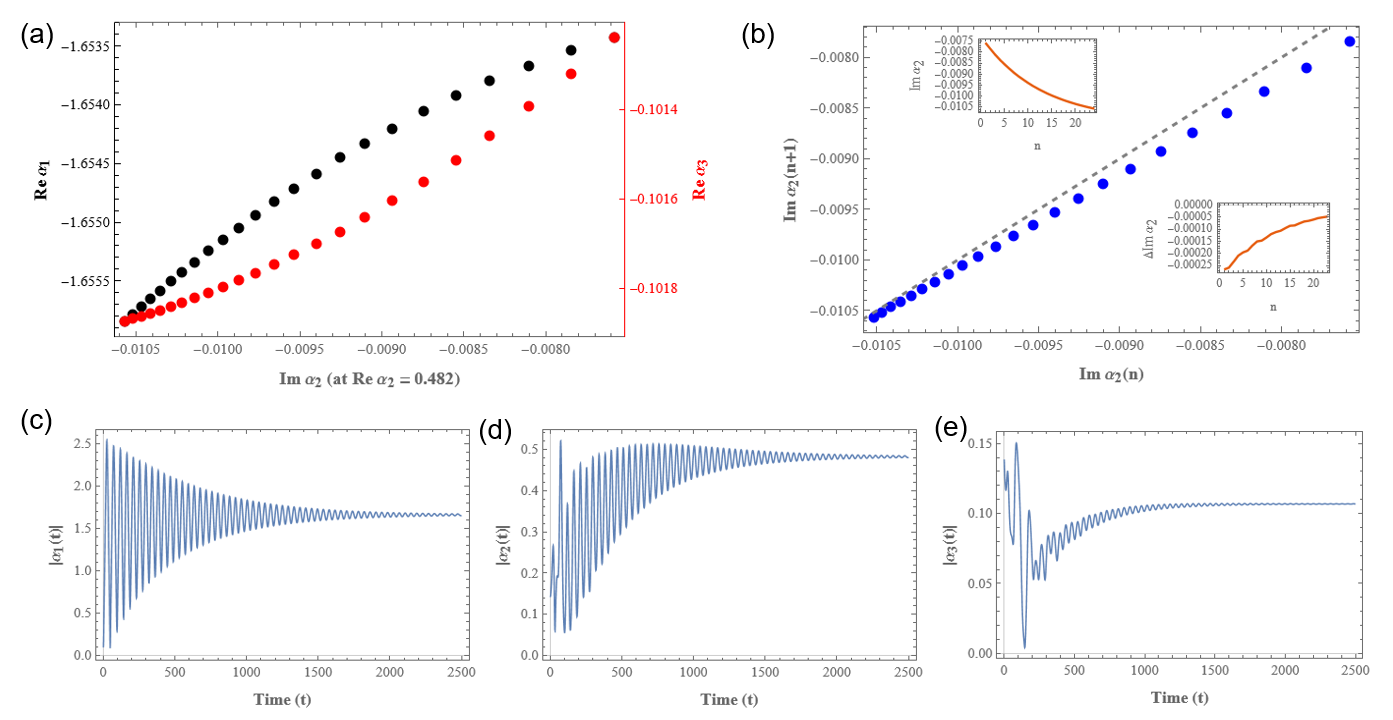}
	\caption{
Quantitative stability analysis via Poincaré section and return maps with $F=6, U=2.5, \gamma_{1,2}=0.2,\gamma_{3}=2$.
(a) Combined Poincaré section at $\text{Re}[\alpha_2]=c$ showing the synchronized drift of site 1 (black) and site 3 (red). (b) The return map for site 2 displays points along the diagonal, indicating a quasi-steady-state evolution. The upper inset is the time-series of the crossing points ${\rm Im}\alpha_{2}(n)$ which decrease monotonically.
This implies that the time sequence of return map is from right to left.
The lower inset is the increment ${\rm Im}\alpha_{2}(n+1)-{\rm Im}\alpha_{2}(n)$ which is negative and approaches zero. These results confirm that the deterministic dynamics represent an underdamped approach to a stable fixed-point attractor, effectively ruling out chaotic instabilities in this parameter regime.
(c) The evolution of the field amplitudes $|\alpha_i(t)|$. This confirms the fixed point type attractor, where $\frac{d\alpha_j}{dt} = 0$ and $\alpha_j(t) = |\alpha_j| e^{i\phi_j}$.
Note that this figure and Figs.\ref{phase},\ref{noiseless} consider the continuous semiclassical ODE evolution due to the TWA deterministic drift, they are independent of the local Hilbert space (Fock space) dimension $N_{max}$ (Gutzwiller truncation).}
\label{poin}
\end{figure}

		\begin{figure}
	\centering
	\includegraphics[width=0.5\linewidth]{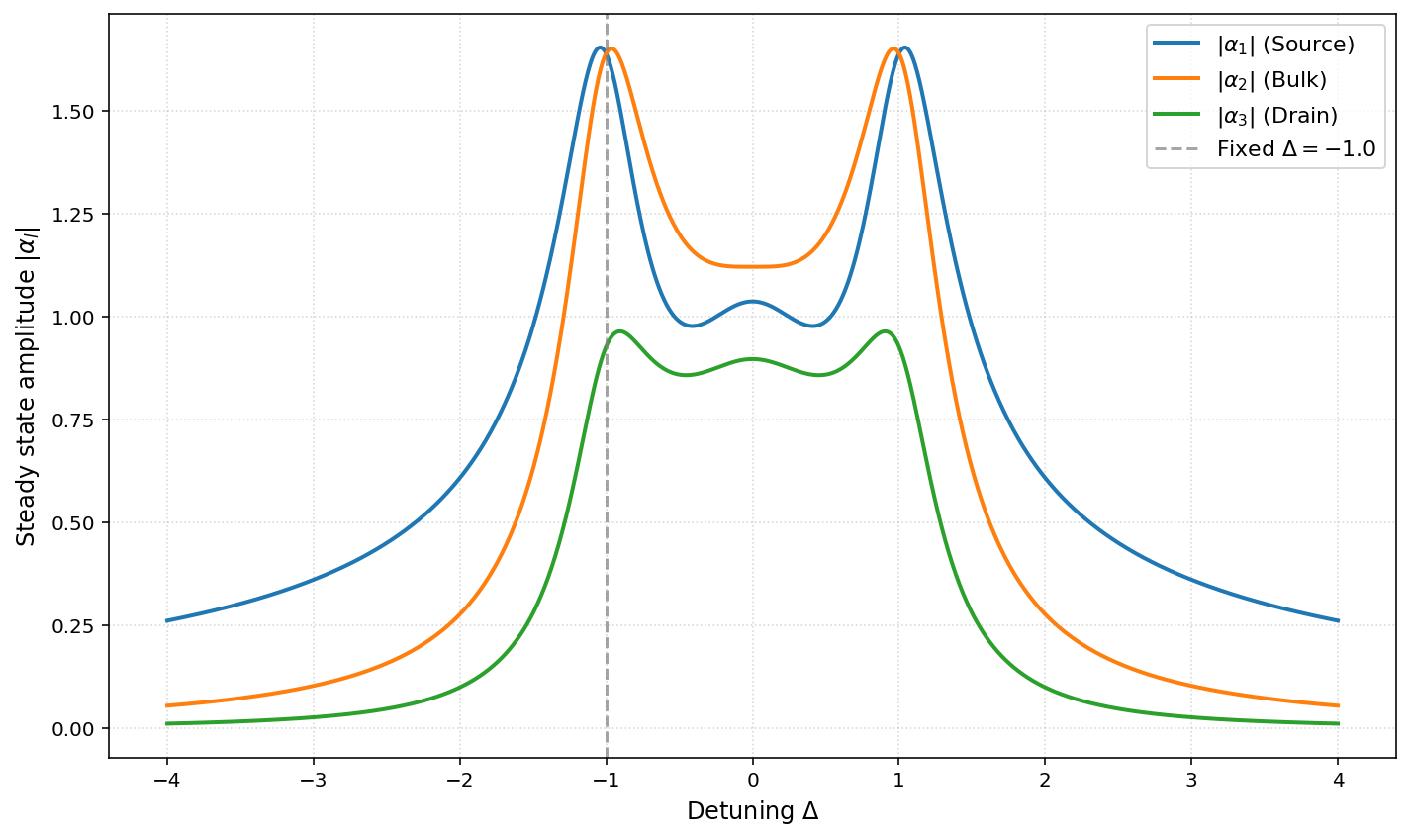}
	\caption{
Amplitudes as a function of driving frequencies (detuning $\Delta$) while $J$ and $F$ are held constant ($J=0.8, F=1.0$).
The spectrum shows three distinct resonance peaks. This is the hallmark of a 3-site system, where the coupling $J$ splits the single-site resonance into three normal modes (eigenfrequencies).
The blue ($|\alpha_1|$) and orange ($|\alpha_2|$) lines show sharp, high peaks because their local decay rates are small ($\gamma = 0.2$).The green line ($|\alpha_3|$) is much lower and the peaks are broader. This is caused by the strong dissipation at the drain site.
The dashed grey line marks the specific detuning $\Delta=-1$.
At this point, the system is near one of its side-resonances.
$|\alpha_2|$ actually slightly exceeds $|\alpha_1|$  here, showing an efficient transfer of energy into the center of the chain.$|\alpha_3|$ remains the lowest, at approximately $0.93$.
The master equation is mapped to the Fokker-Planck Equation (FPE) of Wigner function via Wigner mapping. The resulting PDE contains 1st order derivatives (drift) and 2nd order (diffusion), while the 3rd order derivatives (quantum corrections to the drift) are discarded. Each FPE is equivalents to a Lagevin equation\cite{Solanki}.}
\label{ampl2}
\end{figure}
		\begin{figure}
	\centering
	\includegraphics[width=0.9\linewidth]{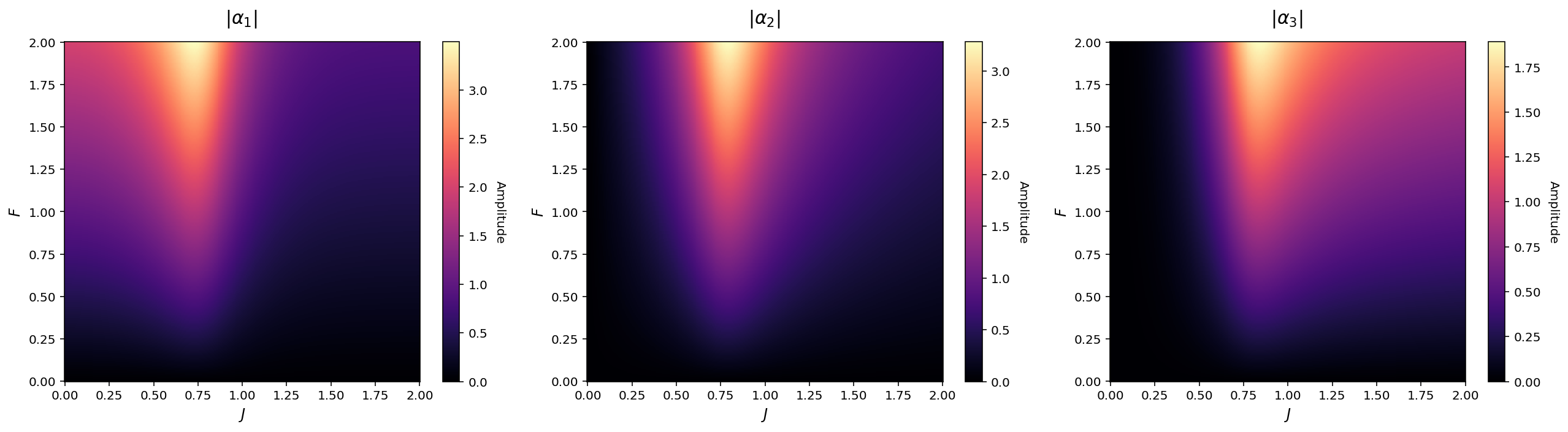}
	\caption{
Amplitude at each site as a function of coupling  ($J$) and the external driving strength ($F$), with the detuning fixed at $\Delta = -1$.
In all three sites, the amplitude increases linearly as the driving force $F$ increases. 
There is a distinct vertical bright region in all sites, most notably around $J \approx 0.75 - 1.0$. This represents the optimal coupling strength where the external driving frequency matches the system's normal modes.
$|\alpha_1|$ and $|\alpha_2|$ reach similar maximum values (around $3.5$ at $F=2.0$).
$|\alpha_3|$ has a significantly lower maximum amplitude (around $1.75$ at $F=2.0$). This is consistent with site 3 having a much higher decay rate ($\gamma_3 = 2.0$), which drains the energy faster than it can accumulate.
	}
	\label{map}
\end{figure}

In the absence of quantum noise (classical mean-field limit or deterministic limit),
the stochastic differential equation is reduced to the deterministic ordinary differential equation
\begin{equation}
	\frac{d\alpha_l}{dt} = \mathcal{A}_l(\{\alpha\})
= -i \left[ \Delta \alpha_l + U(|\alpha_l|^2 - 1)\alpha_l - J(\alpha_{l-1} + \alpha_{l+1}) + F\delta_{l,1} \right] - \frac{\gamma_l}{2} \alpha_l
\end{equation}
which is the damped Gross-Pitaevskii equation (GPE). 
The drift term $\mathcal{A}_l$ in TWA contains quantum bias inside the nonlinearity $U(|\alpha_l|^2 - 1)\alpha_l$ (quantum correction of the symmetric ordering used in Wigner-Weyl calculus). 
While a purely classical GPE would typically use $U|\alpha_l|^2\alpha_l$. 
The TWA drift accounts for the fact that even in the average motion, the particles are interacting with their own symmetric vacuum energy.
The initial fluctuations can be sampled from $\alpha_l(0) \sim \mathcal{N}(0, 1/2) + i\mathcal{N}(0, 1/2)$, which ensures the uncertainty required by Heisenberg principle even at $t=0$.
While the stochastic term $\sum_k \mathcal{B}_{lk} d\mathcal{W}_k$ cause the quantum noise entering from the environment and leads to dynamical fluctuations, which prevents the dissipative system to lose quantum uncertainty and collapse to a purely classical state with the deterministic drift $\mathcal{A}_l$.
The stochastic term also ensuring that the field $\alpha$ maintains a minimum variance of $1/2$ (the vacuum noise) even as it relaxes toward NESS.
Whether that NESS is coherent or chaotic depends on whether the variance stays at $1/2$ or blows up due to the nonlinear scrambling of those trajectories.

A challenge in TWA for nonlinear systems is the  divergence of trajectories for large $|\alpha_l|$.
The cubic term $U|\alpha_l|^2\alpha_l$ can lead to numerical overflows. We implement a fixed-step Heun's method (a predictor-corrector algorithm for SDEs) to enhance stability and utilize a sufficiently small time step $\Delta t$ to ensure that the approximation captures the competition between parametric gain and dissipation accurately.

The Wigner-Weyl correspondence leads to the following drift equation for site $l$:$$i \frac{\partial \alpha_l}{\partial t} = -f(\alpha_l) - J \sum_{k \in \text{neigh}} \alpha_k + F \delta_{l,1} - i \frac{\gamma_l}{2} \alpha_l$$
where $f(\alpha_l) = -\Delta \alpha_l - U(|\alpha_l|^2 - 1)\alpha_l$. Note the $-1$ correction in the Kerr term, which arises from the symmetric ordering of the operators.
The single-photon loss at rate $\gamma_l$ introduces Gaussian white noise $\xi_l(t)$. The full SDE (Langevin equation) in the Itô sense is
\begin{equation} 
d\alpha_l = \left[ -i(\Delta \alpha_l + U(|\alpha_l|^2-1)\alpha_l - J \sum_{k} \alpha_k + F \delta_{l,1}) - \frac{\gamma_l}{2} \alpha_l \right] dt + \sqrt{\frac{\gamma_l}{2}} d\mathcal{W}_l(t)
\end{equation} 
where $d\mathcal{W}_l$ is a complex Wiener process such that $\langle d\mathcal{W}_l d\mathcal{W}_k^* \rangle = \delta_{lk} dt$.
The diffusion coefficient $\sqrt{\gamma/2}$ appears in the stochastic update step of the Langevin equation is the pre-factor of the random noise (Wiener process) $dW$, $\alpha(t+\Delta t) = \alpha_{drift} + \sqrt{\frac{\gamma}{2}} d\mathcal{W}$, reflecting the effect from environment that balance the damping term ($-\gamma/2$) and maintaining the uncertainty principle.

\begin{algorithm}[H]
	\caption{Stabilized TWA for Driven-Dissipative NESS}
	\begin{algorithmic}[1]
		\STATE \textbf{Input:} Parameters $U, J, \Delta, \gamma, F$, time step $\Delta t$, total time $T_{max}$, ensemble size $M$.
		\STATE \textbf{Initialize:} 
		\STATE \quad Create an ensemble of $M$ trajectories.
		\STATE \quad \FOR{each trajectory $m=1$ to $M$}
		\STATE \quad \quad Sample $\alpha_l^{(m)}(0) \sim \mathcal{N}(0, 1/2) + i\mathcal{N}(0, 1/2)$ for all sites $l \in \{1, \dots, L\}$.
		\STATE \quad \ENDFOR
		\STATE \textbf{Time Evolution (Loop $t = 0$ to $T_{max}$ in steps of $\Delta t$):}
		\FOR{each trajectory $m=1$ to $M$}
		\STATE \textbf{1. Generate Noise:}
		\STATE \quad Sample complex Gaussian noise $dW_l$ such that $\text{Var}(\text{Re}, \text{Im}) = \Delta t/2$.
		
		\STATE \textbf{2. Predictor Step (Euler):}
		\STATE \quad $\text{hop}_l = \alpha_{l-1} + \alpha_{l+1}$
		\STATE \quad $\mathcal{A}_l(\alpha) = -i \left[ \Delta \alpha_l + U(|\alpha_l|^2 - 1)\alpha_l - J \cdot \text{hop}_l + F\delta_{l,1} \right] - \frac{\gamma_l}{2} \alpha_l$
		\STATE \quad $\alpha_l^{temp} = \alpha_l(t) + \mathcal{A}_l(\alpha(t)) \Delta t + \sqrt{\gamma_l/2} dW_l$
		
		\STATE \textbf{3. Corrector Step (Heun/RK2):}
		\STATE \quad $\alpha_l(t+\Delta t) = \alpha_l(t) + \frac{1}{2} \left[ \mathcal{A}_l(\alpha(t)) + \mathcal{A}_l(\alpha^{temp}) \right] \Delta t + \sqrt{\gamma_l/2} dW_l$
		
		\STATE \textbf{4. Stability Check:}
		\STATE \quad \textbf{IF} $\max|\alpha| > \text{Threshold}$ \textbf{THEN} discard trajectory.
		\ENDFOR
		\STATE \textbf{Observables (Ensemble Average):}
		\STATE \quad Compute mean occupation: $\langle n_l(t) \rangle = \frac{1}{M_{valid}} \sum_{m} (|\alpha_l^{(m)}(t)|^2 - 1/2)$.
		\STATE \quad Compute coherence: $\langle a_l^\dagger a_k \rangle = \frac{1}{M} \sum \alpha_l^* \alpha_k - \frac{1}{2}\delta_{lk}$.
		\STATE \textbf{Output:} Time-dependent population and steady-state correlation functions.
	\end{algorithmic}
\end{algorithm}

Thus the quantum bias $-1$ term in the drift and the $-1/2$ term in the occupation calculation are crucial to account for the commutation relations in the Wigner representation.
The Heun's metho using a predictor-corrector scheme for stochastic convergence in the presence of $U|\alpha|^2\alpha$ nonlinearities.
The vacuum Noise term $\sqrt{\gamma/2}$ ensures that the system satisfies the fluctuation-dissipation theorem, allowing it to relax to the correct NESS.

\section{Quantum metric framework}

For this time-independent linearized stability non-Hermitian matrix $\mathbf{L}_{\text{eff}}$ in Eq.(\ref{77}),
we can decomposition it into Pauli basis
\begin{equation}
	i\mathbf{L}_{\text{eff}}= \frac{-i\gamma}{2}\mathbf{I} + \frac{1}{2}\sum_{j=x,y,z}(\kappa_j + i\lambda_j)\sigma_j,
\end{equation}
where $\vec{\kappa} = (0, -2U \text{Im}(\alpha^2), 2\Omega)$ and $\vec{\lambda} = (0, 2U \text{Re}(\alpha^2), 0)$.
The metric operator $\rho(t)$\cite{Fring,Sim,Mostafazadeh}, which is Hermitian positive-definite and time-dependent to restore unitarity and preserve the normalization of fluctuations, has the form $\rho(t) = A(t)\mathbf{I} + ( \zeta_1(t) \vec{\kappa} + \zeta_2(t) \vec{\lambda} + \zeta_3(t) (\vec{\kappa} \times \vec{\lambda})) \cdot \vec{\sigma}$ with $A(t) =\frac{1}{2}\text{Tr}(\rho(t))$.
Since $-i\mathbf{L}_{\text{eff}}^{\dag}\rho(t)-\rho(t)i\mathbf{L}_{\text{eff}}=i\dot{\rho}(t)$, there is a set of coupled differential equations
\begin{equation}
	\begin{aligned}
		&\frac{d\zeta_1}{dt} = \zeta_3 (\vec{\kappa} \cdot \vec{\lambda}) = 0,\\
		&\frac{d\zeta_2}{dt} = -A(t) - |\vec{\kappa}|^2 \zeta_3(t),\\
		&\frac{d\zeta_3}{dt} = \zeta_2(t),\\
		&\frac{dA}{dt} = -\zeta_1 (\vec{\kappa} \cdot \vec{\lambda}) - |\vec{\lambda}|^2 \zeta_2(t) = -|\vec{\lambda}|^2 \zeta_2(t),
	\end{aligned}
\end{equation}
which can be solved as 
\begin{equation}
	\begin{aligned}
		&\zeta_1(t) = \zeta_1(0),\\
		&\frac{dA}{dt} = -|\vec{\lambda}|^2 \frac{d\zeta_3}{dt},\\
		&A(t) = A(0) - |\vec{\lambda}|^2 (\zeta_3(t) - \zeta_3(0)),\\
		&\ddot{\zeta}_3 + (|\vec{\kappa}|^2 - |\vec{\lambda}|^2) \zeta_3 = -A(0) - |\vec{\lambda}|^2 \zeta_3(0).
	\end{aligned}
\end{equation}
The eigenvalues of $\rho(t)$ are positive, thus $A(t) \pm |( \zeta_1(t) \vec{\kappa} + \zeta_2(t) \vec{\lambda} + \zeta_3(t) (\vec{\kappa} \times \vec{\lambda}))|>0$.
The eigenvalues of $i\mathbf{L}_{\text{eff}}$ are
$\lambda_{\pm} = -i \frac{\gamma}{2} \pm \sqrt{(\Delta + 2U|\alpha|^2 - U)^2 - U^2 |\alpha|^4}$, with the corresponding eigenvectors $\langle u_\pm|= 
(U \alpha^2 , \pm \sqrt{(\Delta + 2U|\alpha|^2 - U)^2 - U^2|\alpha|^4} - (\Delta + 2U|\alpha|^2 - U))$.
At exceptional point where $(\Delta + 2U|\alpha|^2 - U)^2  = U^2 |\alpha|^4$,
both the eigenvectors and eigenvalues coalesce at $\lambda = -i\gamma/2$, with $\langle u_\pm|=(U \alpha^2, - (\Delta + 2U|\alpha|^2 - U))$
and $\langle u_{+}|u_{-}\rangle=U^2 |\alpha|^4
+(\Delta + 2U|\alpha|^2 - U)^2=2U^2 |\alpha|^4$ can be normalized by $\sqrt{\langle u_{+}|u_{+}\rangle}\sqrt{\langle u_{-}|u_{-}\rangle}$,
where 
\begin{equation}
	\begin{aligned}
&\langle u_+ | u_+ \rangle = 2(\Delta + 2U|\alpha|^2 - U)^2 - 2(\Delta + 2U|\alpha|^2 - U)\sqrt{(\Delta + 2U|\alpha|^2 - U)^2 - U^2|\alpha|^4},\\
&\langle u_- | u_- \rangle = 2(\Delta + 2U|\alpha|^2 - U)^2 + 2(\Delta + 2U|\alpha|^2 - U)\sqrt{(\Delta + 2U|\alpha|^2 - U)^2 - U^2|\alpha|^4}.
\end{aligned}
\end{equation}

For $s=4(\Delta + 2U|\alpha|^2 - U)^2-4U^2 |\alpha|^4>0$, in which case the eigenvalues relative to the global decay $-i\gamma/2$ are real and thus corresponds to pure oscillations
$\zeta_3(t) = C_1 \cos(\sqrt{s}t) + C_2 \sin(\sqrt{s}t) + C_3$, we define $H_{PT}:=	i\mathbf{L}_{\text{eff}}+ \frac{i\gamma}{2}\mathbf{I}$
and $\rho(t)=e^{\gamma t}\rho_{0}$ with $\rho_{0}$ the geometric core,
then
the quasi-Hermiticity condition $i\dot{\rho}_{0} =  H_{PT}^{\dag}\rho_{0}- \rho_{0} H_{PT}=H_{PT}^{\dag}\rho(t)- \rho(t) H_{PT} = 0$
($(i\mathbf{L}_{\text{eff}})^{\dagger}\rho(t) - \rho(t) (i\mathbf{L}_{\text{eff}}) = i\gamma\rho(t)$) signifying a time-independent metric.
A static metric implies a static energy landscape\cite{Bongini}.
For a positive-definite metric $\rho=(\rho^{1/2})^{\dag}\rho^{1/2}$ (with unique Hermitian square root $\rho^{1/2}$) that meets the quasi-Hermiticity condition $H_{PT}^{\dag}=\rho_{0} H_{PT} \rho_{0}^{-1}=\rho(t) H_{PT} \rho^{-1}(t)$, $H_{PT}$ is similar to a Hermitian operator $\rho^{1/2}_{0} H_{PT} \rho^{-1/2}_{0}$.
The eigenvectors in mapped space are $|w_\pm\rangle = \rho_{0}^{1/2} |u_\pm\rangle$. 
The orthogonality $	\langle w_+ | w_- \rangle = \langle u_+ | \rho_{0} | u_- \rangle=\langle u_+ | \rho(t) | u_- \rangle=0$
where $|w_\pm\rangle$ are the eigenvectors of the Hermitian Hamiltonian $\rho_{0}^{1/2} H_{PT} \rho_{0}^{-1/2}$, and Hermitian operators always have orthogonal eigenvectors for distinct eigenvalues.
In the $\mathcal{PT}$-symmetric phase with stationary metric, it is constructed to satisfy the biorthogonality condition $\langle u_m | \rho | u_n \rangle = \delta_{mn}$. 
The stationary metric can be constructed into Hermitian form
$\rho_{0} = \sum_{n=\pm} |l_n\rangle\langle l_n|$,
where $\langle l_n |$ are the left eigenvectors of $H_{PT}$ or $i\mathbf{L}_{\text{eff}}$ (also, right eigenvectors of $H_{PT}^{\dag}$ or $i\mathbf{L}_{\text{eff}}$) satisfying $\langle l_m | u_n \rangle = \delta_{mn}$ (this  orthogonality condition is valid only in the PT-symmetric phase where the Hamiltonian is quasi-Hermitian). This construction ensures the metric is the inverse of overlap matrix of the right eigenvectors.
While in the PT-broken phase,
$\rho = \sum_n |l_n\rangle\langle l_n|$ results in a metric that is not positive-definite and corresponds to a non-unitary mapping. To preserve a consistent probability interpretation in the broken phase, the metric must be time-dependent.
The above similar transformation can be reproduced by the Dyson map 
\begin{equation}
\rho^{1/2}(t) i\mathbf{L}_{\text{eff}} \rho^{-1/2}(t) + i \dot{\rho^{1/2}}(t) \rho^{-1/2}(t)
= \rho_0^{1/2} H_{PT} \rho_0^{-1/2}
\end{equation}
where $\rho^{1/2}(t) i\mathbf{L}_{\text{eff}} \rho^{-1/2}(t) = (e^{\gamma t/ 2} \rho_0^{1/2}) (H_{PT} - i\frac{\gamma}{2}\mathbf{I}) (e^{-\gamma t/ 2} \rho_0^{-1/2}) = \rho_0^{1/2} H_{PT} \rho_0^{-1/2} - i\frac{\gamma}{2}\mathbf{I}$,
$i \dot{\rho^{1/2}}(t) \rho^{-1/2}(t) = i (\frac{\gamma}{2} \rho^{1/2}(t)) \rho^{-1/2}(t) = i\frac{\gamma}{2}\mathbf{I}$.

In $\mathcal{PT}$-Symmetric phase, the time-dependence of metric is only due to the global decay, $i\dot{\rho}(t)=i\gamma\rho(t)$,
and the state vector is
$|\mathbf{v}(t)\rangle = |\mathbf{v}(t)\rangle = (\delta \hat{a}(t) , \delta \hat{a}^\dagger(t))^T=e^{(-\gamma/2)t}(c_+ e^{i(-\sqrt{s}/2)t} |u_+\rangle + c_- e^{i(\sqrt{s}/2)t} |u_-\rangle)$,
such that 
\begin{equation}
\begin{aligned}
&\delta \hat{a}(t) = e^{-\frac{\gamma}{2}t} (c_+ e^{-i(\sqrt{s}/2) t} + c_- e^{i(\sqrt{s}/2) t}) U \alpha^2,\\
&\delta \hat{a}^\dagger(t) = e^{-\frac{\gamma}{2}t} \left[ c_+ e^{-i(\sqrt{s}/2) t}(\sqrt{s}/2 - (\Delta + 2U|\alpha|^2 - U)) - c_- e^{i(\sqrt{s}/2) t}(\sqrt{s}/2 + (\Delta + 2U|\alpha|^2 - U)) \right],\\
&|\delta a|^2 + |\delta a^\dagger|^2 = e^{-\gamma t} \{ |c_+|^2 \left[ |U \alpha^2|^2 + (\sqrt{s}/2 -  (\Delta + 2U|\alpha|^2 - U))^2 \right]
 + |c_-|^2 \left[ |U \alpha^2|^2 + (\sqrt{s}/2 +  (\Delta + 2U|\alpha|^2 - U))^2 \right]\\
&  +2\text{Re} \left( c_+ c_-^* e^{-2i(\sqrt{s}/2) t} \left[ |U \alpha^2|^2 - ((\sqrt{s}/2)^2 -  (\Delta + 2U|\alpha|^2 - U)^2) \right] \right) \}
\end{aligned}
\end{equation}
The Euclidean norm reads
\begin{equation}
	\langle \mathbf{v}(t) | \mathbf{v}(t) \rangle =
	e^{-\gamma t}[ |c_+|^2 \langle u_+ | u_+ \rangle + |c_-|^2 \langle u_- | u_- \rangle + 2\text{Re}(c_+^* c_- e^{2i(\sqrt{s}/2) t} \langle u_+ | u_- \rangle)].
\end{equation}
Because the eigenvectors are non-orthogonal ($\langle u_+ | u_- \rangle \neq 0$), the Euclidean norm oscillates in time even though the system is stable. 
The metric framework is used to define a metric that satisfies
$\langle \mathbf{v}(t) | \rho_0 | \mathbf{v}(t) \rangle = e^{-\gamma t} (|c_+|^2 + |c_-|^2)=e^{-\gamma t} $ and
$\langle \mathbf{v}(t) | \rho(t) | \mathbf{v}(t) \rangle = \langle \mathbf{v}(t) | e^{\gamma t} \rho_0 | \mathbf{v}(t) \rangle = |c_+|^2 + |c_-|^2 = 1$
(where we use $\langle u_{m}|\rho_{0}|u_{n}\rangle=\delta_{mn}$).

\begin{equation}
	\begin{aligned}
&\frac{d}{dt} \langle \mathbf{v}(t) | \rho(t) | \mathbf{v}(t) \rangle 
= \left( \frac{d}{dt} \langle \mathbf{v}(t) | \right) \rho |\mathbf{v}(t)\rangle + \langle \mathbf{v}(t) | \left( \frac{d}{dt} \rho \right) |\mathbf{v}(t)\rangle + \langle \mathbf{v}(t) | \rho \left( \frac{d}{dt} |\mathbf{v}(t)\rangle \right)\\
&= \langle \mathbf{v} | \mathbf{L}_{\text{eff}}^\dagger \rho |\mathbf{v}\rangle + \langle \mathbf{v} | \dot{\rho} |\mathbf{v}\rangle + \langle \mathbf{v} | \rho \mathbf{L}_{\text{eff}} |\mathbf{v}\rangle\\
&= \langle \mathbf{v} | \left[ (i H_{PT}^\dagger - \frac{\gamma}{2}) \rho + \gamma \rho + \rho (-i H_{PT} - \frac{\gamma}{2}) \right] | \mathbf{v} \rangle
=0.
	\end{aligned}
\end{equation}
The non-orthogonality cause the complex transient dynamics with amplified noise and fluctuations at short time in both the cases of $s<0$ and $s>0$\cite{Chen,Das}.

For $s<0$, in which case the eigenvalues relative to the global decay $-i\gamma/2$ are imaginary and thus corresponds to exponential growth and decay,
$\zeta_3(t) = C_1 \cosh(\sqrt{-s}t) + C_2 \sinh(\sqrt{-s}t) + C_3\sim e^{\sqrt{-s}t}$,
$A(t) \sim -|\vec{\lambda}|^2 \zeta_3(t) \sim - e^{2\Gamma t}$.
If the system enters the PT-broken phase where $\zeta_3(t)$ grows exponentially, $A(t)$ will evolve accordingly to ensure that the norm of the fluctuation vector $\langle \mathbf{v}(t) | \rho(t) | \mathbf{v}(t) \rangle =1$, which satisfies the time-dependent Schrodinger equation 
\begin{equation}
i \partial_t  |\mathbf{v}(t)\rangle  = -i\frac{\gamma}{2} |\mathbf{v}(t)\rangle + ie^{-\gamma t/2} \left( c_+ \Gamma e^{\Gamma t} |u_+\rangle - c_- \Gamma e^{-\Gamma t} |u_-\rangle \right)
=i\mathbf{L}_{\text{eff}}  |\mathbf{v}(t)\rangle
= i\mathbf{L}_{\text{eff}} e^{-\gamma t/2} \left( c_+ e^{\Gamma t} |u_+\rangle + c_- e^{-\Gamma t} |u_-\rangle \right).
\end{equation}
with 
\begin{equation}
		\begin{aligned}
	\mathbf{L}_{\text{eff}} |\mathbf{v}(t)\rangle &= e^{-\gamma t/2} \left( c_+ e^{\Gamma t} \mathbf{L}_{\text{eff}} |u_+\rangle + c_- e^{-\Gamma t} \mathbf{L}_{\text{eff}} |u_-\rangle \right) \nonumber \\
	&= e^{-\gamma t/2} \left[ c_+ e^{\Gamma t} \left( \Gamma - \frac{\gamma}{2} \right) |u_+\rangle + c_- e^{-\Gamma t} \left( -\Gamma - \frac{\gamma}{2} \right) |u_-\rangle \right] \nonumber \\
	&= -\frac{\gamma}{2} e^{-\gamma t/2} (c_+ e^{\Gamma t} |u_+\rangle + c_- e^{-\Gamma t} |u_-\rangle) + e^{-\gamma t/2} (c_+ \Gamma e^{\Gamma t} |u_+\rangle - c_- \Gamma e^{-\Gamma t} |u_-\rangle)
		\end{aligned}
\end{equation}
For $s<0$ we further have, 
\begin{equation}
	\begin{aligned}
&	i\mathbf{L}_{\text{eff}} |u_+\rangle = \left( -\sqrt{(\Delta+2U|\alpha|^2-U^2)^2-U^2 |\alpha|^4} - i\frac{\gamma}{2} \right) |u_+\rangle
 = \left( i\frac{\sqrt{-s}}{2} - i\frac{\gamma}{2} \right) |u_+\rangle,\\
&	i\mathbf{L}_{\text{eff}} |u_-\rangle = \left( \sqrt{(\Delta+2U|\alpha|^2-U^^2)^2-U^2 |\alpha|^4} - i\frac{\gamma}{2} \right) |u_-\rangle 
= \left( -i\frac{\sqrt{-s}}{2}- i\frac{\gamma}{2} \right) |u_-\rangle,\\
&-\langle u_+ | i\mathbf{L}_{\text{eff}}^\dagger = \left( -i\frac{\sqrt{-s}}{2} +i\frac{\gamma}{2} \right) \langle u_+ |,\\
&-\langle u_- | i\mathbf{L}_{\text{eff}}^\dagger = \left( i\frac{\sqrt{-s}}{2} +i \frac{\gamma}{2} \right) \langle u_- |
	\end{aligned}
\end{equation}
The dynamics of metric elements read
\begin{equation}
	\begin{aligned}
  \langle u_{+}|\dot{\rho}(t)|u_{+}\rangle &= - \langle u_+ | (\mathbf{L}_{\text{eff}}^\dagger \rho + \rho \mathbf{L}_{\text{eff}}) | u_+ \rangle = -2(\frac{\sqrt{-s}}{2} - \gamma/2)  \langle u_{+}|\rho(t)|u_{+}\rangle \implies 
\langle u_{+}|\rho(t)|u_{+}\rangle = \langle u_{+}|\rho(0)|u_{+}\rangle e^{-(2\frac{\sqrt{-s}}{2} - \gamma)t}, \\
\langle u_{-}|\dot{\rho}(t)|u_{-}\rangle &= - \langle u_- | (\mathbf{L}_{\text{eff}}^\dagger \rho + \rho \mathbf{L}_{\text{eff}}) | u_- \rangle = -2(-\frac{\sqrt{-s}}{2} - \gamma/2) \langle u_{-}|\rho(t)|u_{-}\rangle  \implies 
\langle u_{-}|\rho(t)|u_{-}\rangle  = \langle u_{-}|\rho(0)|u_{-}\rangle e^{(2\frac{\sqrt{-s}}{2} + \gamma)t},\\
  \langle u_{+}|\dot{\rho}(t)|u_{-}\rangle & = - \langle u_+ | (\mathbf{L}_{\text{eff}}^\dagger \rho + \rho \mathbf{L}_{\text{eff}}) | u_- \rangle = - [ (\frac{\sqrt{-s}}{2} - \gamma/2) + (-\frac{\sqrt{-s}}{2} - \gamma/2) ] \langle u_{+}|\rho(t)|u_{-}\rangle  = \gamma   \langle u_{+}|\rho(t)|u_{-}\rangle  \\
  &\implies 
 \langle u_{+}|\rho(t)|u_{-}\rangle =  \langle u_{+}|\rho(0)|u_{-}\rangle e^{\gamma t},\\
   \langle u_{-}|\dot{\rho}(t)|u_{+}\rangle & = - \langle u_- | (\mathbf{L}_{\text{eff}}^\dagger \rho + \rho \mathbf{L}_{\text{eff}}) | u_+ \rangle 
   = - [ (-\frac{\sqrt{-s}}{2} - \gamma/2) + (\frac{\sqrt{-s}}{2} - \gamma/2) ] \langle u_{-}|\rho(t)|u_{+}\rangle  = \gamma   \langle u_{-}|\rho(t)|u_{+}\rangle  \\
 &\implies 
 \langle u_{-}|\rho(t)|u_{+}\rangle =  \langle u_{-}|\rho(0)|u_{+}\rangle e^{\gamma t},\\
	\end{aligned}
\end{equation}
The norm reads
\begin{equation}
	\begin{aligned}
\langle \mathbf{v}(t) | \rho(t) | \mathbf{v}(t) \rangle 
	&= e^{-\gamma t} \left( c_+^* e^{\Gamma t} \langle u_+ | + c_-^* e^{-\Gamma t} \langle u_- | \right) \rho(t) \left( c_+ e^{\Gamma t} |u_+\rangle + c_- e^{-\Gamma t} |u_-\rangle \right)\\
	 &= e^{-\gamma t} [ |c_+|^2 e^{2\Gamma t} R_{++}(t) + |c_-|^2 e^{-2\Gamma t} R_{--}(t) + c_+^* c_- R_{+-}(t) + c_-^* c_+ R_{-+}(t) ]\\
	&= |c_+|^2 R_{++}(0) + |c_-|^2 R_{--}(0) + c_+^* c_- R_{+-}(0) + c_-^* c_+ R_{-+}(0) = \langle \mathbf{v}(0) | \rho(0) | \mathbf{v}(0) \rangle =1,
	\end{aligned}
\end{equation}
where the initial state is normalized under the initial metric such that $\langle \mathbf{v}(t) | \rho(t) | \mathbf{v}(t) \rangle =\langle \mathbf{v}(0) | \rho(0) | \mathbf{v}(0) \rangle = 1$.

The Euclidean norm $\langle \mathbf{v}(t) | \mathbf{v}(t) \rangle = |\delta a|^2 + |\delta a^\dagger|^2$ growing exponentially toward infinity,
representing the instability due to the drive, making the probability non-conserved.
The metric norm reads
$\langle \mathbf{v}(t) | \rho_0 | \mathbf{v}(t) \rangle = e^{-\gamma t} (|c_+|^2 e^{2\Gamma t} + |c_-|^2 e^{-2\Gamma t})$ which grows exponentially as $e^{(2\Gamma - \gamma)t}$.
$\rho(t)$ is now no longer $e^{\gamma t}\rho_0$ with static $\rho_0$. It follows $\rho(t) = e^{\gamma t} \rho_{core}(t)$, where $\rho_{core}$ evolves to cancel $e^{\pm 2\Gamma t}$, such that
$\langle \mathbf{v}(t) | \rho(t) | \mathbf{v}(t) \rangle  = 1$.
The unitarity restoration condition $-i\mathbf{L}_{\text{eff}}^{\dag}\rho(t)-\rho(t)i\mathbf{L}_{\text{eff}}=i\dot{\rho}(t)$ ensures that the metric $\rho(t)$ evolves against the non-Hermitian Langevin drift of $\mathbf{L}_{\text{eff}}$ to keep the norm of the fluctuations invariant, i.e.,
$\frac{d}{dt} \langle \mathbf{v}(t) | \rho(t) | \mathbf{v}(t) \rangle = 0$, and thus $\langle \mathbf{v}(t) | \mathbf{L}_{\text{eff}}^{\dagger} \rho(t) | \mathbf{v}(t) \rangle + \langle \mathbf{v}(t) | \dot{\rho}(t) | \mathbf{v}(t) \rangle + \langle \mathbf{v}(t) | \rho(t) \mathbf{L}_{\text{eff}} | \mathbf{v}(t) \rangle = 0$ due to the drift equations $\dot{\mathbf{v}} (t)= \mathbf{L}_{\text{eff}} \mathbf{v}(t)$ and $\langle \dot{\mathbf{v}}(t) | = \langle \mathbf{v}(t) | \mathbf{L}_{\text{eff}}^{\dagger}$.
 Because a matrix with complex eigenvalues cannot meet the quasi-Hermiticity condition.
 In the PT broken phase, the system is inherently non-conservative (gaining/losing energy exponentially). 
The metric $\rho(t)$ serves as a dynamical renormalization tool and ensures the conserved probability density (unity metric norm).

The geometric core part satisfies the evolution based on the non-dissipative part $H_{PT}$
\begin{equation}
	\begin{aligned}
&	i\dot{\rho}_{core} = H_{PT}^\dagger \rho_{core} - \rho_{core} H_{PT},
\end{aligned}
\end{equation}
and thus $H_{PT}^{\dagger}\rho(t) - \rho(t) H_{PT} = e^{\gamma t} (H_{PT}^{\dagger}\rho_{core} - \rho_{core} H_{PT}) = e^{\gamma t} i\dot{\rho}_{core}$.
In the $\mathcal{PT}$-symmetric phase, $\rho_{core} = \rho_0$ is static, so $i\dot{\rho}_0 = 0$. In the $\mathcal{PT}$-broken phase, $\rho_{core}(t)$ must evolve to compensate for complex eigenvalues.
For the full scaling metric $\rho(t) = e^{\gamma t} \rho_{core}(t)$, the evolution under the full Hamiltonian $H$ is
\begin{equation}
	i\dot{\rho}(t) = (H_{PT}^\dagger \rho(t) - \rho(t) H_{PT}) + i\gamma \rho(t)
\end{equation}
Substituting $\rho(t) = e^{\gamma t} \rho_{core}(t)$, we find
\begin{equation}
	i (\gamma e^{\gamma t} \rho_{core}(t) + e^{\gamma t} \dot{\rho}_{core}(t)) = e^{\gamma t} (H_{PT}^\dagger \rho_{core}(t) - \rho_{core}(t) H_{PT}) + i\gamma e^{\gamma t} \rho_{core}(t)
\end{equation}
This confirms that the global scaling $e^{\gamma t}$ handles the identity-like dissipation $-i\gamma/2$, while $\rho_{core}(t)$ handles the internal $\mathcal{PT}$ geometry.
Using the Dyson map $\rho(t)^{1/2} = e^{\gamma t / 2} \rho_{core}^{1/2}(t)$, the mapped Hermitian operator reads
\begin{equation}
e^{\gamma t / 2} \rho_{core}^{1/2} (H_{PT} - i\frac{\gamma}{2}) e^{-\gamma t / 2} \rho_{core}^{-1/2} + i (\frac{\gamma}{2} e^{\gamma t / 2} \rho_{core}^{1/2} + e^{\gamma t / 2} \dot{\rho}_{core}^{1/2}) (e^{-\gamma t / 2} \rho_{core}^{-1/2})
 = \rho_{core}^{1/2}(t) H_{PT} \rho_{core}^{-1/2}(t) + i \dot{\rho}_{core}^{1/2}(t) \rho_{core}^{-1/2}(t)
\end{equation}

For $s<0$, the eigenvalues of $i\mathbf{L}_{\text{eff}}$ is
$\lambda_{\pm} =-i\gamma/2 \pm i \sqrt{ U^2 |\alpha|^4-(\Delta + 2U|\alpha|^2 - U)^2}$ and the corresponding eigenvectors are $|u_{\pm}\rangle$. Then
the fluctuation vector $|\mathbf{v}(t)\rangle$ evolves as a superposition of these two eigenvectors
$|\mathbf{v}(t)\rangle =e^{(-\gamma/2)t}( c_+ e^{(\sqrt{-s}/2) t} |u_+\rangle + c_- e^{(-\sqrt{-s}/2) t} |u_-\rangle$,
where $c_{\pm}$ are expansion coefficients determined by initial conditions at $t=0$ ($|\mathbf{v}(0)\rangle = c_+ |u_+\rangle + c_- |u_-\rangle$).
In non-Hermitian systems, these two eigenvectors are generally non-orthogonal.
The Euclidean norm behaves as 
\begin{equation}
	\langle \mathbf{v}(t) | \mathbf{v}(t) \rangle = e^{-\gamma t} \left[ |c_+|^2 e^{2(\sqrt{-s}/2) t} \langle u_+ | u_+ \rangle + |c_-|^2 e^{-2(\sqrt{-s}/2) t} \langle u_- | u_- \rangle + 2\text{Re}(c_+^* c_- \langle u_+ | u_- \rangle) \right]
\end{equation}
which $\sim e^{(-\gamma+\sqrt{-s}) t}$ at long time.
The individual instantaneous eigenvectors $|u_\pm\rangle$ with complex eigenvalues do not maintain a simple biorthogonal relationship under a single evolving metric.

In Hermitian limit ($\gamma \to 0$ and $U \to 0$ in the relevant terms), the metric $\rho(t)$ and Dyson map tend to the identity operator.
As the non-Hermitian contribution to the Hamiltonian vanishes, the right eigenvectors $|u_\pm\rangle$ become orthogonal ($\langle u_+ | u_- \rangle \to 0$), and since the left and right eigenvectors coincide ($|l_n\rangle = |u_n\rangle$), the metric simplifies to $\rho = \sum_n |u_n\rangle\langle u_n| = \mathbb{I}$.


\section{Conclusion}

In summary, we have systematically investigated the non-equilibrium steady states in a driven-dissipative three-site Bose-Hubbard chain. By employing a self-consistent Gutzwiller mean-field approach, we identify a distinct intermediate stable nonlinear regime situated between the quasilinear and chaotic phases. This regime is uniquely characterized by the restoration of $U(1)$ symmetry and significant phase dephasing, where the operator OTOC exhibits a transient increase followed by an exponential decay to zero, signaling a stable but non-coherent attractor. In contrast, the transition to the genuine chaotic regime is marked by a non-vanishing Liouvillian gap and the macroscopic saturation of the OTOC, providing a clear signature of persistent information scrambling. Comparison between the mean-field approximation and exact many-body solvers further reveals that while local fluctuations are well-captured by the Gutzwiller approach, inter-site correlations and spatial information spreading are essential. Our choice of the three-site model proves critical, as it provides the minimal lattice depth necessary to observe bulk phase-decoupling while remaining below the threshold for global many-body instability. 

Our quantitative stability analysis, particularly through Poincaré sections and return maps, reveals that in the high-drive regime ($F=6.0$), the deterministic dynamics correspond to a stable fixed-point attractor rather than a permanent limit cycle. The observed phase-space spreading in the TWA ensemble is identified as a consequence of nonlinear phase shearing of initial Wigner fluctuations during an underdamped approach to the NESS, where coherent oscillations are gradually suppressed by the environment. Furthermore, our results demonstrate that local Kerr nonlinearity $U$ can effectively restore $U(1)$ symmetry at the bulk and drain sites by scrambling the phase information inherited from the coherent drive. The decay of the OTOC to zero in these stable regimes provides a clear signature that information scrambling is eventually overtaken by dissipative damping. 
Thus the effect of $U$ is to cause the stable dephasing/incoherent nonlinear regime with stable contractive/dephased NESS (due to the dissipation which damp the perturbations)
instead of the scrambling-dominated chaotic regime
since the deterministic instability is suppressed by the dissipation.
This numerical framework and the diagnostic methods developed here offer a versatile pathway for exploring multi-body correlations and stability in larger-scale open photonic or superconducting lattices. These findings offer a precise framework for diagnosing the boundaries between stable nonlinear dynamics and many-body chaos in open quantum systems.

\section{Appendix.A: Relationship between Degrees of Freedom, Lattice Size, and the Transition to Chaos}

The semiclassical dynamics are governed by $2L=6$ coupled nonlinear ordinary differential equations (ODEs). Each site $j$ is described by a complex amplitude $\alpha_j$, leading to a real phase-space dimension of $2L=6$. The evolution of the system state vector $\mathbf{X} = (\text{Re}\,\alpha_1, \text{Im}\,\alpha_1, \dots, \text{Re}\,\alpha_L, \text{Im}\,\alpha_L)^T$ can be written as $\dot{\mathbf{X}} = \mathbf{f}(\mathbf{X})$.
The local stability of NESS is determined by the $2L \times 2L$ Jacobian matrix $\mathbf{J}$, where $J_{ik} = \partial f_i / \partial X_k$. 
For a Lindblad-type dissipative system, the instantaneous rate of phase-space volume contraction is given by the trace of the Jacobian
\begin{equation}
\text{Tr}(\mathbf{J}) = \sum_{j=1}^L \left( \frac{\partial \dot{\alpha}_j}{\partial \alpha_j} + \frac{\partial \dot{\alpha}_j^*}{\partial \alpha_j^*} \right) =\sum_{i=1}^{2L} \lambda_i= - \sum_{j=1}^L \gamma_j<0
\end{equation}
where $\gamma_j$ represents the local dissipation rate at site $j$. The coherent drive $F$ and Kerr nonlinearity $U$ contribute only to the conservative (imaginary) part of the drift, which is divergence-free. Thus the total phase-space volume undergoes a global contraction at a constant rate determined solely by the dissipation $\gamma_j$.
This identity implies that the system is dissipative and the phase-space volume contracts globally at a constant rate $\sum \gamma_j$, forcing trajectories toward a lower-dimensional attractor (such as a fixed point, limit cycle, or strange attractor).


The sensitivity of the system to initial conditions is quantified by the spectrum of Lyapunov exponents $\{\lambda_i\}_{i=1}^{2L}$ of the semiclassical dissipative flow (not the Lyapunov exponents of full quantum Liouvillian dynamics). These exponents represent the long-term average rates of exponential divergence or convergence of nearby trajectories. According to the Oseledec theorem, 
the sum of the Lyapunov spectrum must satisfy the global volume contraction constraint
$ \text{Tr}(\mathbf{J}) \rangle_t <0$.
Deterministic chaos occurs if the maximum Lyapunov exponent is positive, $\lambda_{\text{max}} > 0$. 
In our three-site driven-dissipative system, a positive maximal Lyapunov exponent (unstable directions driven by nonlinearity $U$ and drive $F$) does not exist due to the strong dissipative contraction in the remaining stable directions to satisfy the global volume-contraction constraint. 


The transition to chaos becomes increasingly favorable as the number of sites $L$ increases.
The number of collective modes and parametric instabilities increases with system size.
The number of nonzero off-diagonal coupling terms in the $2L \times 2L$ Jacobian increases as $L$. As $L$ grows, $U$ at each site facilitates complex feedback loops and mode-mixing along the chain. These increased interaction channels provide more pathways for parametric instabilities to develop.

As $L$ increases, the number of collective modes and available phase-space routes for nonlinear mixing and instability also increases,  
and the dissipation is distributed over a broader spectrum, allowing localized and collective nonlinear excitations to realize $\lambda_{\text{max}}>0$,
thereby broadening the parameter region where chaos may appear.
Stable OTOC decay observed in the present three-site chain is consistent with a strongly contractive low-dimensional stable nonlinear regime rather than a fully developed chaotic attractor,
where $F$ confining the system to a contractive manifold that collapses into a stable fixed point attractor at NESS and the dissipation completely dominates the nonlinear expansion.

The complexity of the resulting attractor can be characterized by the Kaplan-Yorke dimension $D_{KY}$:$$D_{KY} = k + \frac{\sum_{i=1}^k \lambda_i}{|\lambda_{k+1}|}$$where $k$ is the largest integer such that the sum of the first $k$ exponents is non-negative. 
The dissipation completely dominates the nonlinear expansion, leading to $D_{KY} \to 0$ (a fixed-point attractor). As $L$ increases toward the many-body limit, the emergence of $\lambda_{\text{max}} > 0$ leads to a fractal $D_{KY} > 2$, signaling the onset of quantum chaos and the corresponding exponential growth of the OTOC.

\section{Appendix A: Detailed Derivation of the Short-time Perturbative Expansion}

To obtain the self-consistent solution of NESS,
we assume the chain starts in vacuum state
$\rho_{r}(0)=|0\rangle_{r}\langle 0|_{r}$,
whose expectation is $\Psi_{r}(0)={\rm Tr}[a_{r}\rho_{r}(0)]=0$.
For the drive site $r=1$,
$\rho_{1}(0)=|0\rangle_{1}\langle 0|_{1}$,
\begin{equation} 
	\begin{aligned}
		&	\mathcal{L}_1(\rho_1(0)) = -i[H_{{\rm MF},1}(0), \rho_1(0)] + \mathcal{D}_1[\rho_1(0)]
		\\&	=-i[\Delta \hat{a}_1^\dagger \hat{a}_1 + \frac{U}{2} a_1^\dagger \hat{a}_1^\dagger a_1 a_1
		-J ( \Psi_{2} (0) a_r^\dagger + 
		\Psi_{2}^* (0) a_r )+F(a_{1}^{\dag}+a_{1})
		, \rho_1(0)]\\
		&= -iF \left( (\hat{a}^\dagger_1 + \hat{a}_1)|0\rangle_{1}\langle 0|_{1} - |0\rangle_{1}\langle 0|_{1}(a^\dagger_1 + a_1) \right)\\
		&= -iF \left( |1\rangle_{1}\langle 0|_{1} - |0\rangle_{1}\langle 1|_{1} \right)
	\end{aligned}
\end{equation}
where we use $\mathcal{D}_1[\rho_1(0)]=0$,
$[\Delta \hat{a}_1^\dagger \hat{a}_1 + \frac{U}{2} a_1^\dagger \hat{a}_1^\dagger a_1 a_1, \rho_1(0)]=0$,
$a_1|0\rangle_{1}=0$, $a_1|1\rangle_{1}=|0\rangle_{1}$, $a^\dagger_1|0\rangle_{1}=|1\rangle_{1}$,
$a^\dagger_1|1\rangle_{1}=\sqrt{2}|2\rangle_{1}$,
$\langle 0|_{1}a^{\dag}_{1}=(a_{1}|0\rangle_{1})^{\dag}=0$,
$\langle 1|_{1}a^{\dag}_{1}=(a_{1}|1\rangle_{1})^{\dag}=\langle 0|$,
$\langle 1|_{1}a_{1}=(a^{\dag}_{1}|1\rangle_{1})^{\dag}
=\sqrt{2}\langle 2|$.
For number (Fock) basis $\{|k\rangle\}$,
${\rm Tr}[a|n\rangle\langle m|]
=\sum_{k=0}^{\infty}\langle k|a|n\rangle\langle m|k\rangle
=\sum_{k=0}^{\infty}\sqrt{n}\delta_{k,n-1}\delta_{m,k}=\sqrt{n}\delta_{m,n-1}$. Similarly,
$\text{Tr}[\hat{a}_l \sum_{n,m} c_{nm} |n\rangle\langle m|] = \sum_{n,m} c_{nm} \text{Tr}[\hat{a}_l |n\rangle\langle m|]= \sum_{n,m} c_{nm} (\sqrt{n} \delta_{m, n-1})= \sum_{n=1}^{\infty} c_{n, n-1} \sqrt{n}$.
Thus
\begin{equation} 
	\begin{aligned}
		\rho_1(dt) = \rho_1(0) + \mathcal{L}_1(\rho_1(0))  dt
		= |0\rangle_{1}\langle 0|_{1} - iF dt  \left( |1\rangle_{1}\langle 0|_{1} - |0\rangle_{1}\langle 1|_{1} \right)
	\end{aligned}
\end{equation}
where we use the Euler method approximation
\begin{equation} 
	\begin{aligned}
		\rho_r(dt) = \rho_r(0) + \left( \frac{d\rho_r}{dt} \bigg|_{t=0} \right)  dt
		= \rho_r(0) + \mathcal{L}_l(\rho_r(0))  dt.
	\end{aligned}
\end{equation}
The corresponding condensate amplitude reads
\begin{equation} 
	\begin{aligned}
		\Psi_1(dt) = \text{Tr}\left[ \hat{a}_1 \cdot \left( |0\rangle_{1}\langle 0|_{1} - iF dt (|1\rangle_{1}\langle 0|_{1} - |0\rangle_{1}\langle 1|_{1}) \right) \right]
		=\text{Tr}[ -iF dt  |0\rangle_{1}\langle 0|_{1}]
		= -i F  dt
	\end{aligned}
\end{equation}

For bulk site $1<r<L$, $\rho_{r}(0)=|0\rangle_{r}\langle 0|_{r}$, and
$\mathcal{L}_r(\rho_r(0)) = -i[
\Delta \hat{a}_r^\dagger \hat{a}_r + \frac{U}{2} a_r^\dagger \hat{a}_r^\dagger a_r a_r, \rho_r(0)] + \mathcal{D}_r[\rho_r(0)]=0$ since there is no loss in bulk $\mathcal{D}_r[\rho_r(0)] = 0$.
Thus $\rho_r(dt) = |0\rangle_{r}\langle 0|_{r}$ and $\Psi_{r}(dt)={\rm Tr}[a_{r}\rho_{r}(dt)]=0$, i.e., it remains in vacuum state in the first time step.
For the loss site we have
\begin{equation} 
	\begin{aligned}
		&\mathcal{L}_L(\rho_L(0)) = -i[H_{{\rm MF},L}(0), \rho_L(0)] + \mathcal{D}_L[\rho_L(0)]
		=  \mathcal{D}_L[\rho_L(0)]\\
		&= \gamma \hat{a}_L |0\rangle_{L}\langle 0|_{L} \hat{a}_L^\dagger - \frac{\gamma}{2}\{\hat{a}_L^\dagger \hat{a}_L, |0\rangle_{L}\langle 0|_{L} \}=0
	\end{aligned}
\end{equation}
where we use $a_L |0\rangle = 0$ and $a_L^\dagger a_L |0\rangle = 0$. Thus $\rho_L(dt)  = |0\rangle_{L}\langle 0|_{L}$ and $\Psi_{L}(dt)=0$.
Now the mean-field single-site Hamiltonian for bath reads 
\begin{equation} 
	\begin{aligned}
		&		H_{{\rm MF},1}(dt) = H_{loc,1} + F(\hat{a}^\dagger_1 + \hat{a}_1),\\
		&H_{{\rm MF},2}(dt) = H_{loc,2} - J[(\Psi_{1}(dt) + \Psi_{3}(dt))a_2^\dagger + h.c.]
		= H_{loc,2} - J[(-iF  dt)a_2^\dagger + (iF  dt)a_2],\\
		&H_{{\rm MF},r}(dt) = H_{loc,r}, (for\ 3\le r\le L).
	\end{aligned}
\end{equation}

At $2dt$, the densities read
\begin{equation} 
	\begin{aligned}
		&\rho_{1}(2dt)=\rho_{1}(dt)+\mathcal{L}_{1}(\rho_{1}(dt))dt\\
		&= |0\rangle_{1}\langle 0|_{1} - iF dt  \left( |1\rangle_{1}\langle 0|_{1} - |0\rangle_{1}\langle 1|_{1} \right)
		-i[H_{{\rm MF},1}(dt),\rho_{1}(dt)]dt+\mathcal{D}_{1}[\rho_{1}(dt)]dt\\
		&= |0\rangle_{1}\langle 0|_{1} - iF dt  \left( |1\rangle_{1}\langle 0|_{1} - |0\rangle_{1}\langle 1|_{1} \right)
		-i[H_{loc,1}+F(a_{1}^{\dag}+a_{1}),\rho_{1}(dt)]dt
		+\mathcal{D}_{1}[\rho_{1}(dt)]dt\\
		&= |0\rangle_{1}\langle 0|_{1} - 2iFdt (|1\rangle_{1}\langle 0|_{1} - |0\rangle_{1}\langle 1|_{1}) - F\Delta(dt)^2 (|1\rangle_{1}\langle 0|_{1} + |0\rangle_{1}\langle 1|_{1})\\
		& - F^2(dt)^2 \left[ 2|0\rangle_{1}\langle 0|_{1} - 2|1\rangle_{1}\langle 1|_{1} + \sqrt{2}(|2\rangle_{1}\langle 0|_{1} + |0\rangle_{1}\langle 2|_{1}) \right],\\
		&\rho_{2}(2dt)=\rho_{2}(dt)+\mathcal{L}_{2}(\rho_{2}(dt))dt
		= |0\rangle_{2}\langle 0|_{2}
		-i[H_{{\rm MF},2}(dt),\rho_{2}(dt)]dt+\mathcal{D}_{2}[\rho_{2}(dt)]dt\\
		&= |0\rangle_{2}\langle 0|_{2}
		-i[H_{loc,2} - J[(-iF  dt)a_2^\dagger + (iF  dt)a_2],\rho_{2}(dt)]dt\\
		& = |0\rangle_{2}\langle 0|_{2} +i J \left[ ((-iF  dt) \hat{a}_2^\dagger + (iF  dt) \hat{a}_2), |0\rangle\langle 0| \right]dt\\
		&= |0\rangle_{2}\langle 0|_{2}+ iJ(-iF  dt)  (|1\rangle_{2}\langle 0|_{2}+ |0\rangle_{2}\langle 1|_{2})dt\\
		&= |0\rangle_{2}\langle 0|_{2}+JF  (dt)^2  (|1\rangle_{2}\langle 0|_{2}+ |0\rangle_{2}\langle 1|_{2}),\\
		&\rho_{3}(2dt)=\rho_{3}(dt)+\mathcal{L}_{3}(\rho_{3}(dt))dt
		= |0\rangle_{3}\langle 0|_{3}-i[H_{loc,3}- J[(\Psi_2(dt) + \Psi_4(dt))\hat{a}_3^\dagger + h.c.],\rho_{3}(dt)]dt+\mathcal{D}_{3}[\rho_{3}(dt)]dt\\
		&= |0\rangle_{3}\langle 0|_{3}-i[H_{loc,3},\rho_{3}(dt)]dt+\mathcal{D}_{3}[\rho_{3}(dt)]dt
		=|0\rangle_{3}\langle 0|_{3},\\
		&\cdots,\\
		&\rho_L(2dt) \approx |0\rangle_{L}\langle 0|_{L},
	\end{aligned}
\end{equation}
where the dissipation rate $\gamma_1=\gamma_L=0$.
We use the following relations
\begin{equation} 
	\begin{aligned}
		&H_{loc,r}|0\rangle_r=\langle 0|_{r}H_{loc,r}=0,\ 
		H_{loc,r}|1\rangle_r=\Delta|1\rangle_r,\ 
		\langle 1|_{r}H_{loc,r}=\Delta \langle 1|_{r},
		H_{loc,r}|2\rangle_r=(2\Delta+U)|2\rangle_r,\\
		&[H_{loc,1},|0\rangle_1\langle 0|_1]=0,
		[F(a_{1}^{\dag}+a_{1}),|0\rangle_1\langle 0|_1]=
		F(|1\rangle_1\langle 0|_1-|0\rangle_1\langle 1|_1),\\
		&[H_{loc,1},-iFdt(|1\rangle_1\langle 0|_1-|0\rangle_1\langle 1|_1)]=-iFdt\Delta( |1\rangle_1\langle 0|_1+|0\rangle_1\langle 1|_1),\\
		& [F(\hat{a}^\dagger_1+\hat{a}_1), -iFdt (|1\rangle_{1}\langle 0|_{1} - |0\rangle_{1}\langle 1|_{1})]
		= -iF^2 dt \left[ 2|0\rangle_{1}\langle 0|_{1} - 2|1\rangle_{1}\langle 1|_{1} + \sqrt{2}\left( |2\rangle_{1}\langle 0|_{1} + |0\rangle_{1}\langle 2|_{1} \right) \right],\\
		&[H_{loc,1}+F(a_{1}^{\dag}+a_{1}),\rho_{1}(dt)]
		=F(|1\rangle_1\langle 0|_1-|0\rangle_1\langle 1|_1)
		-iFdt\Delta( |1\rangle_1\langle 0|_1+|0\rangle_1\langle 1|_1)\\
		&-iF^2 dt \left[ 2|0\rangle\langle 0| - 2|1\rangle\langle 1| + \sqrt{2}\left( |2\rangle\langle 0| + |0\rangle\langle 2| \right) \right],\\
		&H_{loc,r}|n\rangle_{r}=\Delta n|n\rangle_{r}+\frac{U}{2}n(n-1)|n\rangle_{r},\\
		&\langle n|_{r}H_{loc,r}=\Delta n\langle n|_{r}+\frac{U}{2}n(n-1)\langle n|_{r},
	\end{aligned}
\end{equation}
Note that even for $\gamma_{1}\neq 0$,
\begin{equation} 
	\begin{aligned}
		&	\mathcal{D}_{1}[\rho_{1}(dt)]
		=\gamma_{1}a_{1}\rho_{1}(dt)a^{\dag}_{1}
		-\frac{\gamma_{1}}{2}\{a_{1}^{\dag}a_{1},\rho_{1}(dt)\}\\
		&
		=-iFdt \gamma_1 |0\rangle\langle 1|
		-\frac{\gamma_{1}}{2}\{a_{1}^{\dag}a_{1},
		-iFdt \left( |1\rangle_{1}\langle 0|_{1} - |0\rangle_{1}\langle 1|_{1} \right)
		\}=	-iFdt \gamma_1 |0\rangle\langle 1|+
		\frac{iFdt \gamma_1}{2} 
		\left( |1\rangle_{1}\langle 0|_{1} - |0\rangle_{1}\langle 1|_{1} \right),\\
		&	\mathcal{D}_{2}[\rho_{2}(dt)]
		=\gamma_{2}a_{2}|0\rangle_{2}\langle 0|_{2}a^{\dag}_{2}
		-\frac{\gamma_{2}}{2}\{a_{2}^{\dag}a_{2},|0\rangle_{2}\langle 0|_{2}\}=0.
	\end{aligned}
\end{equation}
where 
$\{ a_1^\dagger a_1, |1\rangle\langle 0| \}= |1\rangle\langle 0|$,
$\{a^{\dag}_{1}a_{1},|0\rangle_{1}\langle 0|_{1}\}=|0\rangle_{1}\langle 0|_{1}$.

At $2dt$, the mean-field Hamiltonian read
\begin{equation} 
	\begin{aligned}
		&H_{MF,1}(2dt) = H_{loc,1} + F(a^\dagger_1 + a_1) - J[\Psi_2(2dt)a_1^\dagger + h.c.]\\
		&=H_{loc,1} + F(a^\dagger_1 + a_1) - J[JF(dt)^2 a_1^\dagger + h.c.]\\
		&=H_{loc,1} + F(a^\dagger_1 + a_1) - J^2 F(dt)^2( a_1^\dagger + a_1),\\
		&H_{MF,2}(2dt) = H_{loc,2} - J[(\Psi_1(2dt) + \Psi_3(2dt))a_2^\dagger + h.c.]\\
		&= H_{loc,2} - J[(-2iFdt - F\Delta(dt)^2)a_2^\dagger + h.c.],\\
		&H_{MF,3}(2dt) = H_{loc,3} - J \left[ (\Psi_2(2dt) + \Psi_4(2dt)) a_3^\dagger + h.c. \right]
		= H_{loc,3} - J^2F(dt)^2 (a_3^\dagger + a_3)
	\end{aligned}
\end{equation}
where the amplitudes read
$\Psi_1(2dt) = -2iFdt - F\Delta(dt)^2$, $\Psi_2(2dt) = JF(dt)^2$, $\Psi_3(2dt) = 0$, $\cdots$.

At time $2dt$, we have
\begin{equation} 
	\begin{aligned}
		\mathcal{D}_{1}[\hat{\rho}_1(2dt)] = &
		\mathcal{D}_{1}[	|0\rangle_{1}\langle 0|_{1} - 2iFdt (|1\rangle_{1}\langle 0|_{1} - |0\rangle_{1}\langle 1|_{1}) - F\Delta(dt)^2 (|1\rangle_{1}\langle 0|_{1} + |0\rangle_{1}\langle 1|_{1})\\
		& - F^2(dt)^2 \left[ 2|0\rangle_{1}\langle 0|_{1} - 2|1\rangle_{1}\langle 1|_{1} + \sqrt{2}(|2\rangle_{1}\langle 0|_{1} + |0\rangle_{1}\langle 2|_{1}) \right]]\\
		=&
		-2iFdt(-\frac{\gamma_{1}}{2}|1\rangle_{1}\langle 0|_{1}+\frac{\gamma_{1}}{2}|0\rangle_{1}\langle 1|_{1})\\
		&+
		\frac{F\Delta\gamma_1}{2}(dt)^2 (|1\rangle_1\langle 0|_1 + |0\rangle_1\langle 1|_1) \\
		&+ F^2\gamma_1(dt)^2 \left[ 2|0\rangle_1\langle 0|_1 - 2|1\rangle_1\langle 1|_1 + \sqrt{2}(|2\rangle_1\langle 0|_1 + |0\rangle_1\langle 2|_1) \right],\\
		[H_{MF,1}(2dt), \rho_1(2dt)] =&
		[H_{loc,1} + F(a^\dagger_1 + a_1) - J^2 F(dt)^2( a_1^\dagger + a_1), |0\rangle_{1}\langle 0|_{1} - 2iFdt (|1\rangle_{1}\langle 0|_{1}
		- |0\rangle_{1}\langle 1|_{1}) \\
		&- F\Delta(dt)^2 (|1\rangle_{1}\langle 0|_{1} + |0\rangle_{1}\langle 1|_{1})\\
		& - F^2(dt)^2 \left[ 2|0\rangle_{1}\langle 0|_{1} - 2|1\rangle_{1}\langle 1|_{1} + \sqrt{2}(|2\rangle_{1}\langle 0|_{1} + |0\rangle_{1}\langle 2|_{1}) \right]]\\
		&=(F - J^2F(dt)^2)(|1\rangle_1\langle 0|_1 - |0\rangle_1\langle 1|_1)-2iF\Delta dt(|1\rangle_1\langle 0|_1 + |0\rangle_1\langle 1|_1)\\
		& - 2iF^2dt(2|0\rangle_1\langle 0|_1 - 2|1\rangle_1\langle 1|_1 + \sqrt{2}(|2\rangle_1\langle 0|_1 + |0\rangle_1\langle 2|_1))\\
		&
		-F\Delta^2(dt)^2(|1\rangle_1\langle 0|_1 - |0\rangle_1\langle 1|_1)
		-4F^3 (dt)^2(|1\rangle_1\langle 0|_1 - |0\rangle_1\langle 1|_1) \\
		&- \sqrt{2}F^2(3\Delta+U)(dt)^2(|2\rangle_1\langle 0|_1 - |0\rangle_1\langle 2|_1)\\
		& - 6F^3(dt)^2(|1\rangle_1\langle 0|_1 - |0\rangle_1\langle 1|_1) + O(F^3(dt)^2)\\
		& + 2iF^2 J^2 (dt)^3(2|0\rangle_1\langle 0|_1 - 2|1\rangle_1\langle 1|_1 + \sqrt{2}(|2\rangle_1\langle 0|_1 + |0\rangle_1\langle 2|_1)),\\
		\mathcal{D}_2[\rho_2(2dt)]
		&= JF(dt)^2 \left( -\frac{\gamma_2}{2}|1\rangle_{2}\langle 0|_{2} - \frac{\gamma_2}{2}|0\rangle_{2}\langle 1|_{2} \right),\\
		[H_{MF,2}(2dt), \rho_2(2dt)] =&  \quad 2iJFdt (|1\rangle_2\langle 0|_2 + |0\rangle_2\langle 1|_2) \\
		& + 2JF\Delta(dt)^2 (|1\rangle_2\langle 0|_2 - |0\rangle_2\langle 1|_2) \\
		& + 2iJ^2F^2(dt)^3 (\sqrt{2}|2\rangle_2\langle 0|_2 + |1\rangle_2\langle 1|_2 - \sqrt{2}|0\rangle_2\langle 2|_2)+ O(dt^4),\\
		\mathcal{D}_3[\rho_3(2dt)]&=\mathcal{D}_3[|0\rangle_{3}\langle 0|_{3}]=0,\\
		[H_{MF,3}(2dt), \rho_3(2dt)] &= -J^2F(dt)^2 (|1\rangle_3\langle 0|_3 - |0\rangle_3\langle 1|_3),
	\end{aligned}
\end{equation}
where we use the following results
\begin{equation} 
	\begin{aligned}
		&[H_{loc,2}, |0\rangle_{2}\langle 0|_{2}] = 0,\\
		&[-J(-2iFdt - F\Delta(dt)^2) \hat{a}_2^\dagger+h.c.,|0\rangle_{2}\langle 0|_{2}]
		= J(2iFdt + F\Delta(dt)^2) |1\rangle_2\langle 0|_2 + J(2iFdt - F\Delta(dt)^2) |0\rangle_2\langle 1|_2,\\
		&[H_{loc,2},JF(dt)^2 (|1\rangle_2\langle 0|_2+ |0\rangle_2\langle 1|_2)]=JF(dt)^2 \Delta(|1\rangle_2\langle 0|_2- |0\rangle_2\langle 1|_2),\\
		&[-J(-2iFdt - F\Delta(dt)^2) \hat{a}_2^\dagger+h.c.,JF(dt)^2 (|1\rangle_2\langle 0|_2+ |0\rangle_2\langle 1|_2)]=\\
		&   J(2iFdt + F\Delta(dt)^2) |1\rangle_2\langle 0|_2 + J(2iFdt - F\Delta(dt)^2) |0\rangle_2\langle 1|_2  \\
		&  +  JF\Delta(dt)^2 (|1\rangle_2\langle 0|_2 - |0\rangle_2\langle 1|_2)   
		+  2iJ^2F^2(dt)^3 (\sqrt{2}|2\rangle_2\langle 0|_2 + |1\rangle_2\langle 1|_2 - \sqrt{2}|0\rangle_2\langle 2|_2)  + O(dt^4).
	\end{aligned}
\end{equation}
Then we can obtain the density and the related amplitude as
\begin{equation} 
	\begin{aligned}
		\rho_{1}(3dt) 
		&=\rho_1(2dt) + \mathcal{L}_1(\rho_1(2dt))  dt
		= \rho_1(2dt) + \left( -i[H_{MF,1}(2dt), \rho_1(2dt)] + \mathcal{D}_1[\rho_1(2dt)] \right)  dt\\
		&= |0\rangle_1\langle 0|_1 -3iFdt (|1\rangle_1\langle 0|_1 - |0\rangle_1\langle 1|_1)\\
		&+ \left(-3F\Delta(dt)^2 + iF\gamma_1(dt)^2\right) |1\rangle_1\langle 0|_1 + \left(-3F\Delta(dt)^2 - iF\gamma_1(dt)^2\right) |0\rangle_1\langle 1|_1\\
		&		- 3F^2(dt)^2 \left[ 2|0\rangle_1\langle 0|_1 - 2|1\rangle_1\langle 1|_1 + \sqrt{2}(|2\rangle_1\langle 0|_1 + |0\rangle_1\langle 2|_1) \right]\\
		&		+ \left(iF\Delta^2 + 6iF^3 + iJ^2F + \frac{\gamma_1 F\Delta}{2}\right)(dt)^3 |1\rangle_1\langle 0|_1\\
		&		+ \left(-iF\Delta^2 - 6iF^3 - iJ^2F + \frac{\gamma_1 F\Delta}{2}\right)(dt)^3 |0\rangle_1\langle 1|_1\\
		&+ i\sqrt{2}F^2(3\Delta+U)(dt)^3(|2\rangle_1\langle 0|_1 - |0\rangle_1\langle 2|_1)\\
		&+ \gamma_1 F^2(dt)^3 (2|0\rangle_1\langle 0|_1 - 2|1\rangle_1\langle 1|_1 + \sqrt{2}(|2\rangle_1\langle 0|_1 + |0\rangle_1\langle 2|_1)),\\
		\Psi_1(3dt)& =\text{Tr}[a_1 \rho_1(3dt)]=  -3iFdt  + (-3F\Delta + iF\gamma_1)(dt)^2 \\
		& + \left( i(J^2F + F\Delta^2 + 
		6F^3) + \frac{\gamma_1 F\Delta}{2}  \right) (dt)^3,\\
		\rho_2(3dt)& = \rho_2(2dt) + \mathcal{L}_2(\rho_2(2dt)) dt
		=\rho_2(2dt) + \left( -i[H_{MF,2}(2dt), \rho_2(2dt)] + \mathcal{D}_2[\rho_2(2dt)] \right)  dt\\
		&=  |0\rangle_2\langle 0|_2 
		+ 3JF(dt)^2 (|1\rangle_2\langle 0|_2 + |0\rangle_2\langle 1|_2) \\
		& + \left( -2iJF\Delta - \frac{\gamma_2 JF}{2} \right) (dt)^3 |1\rangle_2\langle 0|_2 + \left(2iJF\Delta - \frac{\gamma_2 JF}{2} \right) (dt)^3 |0\rangle_2\langle 1|_2,\\
		\Psi_2(3dt)& =\text{Tr}[a_2 \rho_2(3dt)]= 3JF(dt)^2 + \left(-2iJF\Delta - \frac{\gamma_2 JF}{2}\right) (dt)^3,\\
		\rho_3(3dt) &= \rho_3(2dt)+\mathcal{L}_3(\rho_3(2dt))  dt
		= |0\rangle_3\langle 0|_3 - i[H_{loc,3} - J^2F(dt)^2 (a_3^\dagger + a_3), |0\rangle_3\langle 0|_3]  dt\\
		&= |0\rangle_3\langle 0|_3 +iJ^2F(dt)^2 (|1\rangle_3\langle 0|_3 - |0\rangle_3\langle 1|_3)  dt,\\
		\Psi_3(3dt) &= \text{Tr}[a_3 \rho_3(3dt)]= iJ^2F(dt)^3.
	\end{aligned}
\end{equation}
Thus $\Psi_3$ is non-zero since $3dt$, and will next activate the Hamiltonian $H_{MF,4}$ and create a non-zero $\Psi_4$.

At time $3dt$, the mean-field Hamiltonians are
\begin{equation} 
	\begin{aligned}
		&H_{MF,1}(3dt) = H_{loc,1} + F(\hat{a}^\dagger_1 + \hat{a}_1) - J[\Psi_2(3dt) \hat{a}_1^\dagger + h.c.],\\
		&H_{MF,2}(3dt) =H_{loc,2} - J[(\Psi_1(3dt) + \Psi_3(3dt)) \hat{a}_2^\dagger + h.c.],\\
		&H_{MF,3}(3dt) =H_{loc,3} - J[ \Psi_3(3dt) \hat{a}_3^\dagger + h.c.]
	\end{aligned}
\end{equation}
the corresponding dissipators and commutators read 
\begin{equation} 
	\begin{aligned}
		&\mathcal{D}_1[\rho_1(3dt)]
		= i\frac{3\gamma_1 F}{2} dt (|1\rangle_1\langle 0|_1 - |0\rangle_1\langle 1|_1)
		+ \left(\frac{3\gamma_1 F\Delta}{2} - \frac{i\gamma_1^2 F}{2}\right)(dt)^2 |1\rangle_1\langle 0|_1 \\
		&+ \left(\frac{3\gamma_1 F\Delta}{2} + \frac{i\gamma_1^2 F}{2}\right)(dt)^2 |0\rangle_1\langle 1|_1\\
		&+ 3\gamma_1 F^2(dt)^2 \left[ 2|0\rangle_1\langle 0|_1 - 2|1\rangle_1\langle 1|_1 + \sqrt{2}(|2\rangle_1\langle 0|_1 + |0\rangle_1\langle 2|_1) \right]\\
		&+ \left(-\frac{i\gamma_1}{2}(F\Delta^2 + 6F^3 + J^2F) - \frac{\gamma_1^2 F\Delta}{4}\right)(dt)^3 |1\rangle_1\langle 0|_1
		+ \left(\frac{i\gamma_1}{2}(F\Delta^2 + 6F^3 + J^2F) - \frac{\gamma_1^2 F\Delta}{4}\right)(dt)^3 |0\rangle_1\langle 1|_1\\
		&+ \left(-i\sqrt{2}\gamma_1 F^2(3\Delta+U) - \sqrt{2}\gamma_1^2 F^2 \right)(dt)^3 |2\rangle_1\langle 0|_1\\
		&+ \left(i\sqrt{2}\gamma_1 F^2(3\Delta+U) - \sqrt{2}\gamma_1^2 F^2 \right)(dt)^3 |0\rangle_1\langle 2|_1
		- 2\gamma_1^2 F^2(dt)^3 (|0\rangle_1\langle 0|_1 - |1\rangle_1\langle 1|_1),\\
		&[H_{MF,1}(3dt), \rho_1(3dt)]= 
		F(|1\rangle_1\langle 0|_1 - |0\rangle_1\langle 1|_1)
		-3iF\Delta dt(|1\rangle_1\langle 0|_1 + |0\rangle_1\langle 1|_1)\\
		& - 3iF^2dt \left( 2|0\rangle_1\langle 0|_1 - 2|1\rangle_1\langle 1|_1 + \sqrt{2}(|2\rangle_1\langle 0|_1 + |0\rangle_1\langle 2|_1) \right)\\
		&
		-3J^2F(dt)^2(|1\rangle_1\langle 0|_1 - |0\rangle_1\langle 1|_1)
		-J\left(-2iJF\Delta - \frac{\gamma_2 JF}{2}\right)(dt)^3 |1\rangle_1\langle 0|_1 + J\left(2iJF\Delta - \frac{\gamma_2 JF}{2}\right)(dt)^3 |0\rangle_1\langle 1|_1,\\
		&\mathcal{D}_2[\rho_2(3dt)]
		= -\frac{3\gamma_2 JF}{2}(dt)^2 (|1\rangle_2\langle 0|_2 + |0\rangle_2\langle 1|_2)\\
		&+ \left( i\gamma_2 JF\Delta + \frac{\gamma_2^2 JF}{4} \right) (dt)^3 |1\rangle_2\langle 0|_2 + \left( -i\gamma_2 JF\Delta + \frac{\gamma_2^2 JF}{4} \right) (dt)^3 |0\rangle_2\langle 1|_2,\\
		& [H_{MF,2}(3dt), \rho_2(3dt)]= 3iJFdt (|1\rangle_2\langle 0|_2 + |0\rangle_2\langle 1|_2)
		+(6JF\Delta - iJF\gamma_1)(dt)^2 |1\rangle_2\langle 0|_2 \\
		&+ (-6JF\Delta - iJF\gamma_1)(dt)^2 |0\rangle_2\langle 1|_2
		+ J(i(2J^2F + F\Delta^2 + 6F^3) + \frac{\gamma_1 F\Delta}{2})(dt)^3 |1\rangle\langle 0| \\
		&+ J(-i(2J^2F + F\Delta^2 + 6F^3) + \frac{\gamma_1 F\Delta}{2})(dt)^3 |0\rangle\langle 1|,\\
		&\mathcal{D}_3[\rho_3(3dt)] = -i\frac{\gamma_3 J^2F}{2}(dt)^3 (|1\rangle_3\langle 0|_3 - |0\rangle_3\langle 1|_3),\\
		& [H_{MF,3}(3dt), \rho_3(3dt)] = [-J(3JF(dt)^2 \hat{a}_3^\dagger + \text{H.c.}), |0\rangle\langle 0|]\\
		& = -3J^2F(dt)^2 (|1\rangle_3\langle 0|_3 - |0\rangle_3\langle 1|_3)
		+ J(-2iJF\Delta - \frac{\gamma_2 JF}{2})(dt)^3 |1\rangle_3\langle 0|_3 - J(2iJF\Delta - \frac{\gamma_2 JF}{2})(dt)^3 |0\rangle_3\langle 1|_3,
	\end{aligned}
\end{equation}
and we can then obtain the amplitudes of the next time step
\begin{equation} 
	\begin{aligned}
		&\Psi_1(4dt) = -4iFdt + (-6F\Delta + i\frac{5\gamma_1 F}{2})(dt)^2 + \left(i(4J^2F + 4F\Delta^2 + 12F^3) + 3\gamma_1 F\Delta -\frac{i\gamma_{1}^2 F}{2}\right)(dt)^3,\\
		&\Psi_2(4dt) = 6JF(dt)^2 + \left(-8iJF\Delta - JF\gamma_1 - 2\gamma_2 JF\right) (dt)^3,\\
		&\Psi_3(4dt) = 4iJ^2F(dt)^3.
	\end{aligned}
\end{equation}

\section{Appendix.B: Green's Functions and Liouvillian Spectrum}

In the limit of strong driving and dissipation, the standard Hamiltonian formalism is insufficient to describe the collective response. We treat the lattice problem by first solving the single-site effective Hamiltonians self-consistently coupled to their neighbors, governed by the non-Hermitian Liouvillian superoperator $\mathcal{L}$. The many-body renormalization effects are then incorporated via a diagrammatic T-matrix expansion
base on the NESS.

In the mean-field approximation, the lattice problem is decoupled into single-site effective Hamiltonians self-consistently coupled to their neighbors. Unlike equilibrium systems governed by Hermitian Hamiltonians, the local physics here is governed by the non-Hermitian Liouvillian superoperator $\mathcal{L}$.

The linear response of the system to a weak probe field is encoded in the local retarded Green's function\cite{Wang}. In the frequency domain, this can be expressed using the resolvent of the effective Liouvillian superoperator
\begin{equation} 
	\label{eq:resolvent}
	\begin{aligned}
		G^{R}_{loc}(\omega) &= \text{Tr} \left[ \hat{a} \frac{1}{-i\omega - \mathcal{L}_{eff}} (\hat{a}^\dagger \hat{\rho}_{SS}) \right] - \text{Tr} \left[ \hat{a}^\dagger \frac{1}{-i\omega - \mathcal{L}_{eff}} (\hat{a} \hat{\rho}_{SS}) \right]^\dagger.
	\end{aligned}
\end{equation}
The effective Liouvillian superoperator $\mathcal{L}_{eff}$,
 which governs the time evolution of fluctuations $\delta \hat{\rho}$ linearized around the NESS density matrix $\hat{\rho}_{SS}$. By utilizing the spectral decomposition of $\mathcal{L}_{eff} = \sum_{\alpha} \lambda_{\alpha} | \breve{\rho}_{\alpha} \rangle \langle \breve{O}_{\alpha} |$, where $\lambda_{\alpha} = \Gamma_{\alpha} - i\omega_{\alpha}$ are the complex eigenvalues (with $\Gamma_{\alpha} \le 0$ for stability), the Green's function is
\begin{equation} 
	\label{eq:resolvent_detailed}
	G^{R}_{loc}(\omega) = \sum_{\alpha} \left[ \frac{\text{Tr}(\hat{a} \breve{\rho}_{\alpha}) \text{Tr}(\breve{O}_{\alpha} \hat{a}^\dagger \hat{\rho}_{SS})}{-i\omega - \lambda_{\alpha}} - \frac{\text{Tr}(\hat{a}^\dagger \breve{\rho}_{\alpha})^* \text{Tr}(\breve{O}_{\alpha} \hat{a} \hat{\rho}_{SS})^*}{i\omega - \lambda_{\alpha}^*} \right].
\end{equation}
The poles of $G^{R}_{loc}(\omega)$ identify the elementary excitations of the NESS. 
and correspond to the complex eigenvalues $\lambda_{\alpha} = \Gamma_\alpha - i\omega_\alpha$ of $\mathcal{L}_{eff}$.


The spectral function (local density of states) is defined as $\rho(\omega) = -\frac{1}{\pi} \text{Im} \left[ G^{R}_{loc}(\omega) \right]$. For the Bose-Hubbard model with Kerr nonlinearity $U$ and drive $F$, the spectral function exhibits distinct features depending on the ratio $F/\gamma$.
The spectral function $\rho(\omega) = -\frac{1}{\pi} \text{Im} [ G^{R}_{loc}(\omega) ]$ characterizes the energy distribution of different modes.

In linear regime with weak drive $F \ll \gamma, U$,
the system remains close to the vacuum state. The non-linearity is negligible, and the Liouvillian spectrum is dominated by the single-particle loss $\gamma \mathcal{D}[\hat{a}]$, leading to a single pole at $\lambda = -\gamma/2 - i\Delta$. This yields the standard Lorentzian spectral function with peak centered at the detuning $\Delta$,
\begin{equation}
	\rho(\omega) \approx \frac{1}{\pi} \frac{\gamma/2}{(\omega - \Delta)^2 + (\gamma/2)^2}.
\end{equation}
In non-linear regime with strong drive $F \gg \gamma$,
as the drive increases, the photon number $n$ grows, and the interaction term $U \hat{n}(\hat{n}-1)$ becomes significant. The system behaves like a driven anharmonic oscillator. The excitation spectrum splits due to the dressing of energy levels by the drive field, leading to a multi-peak structure analogous to the mollow triplet. 

  As the coherent amplitude $\langle \hat{a} \rangle$ increases, the Kerr term $U \hat{n}(\hat{n}-1)$ induces energy-level transitions between dressed states. The Liouvillian spectrum splits into multiple branches. The resulting spectral function exhibits a mollow-triplet-like structure, where the spectral function can be approximated as a sum of resonances,
\begin{equation}
	\rho(\omega) \approx \sum_{j \in \{-1, 0, 1\}} \frac{A_j \Gamma_j}{(\omega - [\tilde{\omega}_0 + j \Omega_{R}])^2 + \Gamma_j^2},
\end{equation}
where $\tilde{\omega}_0 = \Delta + 2U \langle n \rangle$ is the Stark-shifted central frequency reflects the Stark-shifted resonance and $\Omega_{R} \approx 2F \sqrt{\langle n \rangle}$ is the vacuum Rabi splitting (effective Rabi frequency) induced by the drive acting on the non-linear medium.

The spatial coupling $J$ is incorporated by treating the hopping term as a perturbation on the local NESS.
To analyze the stability of the spatial structure, we use the random phase approximation (RPA), and assume the fluctuations at different sites are coupled only via the mean-field dispersion.
 The lattice retarded Green's function is related to the local Green's function via the Dyson equation
\begin{equation} 
	\begin{aligned}
		\label{latticeGreen}
		[G_{latt}^R(k, \omega)]^{-1} = [G^{(0)}_{loc}(\omega)]^{-1} - \epsilon_k,
	\end{aligned}
\end{equation}
where $\epsilon_k = -2J \cos(k)$ is the lattice dispersion relation (in 1D).
This formulation maps the local Liouvillian poles onto collective bands in $k$-space. The dispersion of these collective modes is determined by the condition $\text{Re}[G_{loc}^{-1}(\omega)] = \epsilon_k$, while the imaginary part of $G_{loc}^{-1}$ provides the damping across the Brillouin zone.
 Note that $\epsilon_k$ is simply the momentum-space representation (Fourier transform) of the nearest-neighbor hopping term, necessary here to analytically diagonalize the spatial degrees of freedom.

Beyond the coherent propagation described by RPA, we account for the residual scattering between collective excitations. Adapting the T-matrix (ladder) approximation from polaron theory, we define the scattering of two excitations with total momentum $Q$ and total frequency $\Omega$.
The non-equilibrium pair propagator (driven polarization bubble) $\Pi(Q, \Omega)$ can be obtained by the lattice Green's functions. In the frequency domain, this involves a convolution that accounts for the dissipative character of NESS
\begin{equation}
	\Pi(Q, \Omega) = \frac{i}{2\pi} \int d\omega' \frac{1}{L} \sum_{k} G_{latt}^R(k, \omega') G_{latt}^R(Q-k, \Omega-\omega').
\end{equation}
Unlike equilibrium systems, the integration path is determined by the analytic structure of $G_{latt}^R$ in complex plane, where poles reside in the lower half-plane due to $\Gamma_{\alpha} < 0$.
The interaction is dressed by multiple scattering events to form the renormalized T-matrix $\mathcal{T}(Q, \Omega)$:
\begin{equation}
	\mathcal{T}(Q, \Omega) = \frac{U}{1 - U \Pi(Q, \Omega)}.
\end{equation}
Divergences in $\mathcal{T}(Q, \Omega)$ signal the formation of repulsively bound pairs (doublons) in the lattice. In the driven-dissipative case, the condition $1 = U \text{Re}[\Pi(Q, \Omega)]$ determines the energy of these pairs, while $\text{Im}[\Pi]$ determines their finite lifetime.
The final renormalization of the single-particle excitation spectrum is given by the ladder self-energy $\Sigma_{ladder}$:
\begin{equation}
	\Sigma_{ladder}(k, \omega) = \int \frac{d\Omega}{2\pi} \frac{1}{L} \sum_{q} G_{latt}^R(q-k, \Omega-\omega) \mathcal{T}(q, \Omega).
\end{equation}
The total lattice response is then governed by the fully renormalized Green's function $\tilde{G}(k, \omega) = [G_{latt}^R(k, \omega)^{-1} - \Sigma_{ladder}(k, \omega)]^{-1}$. A key consequence of this calculation is the modification of the spectral weight $Z_k$ and the introduction of collision-induced broadening. If $\text{Im}[\Sigma_{ladder}]$ exceeds the drive-induced gain in specific momentum sectors, the spatial NESS becomes unstable, leading to a breakdown of the homogeneous phase.

		\renewcommand\refname{References}

		\end{small}
	\end{document}